\pdfoutput=1

\documentclass[12pt,a4paper]{article}

\usepackage{ifthen}
\newboolean{pdflatex}
\setboolean{pdflatex}{true}

\newboolean{articletitles}
\setboolean{articletitles}{true}

\newboolean{uprightparticles}
\setboolean{uprightparticles}{false}

\newboolean{inbibliography}
\setboolean{inbibliography}{false}

\usepackage{microtype}
\usepackage{lineno}
\usepackage{xspace}
\usepackage{caption}

\usepackage{graphicx}
\usepackage{color}
\usepackage{colortbl}

\usepackage{amsmath}
\usepackage{amssymb}
\usepackage{amsfonts}
\usepackage{upgreek}

\newcommand*\patchAmsMathEnvironmentForLineno[1]{
\expandafter\let\csname old#1\expandafter\endcsname\csname #1\endcsname
\expandafter\let\csname oldend#1\expandafter\endcsname\csname
end#1\endcsname
 \renewenvironment{#1}
   {\linenomath\csname old#1\endcsname}
   {\csname oldend#1\endcsname\endlinenomath}
}
\newcommand*\patchBothAmsMathEnvironmentsForLineno[1]{
  \patchAmsMathEnvironmentForLineno{#1}
  \patchAmsMathEnvironmentForLineno{#1*}
}
\AtBeginDocument{
\patchBothAmsMathEnvironmentsForLineno{equation}
\patchBothAmsMathEnvironmentsForLineno{align}
\patchBothAmsMathEnvironmentsForLineno{flalign}
\patchBothAmsMathEnvironmentsForLineno{alignat}
\patchBothAmsMathEnvironmentsForLineno{gather}
\patchBothAmsMathEnvironmentsForLineno{multline}
\patchBothAmsMathEnvironmentsForLineno{eqnarray}
}

\usepackage{hyperref}
\usepackage[all]{hypcap}

\usepackage{xspace}
\usepackage{upgreek}

\def\lhcb {\mbox{LHCb}\xspace}

\def\MagUp {\mbox{\em Mag\kern -0.05em Up}\xspace}

\ifthenelse{\boolean{uprightparticles}}
{

 \def\Ppi         {\ensuremath{\uppi}\xspace}

 \def\PDelta      {\ensuremath{\Delta}\xspace}
 \def\PXi      {\ensuremath{\Xi}\xspace}
 \def\PLambda      {\ensuremath{\Lambda}\xspace}
 \def\PSigma      {\ensuremath{\Sigma}\xspace}
 \def\POmega      {\ensuremath{\Omega}\xspace}
 \def\PUpsilon      {\ensuremath{\Upsilon}\xspace}

 \def\PB      {\ensuremath{\mathrm{B}}\xspace}
 
 \def\PD      {\ensuremath{\mathrm{D}}\xspace}

 \def\PK      {\ensuremath{\mathrm{K}}\xspace}

 \def\Pb      {\ensuremath{\mathrm{b}}\xspace}
 \def\Pc      {\ensuremath{\mathrm{c}}\xspace}

 \def\Pi      {\ensuremath{\mathrm{i}}\xspace}

 \def\Pp      {\ensuremath{\mathrm{p}}\xspace}

 \def\Ps      {\ensuremath{\mathrm{s}}\xspace}
 
 \def\Pu      {\ensuremath{\mathrm{u}}\xspace}

}
{

 \def\Ppi         {\ensuremath{\pi}\xspace}

 \mathchardef\PDelta="7101
 \mathchardef\PXi="7104
 \mathchardef\PLambda="7103
 \mathchardef\PSigma="7106
 \mathchardef\POmega="710A
 \mathchardef\PUpsilon="7107
 
 \def\PB      {\ensuremath{B}\xspace}
 
 \def\PD      {\ensuremath{D}\xspace}

 \def\PK      {\ensuremath{K}\xspace}

 \def\Pb      {\ensuremath{b}\xspace}
 \def\Pc      {\ensuremath{c}\xspace}

 \def\Pi      {\ensuremath{i}\xspace}

 \def\Pp      {\ensuremath{p}\xspace}

 \def\Ps      {\ensuremath{s}\xspace}
 
 \def\Pu      {\ensuremath{u}\xspace}

}

\makeatletter
\ifcase \@ptsize \relax
  \newcommand{\miniscule}{\@setfontsize\miniscule{4}{5}}
\or
  \newcommand{\miniscule}{\@setfontsize\miniscule{5}{6}}
\or
  \newcommand{\miniscule}{\@setfontsize\miniscule{5}{6}}
\fi
\makeatother

\DeclareRobustCommand{\optbar}[1]{\shortstack{{\miniscule (\rule[.5ex]{1.25em}{.18mm})}
  \\ [-.7ex] $#1$}}

\def\uquark    {{\ensuremath{\Pu}}\xspace}

\def\squark    {{\ensuremath{\Ps}}\xspace}

\def\cquark    {{\ensuremath{\Pc}}\xspace}

\def\bquark    {{\ensuremath{\Pb}}\xspace}

\def\pion   {{\ensuremath{\Ppi}}\xspace}
\def\piz    {{\ensuremath{\pion^0}}\xspace}

\def\pip    {{\ensuremath{\pion^+}}\xspace}
\def\pim    {{\ensuremath{\pion^-}}\xspace}
\def\pipm   {{\ensuremath{\pion^\pm}}\xspace}
\def\pimp   {{\ensuremath{\pion^\mp}}\xspace}

\def\kaon    {{\ensuremath{\PK}}\xspace}
  \def\Kbar    {{\kern 0.2em\overline{\kern -0.2em \PK}{}}\xspace}

\def\KorKbar    {\kern 0.18em\optbar{\kern -0.18em K}{}\xspace}

\def\Kp      {{\ensuremath{\kaon^+}}\xspace}
\def\Km      {{\ensuremath{\kaon^-}}\xspace}
\def\Kpm     {{\ensuremath{\kaon^\pm}}\xspace}

\def\KS      {{\ensuremath{\kaon^0_{\mathrm{ \scriptscriptstyle S}}}}\xspace}

\def\Kstarz  {{\ensuremath{\kaon^{*0}}}\xspace}
\def\Kstarzb {{\ensuremath{\Kbar{}^{*0}}}\xspace}
\def\Kstar   {{\ensuremath{\kaon^*}}\xspace}
\def\Kstarb  {{\ensuremath{\Kbar{}^*}}\xspace}

\def\Kstarzero   {{\ensuremath{\kaon^*}_0}\xspace}

\def\Kstartwo   {{\ensuremath{\kaon^*}_2}\xspace}

  \def\Dbar    {{\kern 0.2em\overline{\kern -0.2em \PD}{}}\xspace}
\def\D       {{\ensuremath{\PD}}\xspace}

\def\DorDbar {\kern 0.18em\optbar{\kern -0.18em D}{}\xspace}
\def\DorDbarz{\kern 0.18em\optbar{\kern -0.18em D}{}^0\xspace}
\def\Dz      {{\ensuremath{\D^0}}\xspace}
\def\Dzb     {{\ensuremath{\Dbar{}^0}}\xspace}

\def\Dm      {{\ensuremath{\D^-}}\xspace}

\def\Dmp     {{\ensuremath{\D^\mp}}\xspace}
\def\Dstar   {{\ensuremath{\D^*}}\xspace}

\def\Dstarz  {{\ensuremath{\D^{*0}}}\xspace}

\def\DorDstar   {{\ensuremath{\D^{(*)}}}\xspace}

\def\DorDstarzb {{\ensuremath{\Dbar{}^{(*)0}}}\xspace}

\def\Dsm     {{\ensuremath{\D^-_\squark}}\xspace}

\def\B       {{\ensuremath{\PB}}\xspace}
\def\Bbar    {{\ensuremath{\kern 0.18em\overline{\kern -0.18em \PB}{}}}\xspace}

\def\BorBbar    {\kern 0.18em\optbar{\kern -0.18em B}{}\xspace}
\def\Bz      {{\ensuremath{\B^0}}\xspace}
\def\Bzb     {{\ensuremath{\Bbar{}^0}}\xspace}
\def\Bu      {{\ensuremath{\B^+}}\xspace}

\def\Bp      {{\ensuremath{\Bu}}\xspace}

\def\Bpm     {{\ensuremath{\B^\pm}}\xspace}

\def\Bd      {{\ensuremath{\B^0}}\xspace}
\def\Bs      {{\ensuremath{\B^0_\squark}}\xspace}
\def\Bds     {{\ensuremath{\B^0_{(\squark)}}}\xspace}
\def\Bsb     {{\ensuremath{\Bbar{}^0_\squark}}\xspace}

  \def\Y#1S{\ensuremath{\PUpsilon{(#1S)}}\xspace}

\def\proton      {{\ensuremath{\Pp}}\xspace}
\def\antiproton  {{\ensuremath{\overline \proton}}\xspace}

\def\Lz          {{\ensuremath{\PLambda}}\xspace}
\def\Lbar        {{\ensuremath{\kern 0.1em\overline{\kern -0.1em\PLambda}}}\xspace}
\def\LorLbar    {\kern 0.18em\optbar{\kern -0.18em \PLambda}{}\xspace}

\def\Lb      {{\ensuremath{\Lz^0_\bquark}}\xspace}
\def\Lbbar   {{\ensuremath{\Lbar{}^0_\bquark}}\xspace}

\def\to                 {\ensuremath{\rightarrow}\xspace}

\def\CP                {{\ensuremath{C\!P}}\xspace}

\def\AT#1     {\ensuremath{A_{\mathrm{T}}^{#1}}\xspace}

\def\C#1      {\ensuremath{\mathcal{C}_{#1}}\xspace}
\def\Cp#1     {\ensuremath{\mathcal{C}_{#1}^{'}}\xspace}
\def\Ceff#1   {\ensuremath{\mathcal{C}_{#1}^{\mathrm{(eff)}}}\xspace}
\def\Cpeff#1  {\ensuremath{\mathcal{C}_{#1}^{'\mathrm{(eff)}}}\xspace}
\def\Ope#1    {\ensuremath{\mathcal{O}_{#1}}\xspace}
\def\Opep#1   {\ensuremath{\mathcal{O}_{#1}^{'}}\xspace}

\newcommand{\tev}{\ifthenelse{\boolean{inbibliography}}{\ensuremath{~T\kern -0.05em eV}\xspace}{\ensuremath{\mathrm{\,Te\kern -0.1em V}}}\xspace}
\newcommand{\gev}{\ensuremath{\mathrm{\,Ge\kern -0.1em V}}\xspace}
\newcommand{\mev}{\ensuremath{\mathrm{\,Me\kern -0.1em V}}\xspace}
\newcommand{\nspgev}{\ensuremath{\mathrm{Ge\kern -0.1em V}}\xspace}
\newcommand{\nspmev}{\ensuremath{\mathrm{Me\kern -0.1em V}}\xspace}
\newcommand{\kev}{\ensuremath{\mathrm{\,ke\kern -0.1em V}}\xspace}
\newcommand{\ev}{\ensuremath{\mathrm{\,e\kern -0.1em V}}\xspace}
\newcommand{\gevc}{\ensuremath{{\mathrm{\,Ge\kern -0.1em V\!/}c}}\xspace}
\newcommand{\mevc}{\ensuremath{{\mathrm{\,Me\kern -0.1em V\!/}c}}\xspace}
\newcommand{\gevcc}{\ensuremath{{\mathrm{\,Ge\kern -0.1em V\!/}c^2}}\xspace}
\newcommand{\nspmevcc}{\ensuremath{{\mathrm{Me\kern -0.1em V\!/}c^2}}\xspace}
\newcommand{\nspgevcc}{\ensuremath{{\mathrm{Ge\kern -0.1em V\!/}c^2}}\xspace}
\newcommand{\gevgevcccc}{\ensuremath{{\mathrm{\,Ge\kern -0.1em V^2\!/}c^4}}\xspace}
\newcommand{\mevcc}{\ensuremath{{\mathrm{\,Me\kern -0.1em V\!/}c^2}}\xspace}

\def\mum  {\ensuremath{{\,\upmu\mathrm{m}}}\xspace}

\def\invfb   {\ensuremath{\mbox{\,fb}^{-1}}\xspace}

\newcommand{\chisq}{\ensuremath{\chi^2}\xspace}

\def\gsim{{~\raise.15em\hbox{$>$}\kern-.85em
          \lower.35em\hbox{$\sim$}~}\xspace}
\def\lsim{{~\raise.15em\hbox{$<$}\kern-.85em
          \lower.35em\hbox{$\sim$}~}\xspace}

\def\sPlot{\mbox{\em sPlot}\xspace}

\def\ptot       {\mbox{$p$}\xspace}
\def\pt         {\mbox{$p_{\mathrm{ T}}$}\xspace}

\def\evtgen     {\mbox{\textsc{EvtGen}}\xspace}

\def\geant      {\mbox{\textsc{Geant4}}\xspace}

\def\photos     {\mbox{\textsc{Photos}}\xspace}

\def\pythia     {\mbox{\textsc{Pythia}}\xspace}

\def\tell1  {TELL1\xspace}
\def\ukl1   {UKL1\xspace}

\newcommand{\ie}{\mbox{\itshape i.e.}\xspace}

\usepackage{cite}
\usepackage{mciteplus}

\usepackage{amsfonts}

\usepackage{pifont}

\begin{document}

\renewcommand{\thefootnote}{\fnsymbol{footnote}}
\setcounter{footnote}{1}

\begin{titlepage}
\pagenumbering{roman}

\vspace*{-1.5cm}
\centerline{\large EUROPEAN ORGANIZATION FOR NUCLEAR RESEARCH (CERN)}
\vspace*{1.5cm}
\noindent
\begin{tabular*}{\linewidth}{lc@{\extracolsep{\fill}}r@{\extracolsep{0pt}}}
\vspace*{-2.7cm}\mbox{\!\!\!\includegraphics[width=.14\textwidth]{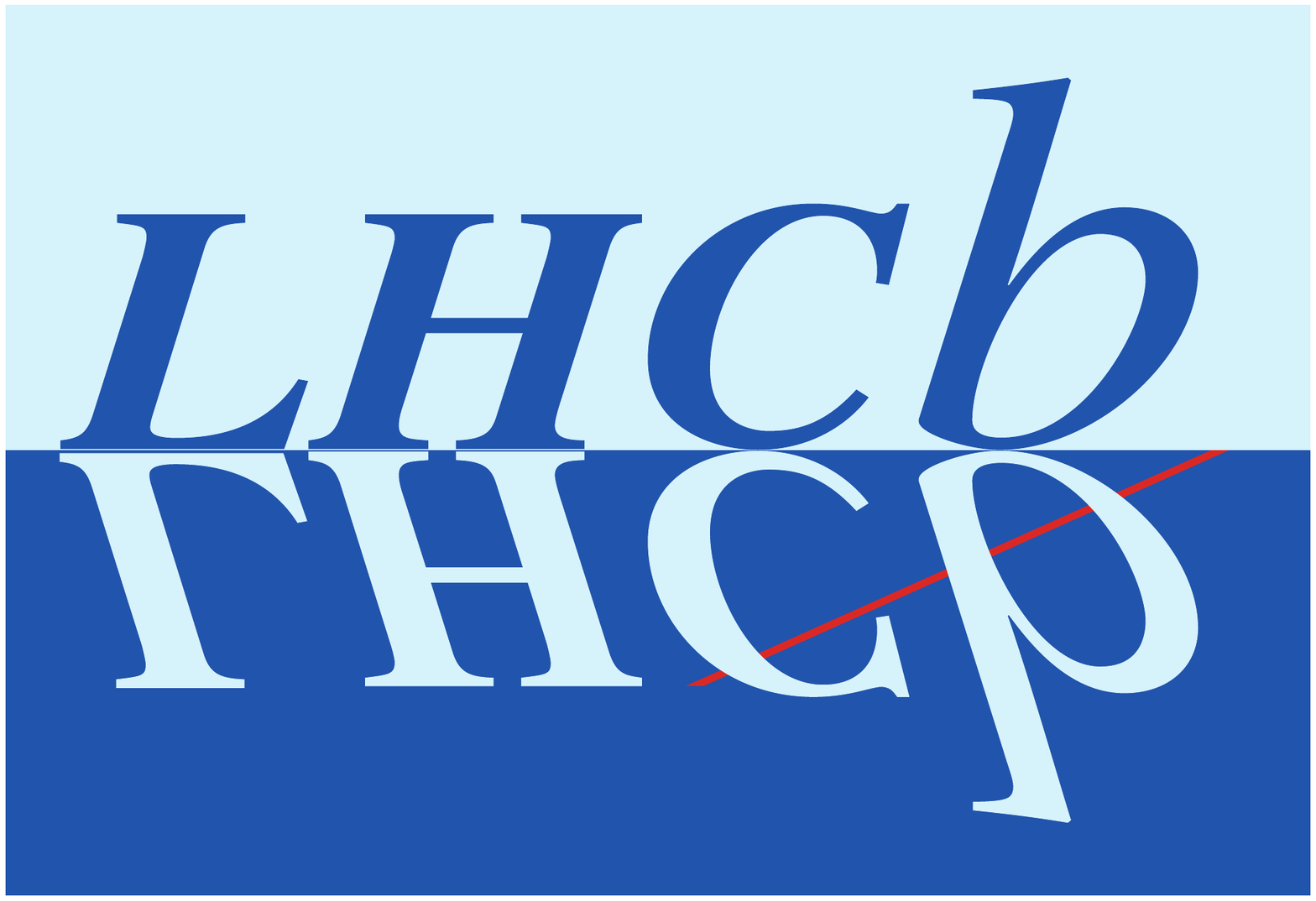}} & &
\\
 & & CERN-EP-2016-024 \\
 & & LHCb-PAPER-2015-059 \\
 & & 11 October 2016 \\
 & & \\
\end{tabular*}

\vspace*{2.0cm}

{\normalfont\bfseries\boldmath\huge
\begin{center}
  Constraints on the unitarity triangle angle $\gamma$ from Dalitz plot analysis of $\Bz \to D\Kp\pim$ decays
\end{center}
}

\vspace*{1.0cm}

\begin{center}
The LHCb collaboration\footnote{Authors are listed at the end of this paper.}
\end{center}

\vspace{\fill}

\begin{abstract}
  \noindent
  The first study is presented of \CP violation with an amplitude analysis of the Dalitz plot of $\Bz \to D\Kp\pim$ decays, with $\D\to\Kp\pim$, $\Kp\Km$ and $\pip\pim$.
  The analysis is based on a data sample corresponding to $3.0\invfb$ of $pp$ collisions collected with the LHCb detector.
  No significant \CP violation effect is seen, and constraints are placed on the angle $\gamma$ of the unitarity triangle formed from elements of the Cabibbo-Kobayashi-Maskawa quark mixing matrix.
  Hadronic parameters associated with the $\Bz \to D\Kstar(892)^0$ decay are determined for the first time.
  These measurements can be used to improve the sensitivity to $\gamma$ of existing and future studies of the $\Bz \to D\Kstar(892)^0$ decay.
\end{abstract}

\vspace*{1.0cm}

\begin{center}
  Submitted to Phys.~Rev.~D.
\end{center}

\vspace{\fill}

{\footnotesize
\centerline{\copyright~CERN on behalf of the \lhcb collaboration, licence \href{http://creativecommons.org/licenses/by/4.0/}{CC-BY-4.0}.}}
\vspace*{2mm}

\end{titlepage}

\newpage
\setcounter{page}{2}
\mbox{~}

\cleardoublepage

\renewcommand{\thefootnote}{\arabic{footnote}}
\setcounter{footnote}{0}

\pagestyle{plain}
\setcounter{page}{1}
\pagenumbering{arabic}

\section{Introduction}
\label{sec:introduction}

One of the most important challenges of physics today is to understand the origin of the matter-antimatter asymmetry of the Universe.
Within the Standard Model (SM) of particle physics, the \CP symmetry between particles and antiparticles is broken only by the complex phase in the Cabibbo-Kobayashi-Maskawa (CKM) quark mixing matrix~\cite{Cabibbo:1963yz,Kobayashi:1973fv}.
An important parameter in the CKM description of the SM flavour structure is $\gamma \equiv \arg\left[-V_{ud}^{}V_{ub}^*/(V_{cd}^{}V_{cb}^*)\right]$, one of the three angles of the unitarity triangle formed from CKM matrix elements~\cite{Wolfenstein:1983yz,Jarlskog:1985ht,Buras:1994ec}.
Since the SM cannot account for the baryon asymmetry of the Universe~\cite{Riotto:1999yt} new sources of \CP violation, that would show up as deviations from the SM, are expected.
The precise determination of $\gamma$ is necessary in order to be able to search for such small deviations.

The value of $\gamma$ can be determined from the \CP-violating interference between the two amplitudes in, for example, $\Bp \to D\Kp$ and charge-conjugate decays~\cite{Gronau:1990ra,Gronau:1991dp,Atwood:1996ci,Atwood:2000ck}.
Here $D$ denotes a neutral charm meson reconstructed in a final state accessible to both \Dzb and \Dz decays,
that is therefore a superposition of the \Dzb and \Dz states produced through $\bquark \to \cquark W$ and $\bquark \to \uquark W$ transitions
(hereafter referred to as $V_{cb}$ and $V_{ub}$ amplitudes).
This approach has negligible theoretical uncertainty in the SM~\cite{Brod:2013sga} but limited data samples are available experimentally.

\begin{figure}[!t]
  \centering
  \includegraphics[width=0.325\textwidth]{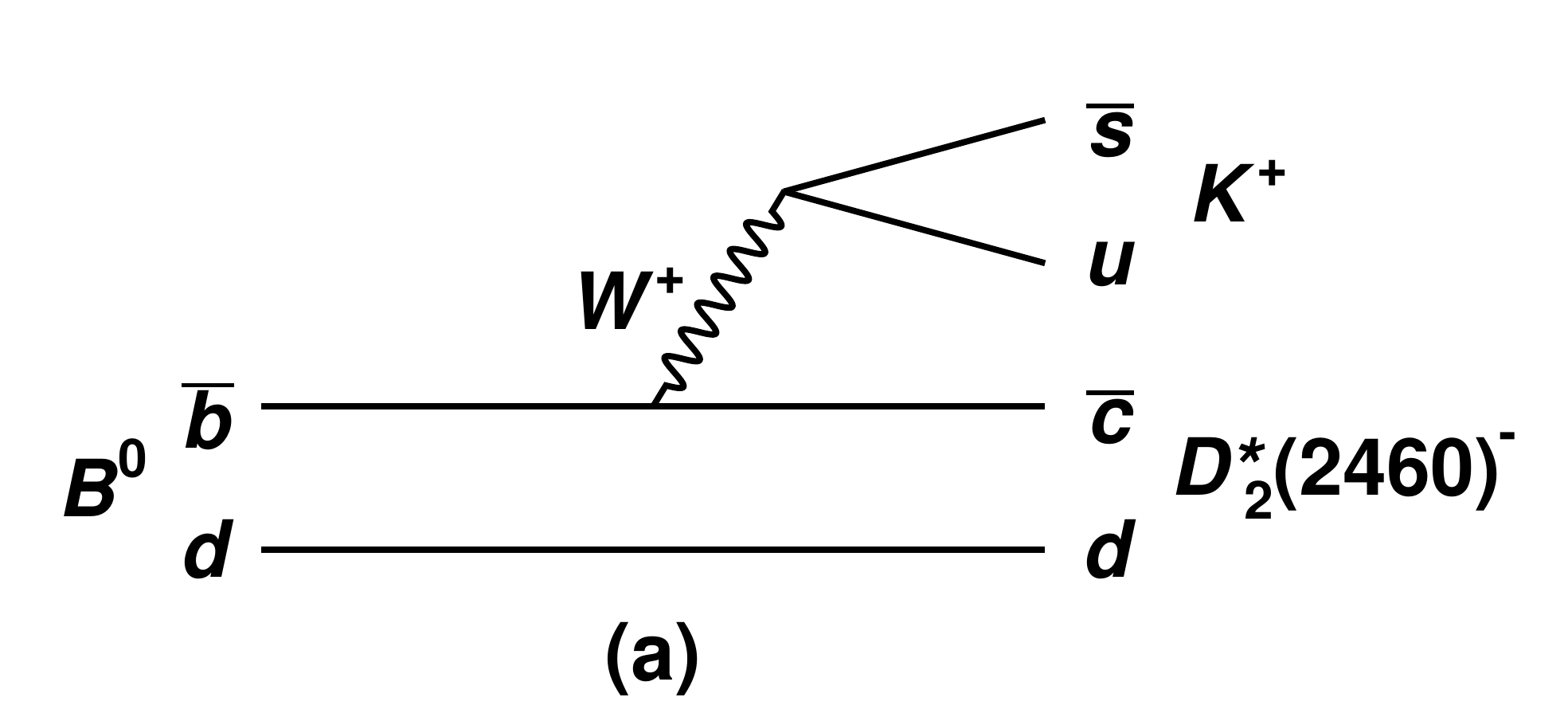}
  \includegraphics[width=0.325\textwidth]{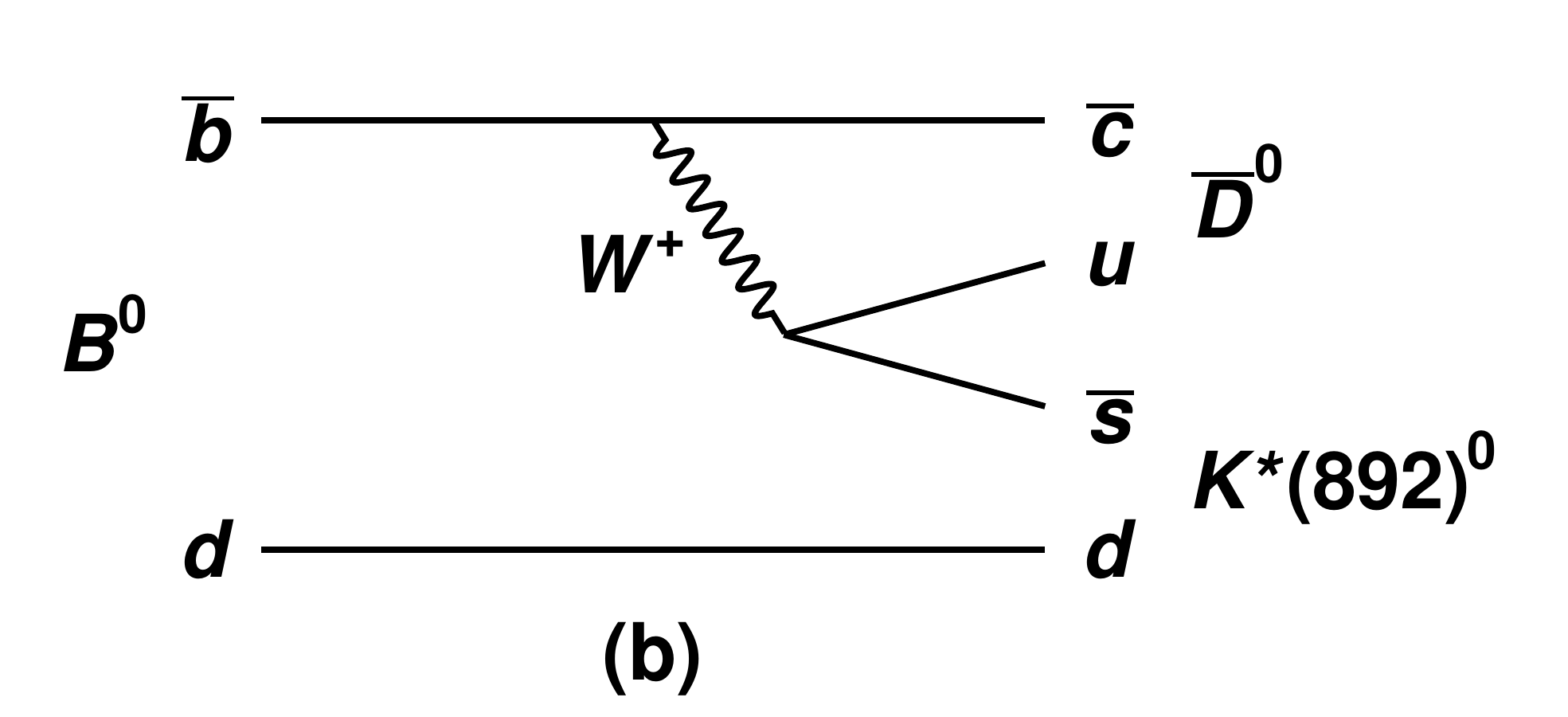}
  \includegraphics[width=0.325\textwidth]{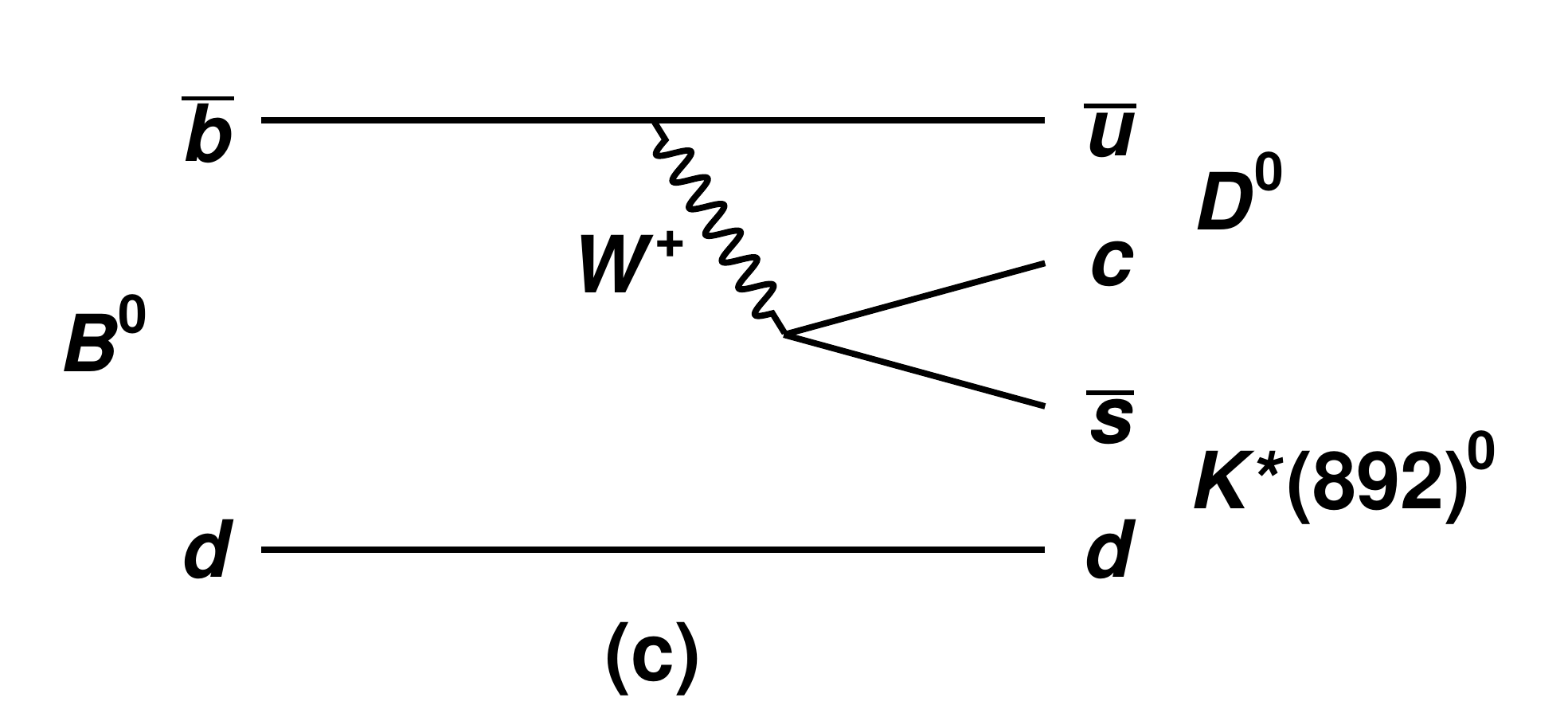}
  \caption{
    Feynman diagrams for the contributions to $\Bd\to D\Kp\pim$ from (a) $\Bz \to D_2^*(2460)^-\Kp$, (b) $\Bz \to \Dzb\Kstar(892)^0$ and (c) $\Bz \to \Dz\Kstar(892)^0$ decays.
  }
  \label{fig:feynman}
\end{figure}

\begin{figure}[!t]
  \centering
  \includegraphics[width=0.48\textwidth]{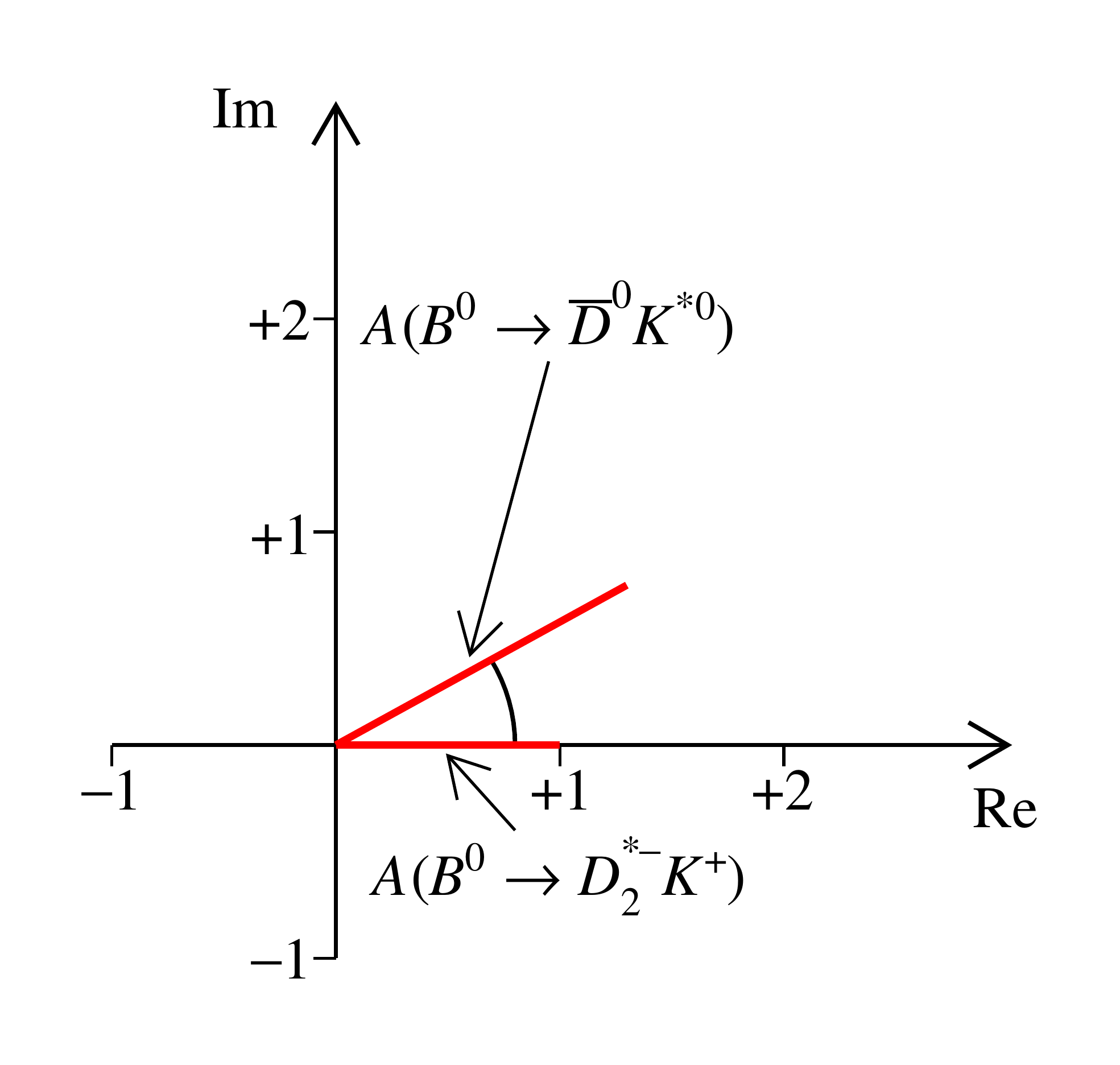}
  \includegraphics[width=0.48\textwidth]{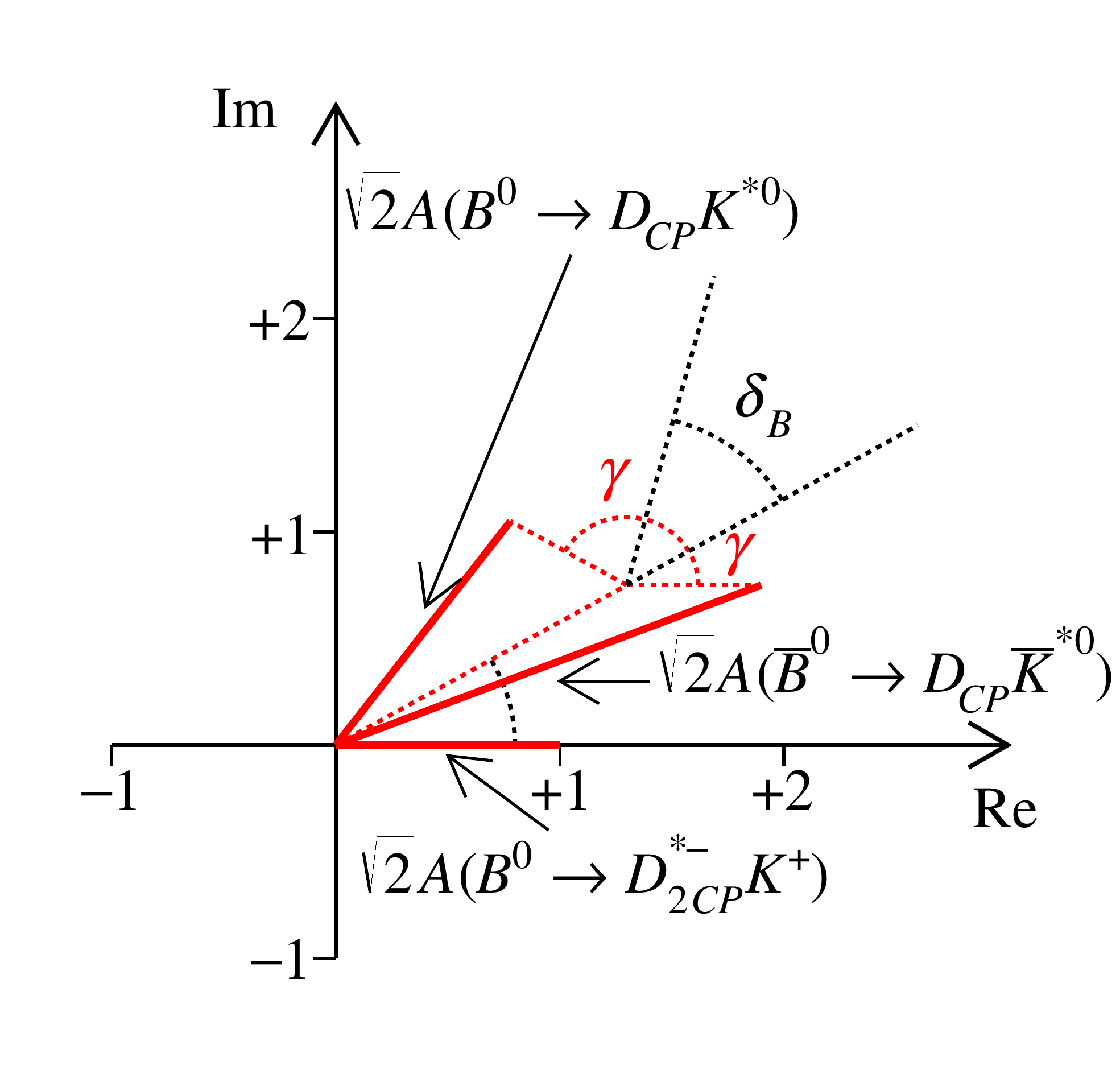}
  \caption{
    Illustration of the method to determine $\gamma$ from Dalitz plot analysis of $\Bd \to \D\Kp\pim$ decays~\cite{Gershon:2008pe,Gershon:2009qc}:
    (left) the $V_{cb}$ amplitude for $\Bz \to \Dzb\Kstarz$ compared to that for $\Bz \to D_2^{*-}\Kp$ decay;
    (right) the effect of the $V_{ub}$ amplitude that contributes to $\Bz\to D_{\CP}\Kstarz$ and $\Bzb\to D_{\CP}\Kstarzb$ decays provides sensitivity to $\gamma$.
    The notation $D_{\CP}$ represents a neutral \D meson reconstructed in a \CP eigenstate, while $D^{*-}_{2\,\CP}$ denotes the decay chain $D_2^{*-}\to \D_{\CP}\pim$, where the charge of the pion tags the flavour of the neutral \D meson, independently of the mode in which it is reconstructed, so there is no contribution from the $V_{ub}$ amplitude.
  }
  \label{fig:gamma-amplitudes}
\end{figure}

A similar method based on $\Bz \to D\Kp\pim$ decays has been proposed~\cite{Gershon:2008pe,Gershon:2009qc} to help improve the precision.
By studying the Dalitz plot (DP)~\cite{Dalitz:1953cp} distributions of \Bzb and \Bz decays, interference between different contributions, such as $\Bz \to D_2^*(2460)^-\Kp$ and $\Bz \to D\Kstar(892)^0$ (Feynman diagrams shown in Fig.~\ref{fig:feynman}), can be exploited to obtain additional sensitivity compared to the ``quasi-two-body'' analysis in which only the region of the DP dominated by the $\Kstar(892)^0$ resonance is selected~\cite{Bigi:1988ym,Dunietz:1991yd,Gronau:2002mu}.
The method is illustrated in Fig.~\ref{fig:gamma-amplitudes}, where the relative amplitudes of the different channels are sketched in the complex plane.
The $\Bz \to \Dzb\Kstarz$ ($V_{cb}$) amplitude is determined, relative to that for $\Bz \to D_2^{*-}\Kp$ decays, from analysis of the Dalitz plot with the neutral $D$ meson reconstructed in a favoured decay mode such as $\Dzb \to \Kp\pim$.
The $V_{ub}$ amplitude can then be obtained from the difference in this relative amplitude compared to the $V_{cb}$ only case when the neutral $D$ meson is reconstructed as a \CP eigenstate.
A non-zero value of $\gamma$ causes different relative amplitudes for \Bz and \Bzb decays, and hence \CP violation.
The method allows the determination of $\gamma$ and the hadronic parameters $r_{B}$ and $\delta_{B}$, which are the relative magnitude and strong (\ie\ \CP-conserving) phase of the $V_{ub}$ and $V_{cb}$ amplitudes for the $\Bz \to \D\Kstarz$ decay, with only \CP-even \D decays required to be reconstructed in addition to the favoured decays.
This feature, which is in constrast to the method of Refs.~\cite{Gronau:1990ra,Gronau:1991dp} that requires samples of both \CP-even and \CP-odd \D decays, is important for analysis of data collected at a hadron collider where reconstruction of \D meson decays to \CP-odd final states such as $\KS\piz$ is challenging.
The Dalitz analysis method also has only a single ambiguity ($\gamma \longleftrightarrow \gamma+\pi$, $\delta_{B} \longleftrightarrow \delta_{B}+\pi$), whereas the method of Refs.~\cite{Gronau:1990ra,Gronau:1991dp} has an eight-fold ambiguity in the determination of $\gamma$.

This paper describes the first study of \CP violation with a DP analysis of $\Bz \to D\Kp\pim$ decays, with a sample corresponding to $3.0\invfb$ of $pp$ collision data collected with the LHCb detector at centre-of-mass energies of $7$ and $8 \tev$.
The inclusion of charge conjugate processes is implied throughout the paper except where discussing asymmetries.

\section{Detector and simulation}
\label{sec:detector}

The \lhcb detector~\cite{Alves:2008zz,LHCb-DP-2014-002} is a single-arm forward spectrometer covering the \mbox{pseudorapidity} range $2<\eta <5$, designed for the study of particles containing \bquark or \cquark quarks.
The detector includes a high-precision tracking system consisting of a silicon-strip vertex detector surrounding the $pp$ interaction region, a large-area silicon-strip detector located upstream of a dipole magnet with a bending power of about $4{\mathrm{\,Tm}}$, and three stations of silicon-strip detectors and straw drift tubes placed downstream of the magnet.
The tracking system provides a measurement of momentum, \ptot, of charged particles with a relative uncertainty that varies from 0.5\% at low momentum to 1.0\% at 200\gevc.
The minimum distance of a track to a primary vertex, the impact parameter, is measured with a resolution of $(15+29/\pt)\mum$, where \pt is the component of the momentum transverse to the beam, in\,\gevc.
Different types of charged hadrons are distinguished using information
from two ring-imaging Cherenkov detectors.
Photons, electrons and hadrons are identified by a calorimeter system consisting of scintillating-pad and preshower detectors, an electromagnetic calorimeter and a hadronic calorimeter.
Muons are identified by a system composed of alternating layers of iron and multiwire proportional chambers.
The online event selection is performed by a trigger, which consists of a hardware stage, based on information from the calorimeter and muon systems, followed by a software stage, in which all charged particles with $\pt>500(300)\mevc$ are reconstructed for 2011 (2012) data.
A detailed description of the trigger conditions is available in Ref.~\cite{LHCb-PUB-2014-046}.

Simulated data samples are used to study the response of the detector and to investigate certain categories of background.
In the simulation, $pp$ collisions are generated using \pythia~\cite{Sjostrand:2007gs,*Sjostrand:2006za} with a specific \lhcb configuration~\cite{LHCb-PROC-2010-056}.
Decays of hadronic particles are described by \evtgen~\cite{Lange:2001uf}, in which final-state radiation is generated using \photos~\cite{Golonka:2005pn}.
The interaction of the generated particles with the detector, and its response, are implemented using the \geant toolkit~\cite{Allison:2006ve, *Agostinelli:2002hh} as described in Ref.~\cite{LHCb-PROC-2011-006}.

\section{Selection}
\label{sec:selection}

Candidate $\Bz \to D\Kp\pim$ decays are selected with the \D meson decaying into the $\Kp\pim$, $\Kp\Km$ or $\pip\pim$ final state.
The selection requirements are similar to those used for the DP analyses of $\Bz \to \Dzb\Kp\pim$~\cite{LHCb-PAPER-2015-017} and $\Bs \to \Dzb\Km\pip$~\cite{LHCb-PAPER-2013-035,LHCb-PAPER-2013-036} decays, where in both cases only the $\Dzb\to\Kp\pim$ mode was used.

The more copious $\Bz\to\D\pip\pim$ modes, with neutral \D meson decays to one of the three final states under study, are used as control channels to optimise the selection requirements.
Loose initial requirements on the final state tracks and the \D and \B candidates are used to obtain a visible peak of $\Bz \to \D\pip\pim$ decays.
The neutral \D meson candidate must satisfy criteria on its invariant mass, vertex quality and flight distance from any PV and from the $B$ candidate vertex.
Requirements on the outputs of boosted decision tree algorithms that identify neutral \D meson decays, in each of the decay chains of interest, originating from \bquark hadron decays~\cite{LHCb-PAPER-2012-025,LHCb-PAPER-2012-050} are also applied.
These requirements are sufficient to reduce to negligible levels potential background from charmless \B meson decays that have identical final states but without an intermediate \D meson.
Vetoes are applied to remove backgrounds from $\Bz \to \Dstar(2010)^-\Kp$, $\Bz \to \Dmp\pipm$, $\Bs \to \Dsm\pip$ and $\Bds \to \Dz\Dzb$ decays,
and candidates consistent with originating from $\Bds \to \Dzb\Kpm\pimp$ decays, where the $\Dzb$ has been reconstructed from the wrong pair of tracks.

Separate neural network (NN) classifiers~\cite{Feindt:2006pm} for each \D decay mode are used to distinguish signal decays from combinatorial background.
The \sPlot\ technique~\cite{Pivk:2004ty}, with the $\Bz\to\D\pip\pim$ candidate mass as the discriminating variable, is used to obtain signal and background weights, which are then used to train the networks.
The networks are based on input variables that describe the topology of each decay channel, and that depend only weakly on the $\B$ candidate mass and on the position of the candidate in the $\B$ decay Dalitz plot.
Loose requirements are made on the NN outputs in order to retain large samples for the DP analysis.

\section{Determination of signal and background yields}
\label{sec:mass-fit}

The yields of signal and of several different backgrounds are determined from an extended maximum likelihood fit, in each mode, to the distributions of candidates in \B candidate mass and NN output.
Unbinned information on the \B candidate mass is used, while each sample is divided into five bins of the NN output that contain a similar number of signal, and varying numbers of background, decays~\cite{LHCb-PAPER-2015-012,LHCb-PAPER-2015-054}.

In addition to $\Bz \to D\Kp\pim$ decays, components are included in the fit to account for $\Bs$ decays to the same final state, partially reconstructed $\Bds \to \DorDstar\Kpm\pimp$ backgrounds,
misidentified $\Bd \to \DorDstar\pip\pim$, $\Bds \to \DorDstar\Kp\Km$, $\Lbbar \to \DorDstar \antiproton \pip$ and $\Lbbar \to \DorDstar \antiproton \Kp$ decays as well as combinatorial background.
The modelling of the signal and background distributions in \B candidate mass is similar to that described in Ref.~\cite{LHCb-PAPER-2015-017}.
The sum of two Crystal Ball functions~\cite{Skwarnicki:1986xj} is used for each of the correctly reconstructed \B decays, where the peak position and core width (\ie\ the narrower of the two widths) are free parameters of the fit, while the \Bs--\Bd mass difference is fixed to its known value~\cite{PDG2014}.
The fraction of the signal function contained in the core and the relative width of the two components are constrained within uncertainties to, and all other parameters are fixed to, their expected values obtained from simulated data, separately for each of the three \D samples.
An exponential function is used to describe combinatorial background, with the shape parameter allowed to vary.
Due to the loose NN output requirement it is necessary, in the $D \to \Kp\pim$ sample, to account explicitly for partially combinatorial background where the final state $D\Kp$ pair originates from a \B decay but is combined with a random pion; this is modelled with a non-parametric function.
Non-parametric functions obtained from simulation based on known DP distributions~\cite{LHCb-PAPER-2014-035,LHCb-PAPER-2014-036,LHCb-PAPER-2014-070,Kuzmin:2006mw,Abe:2004cw,LHCb-PAPER-2012-018,LHCb-PAPER-2013-056} are used to model the partially reconstructed and misidentified \B decays.

The fraction of signal decays in each NN output bin is allowed to vary freely in the fit; the correctly reconstructed \Bs decays and misidentified backgrounds are taken to have the same NN output distribution as signal.
The fractions of combinatorial and partially reconstructed backgrounds in each NN output bin are each allowed to vary freely.
All yields are free parameters of the fit, except those for misidentified backgrounds which are constrained within expectation relative to the signal yield, since the relative branching fractions~\cite{PDG2014} and misidentification probabilities~\cite{LHCb-DP-2012-003} are well known.

\begin{figure}[!tb]
  \centering
  \includegraphics[width=0.427\textwidth]{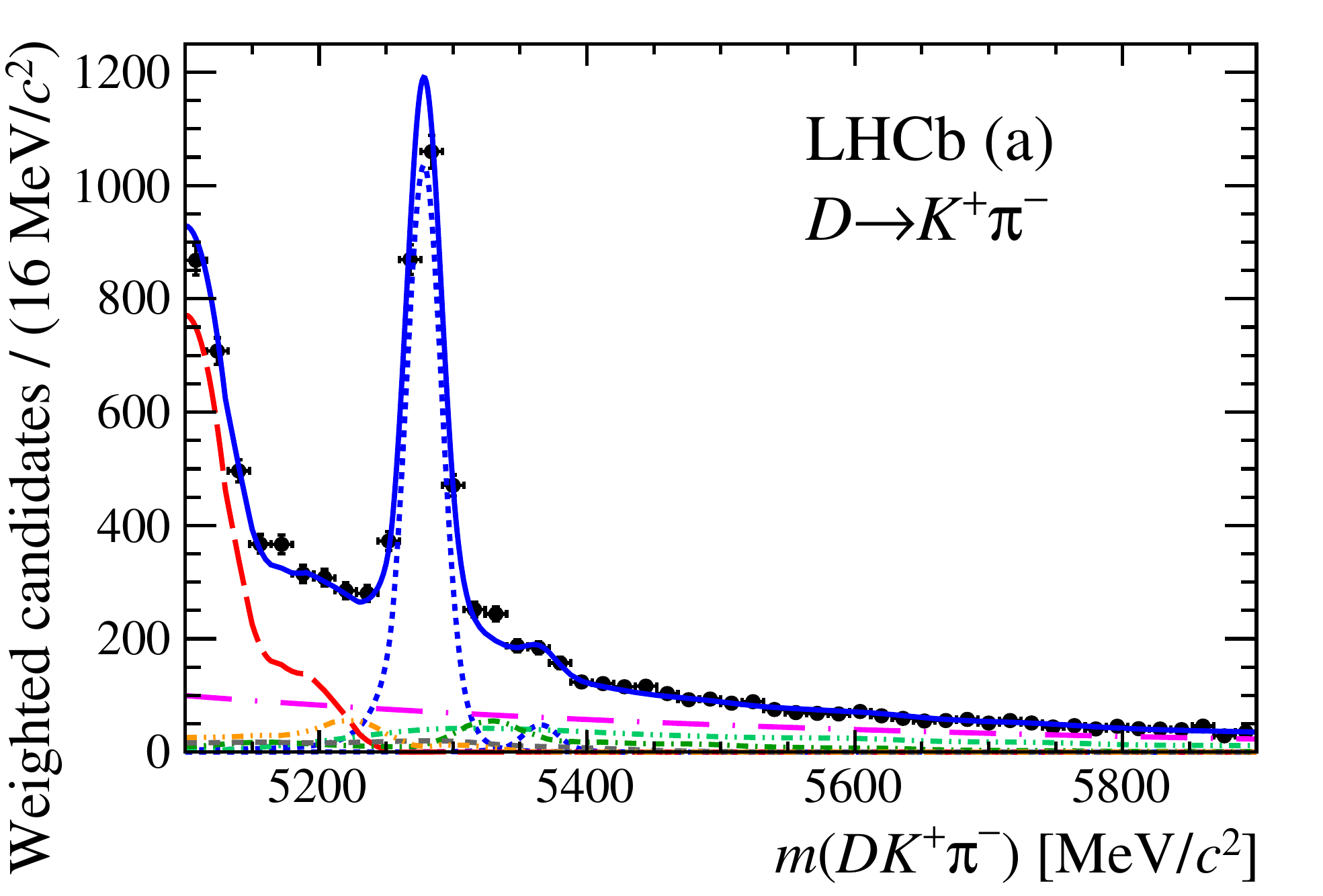}
  \includegraphics[width=0.427\textwidth]{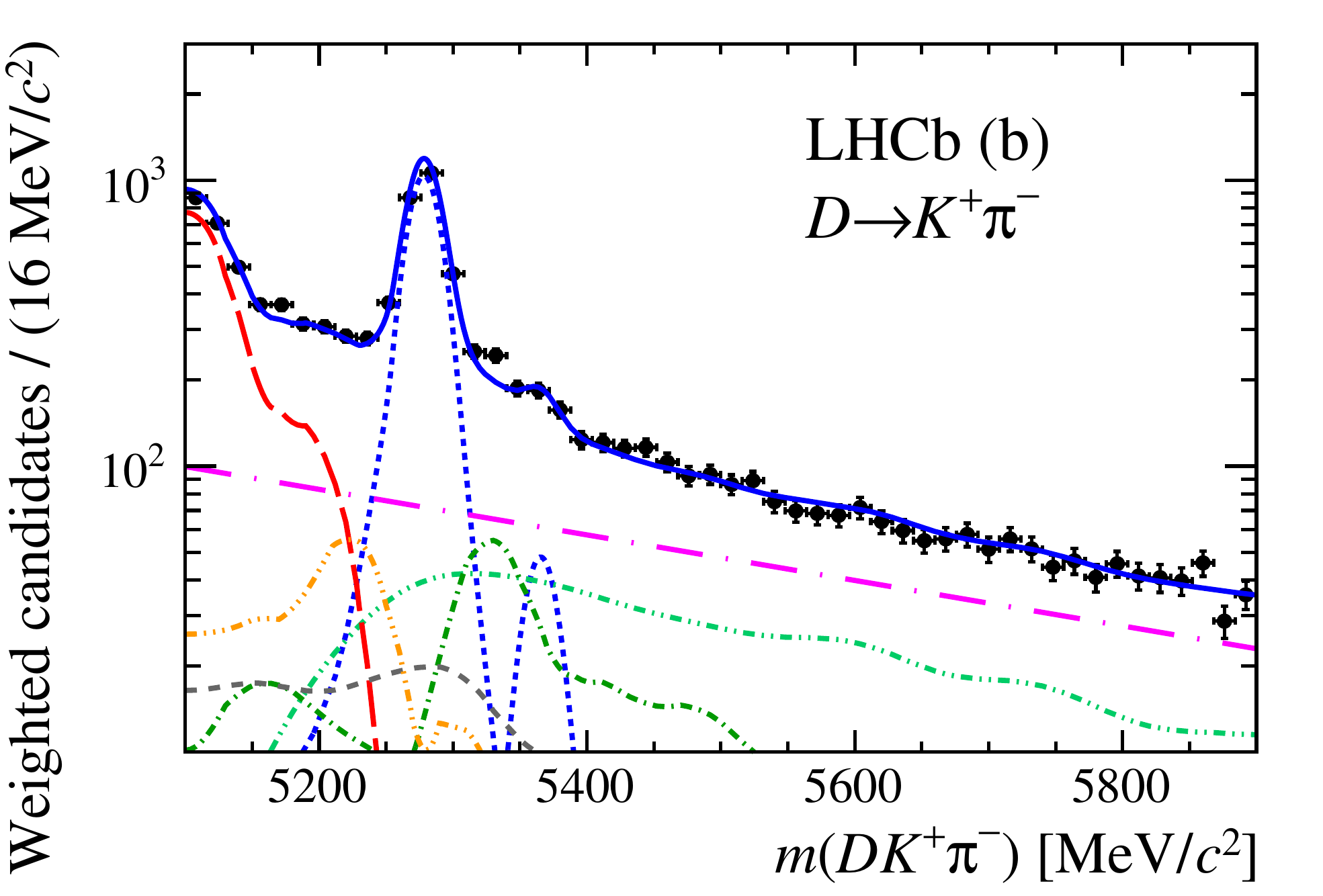}
  \includegraphics[width=0.427\textwidth]{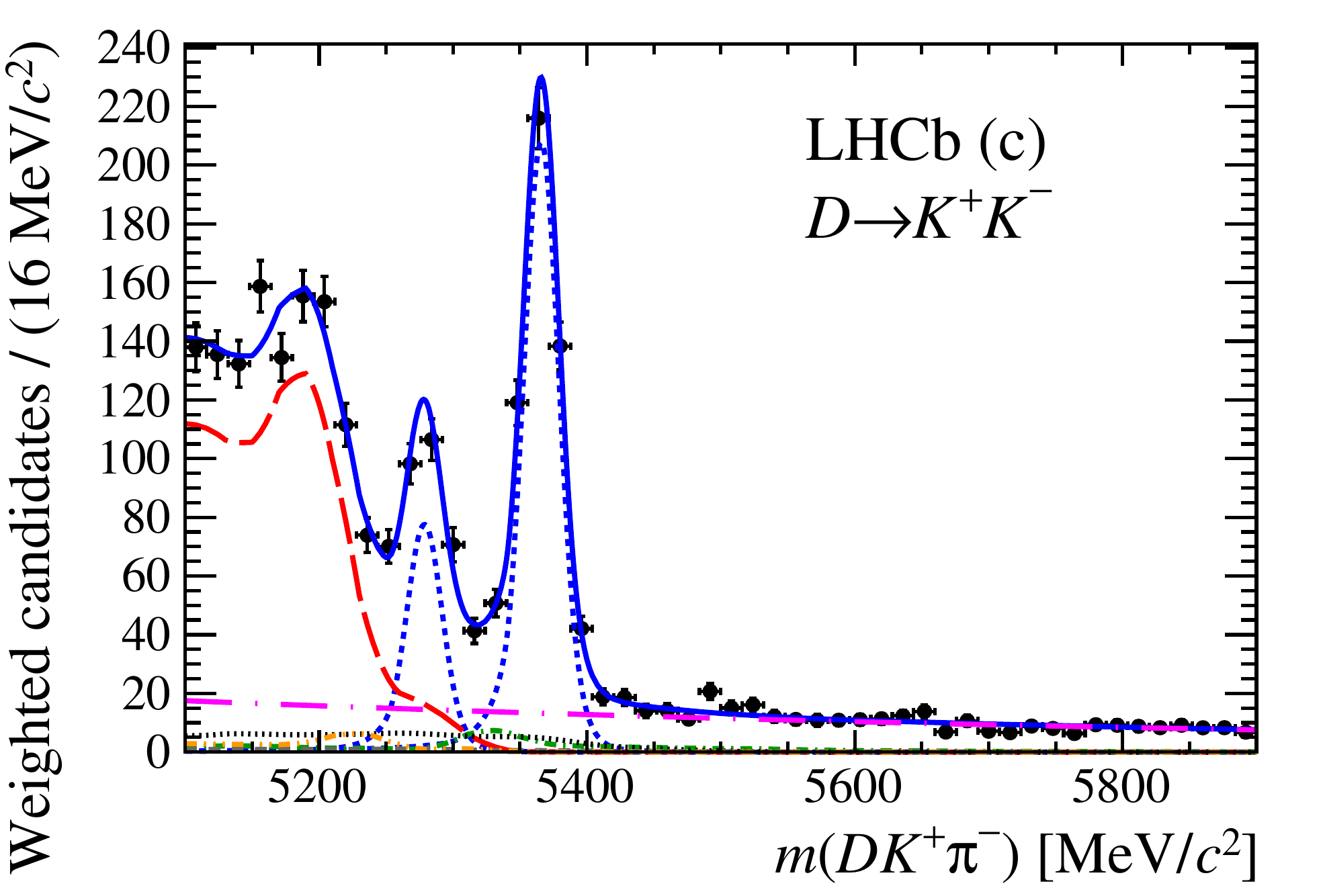}
  \includegraphics[width=0.427\textwidth]{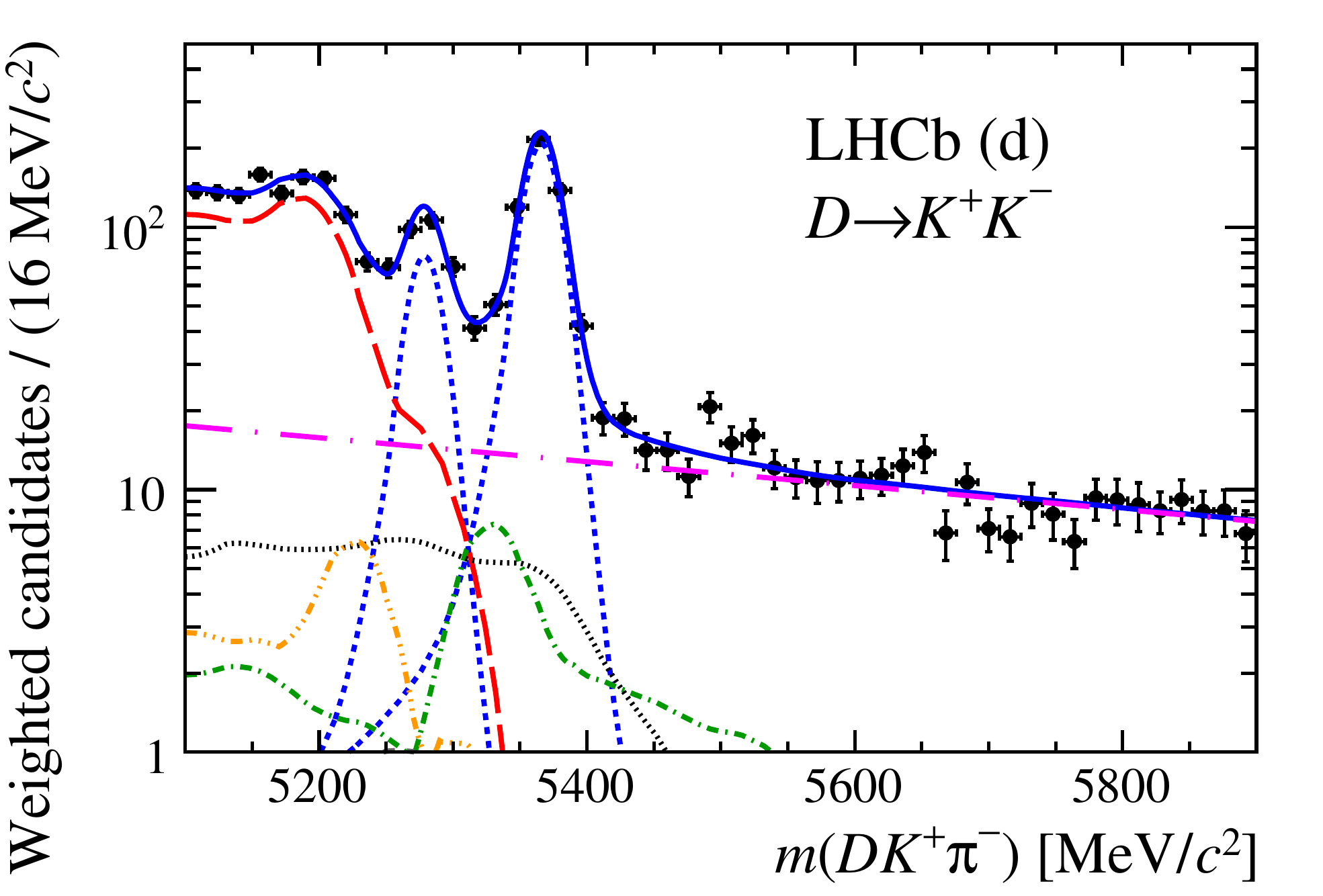}
  \includegraphics[width=0.427\textwidth]{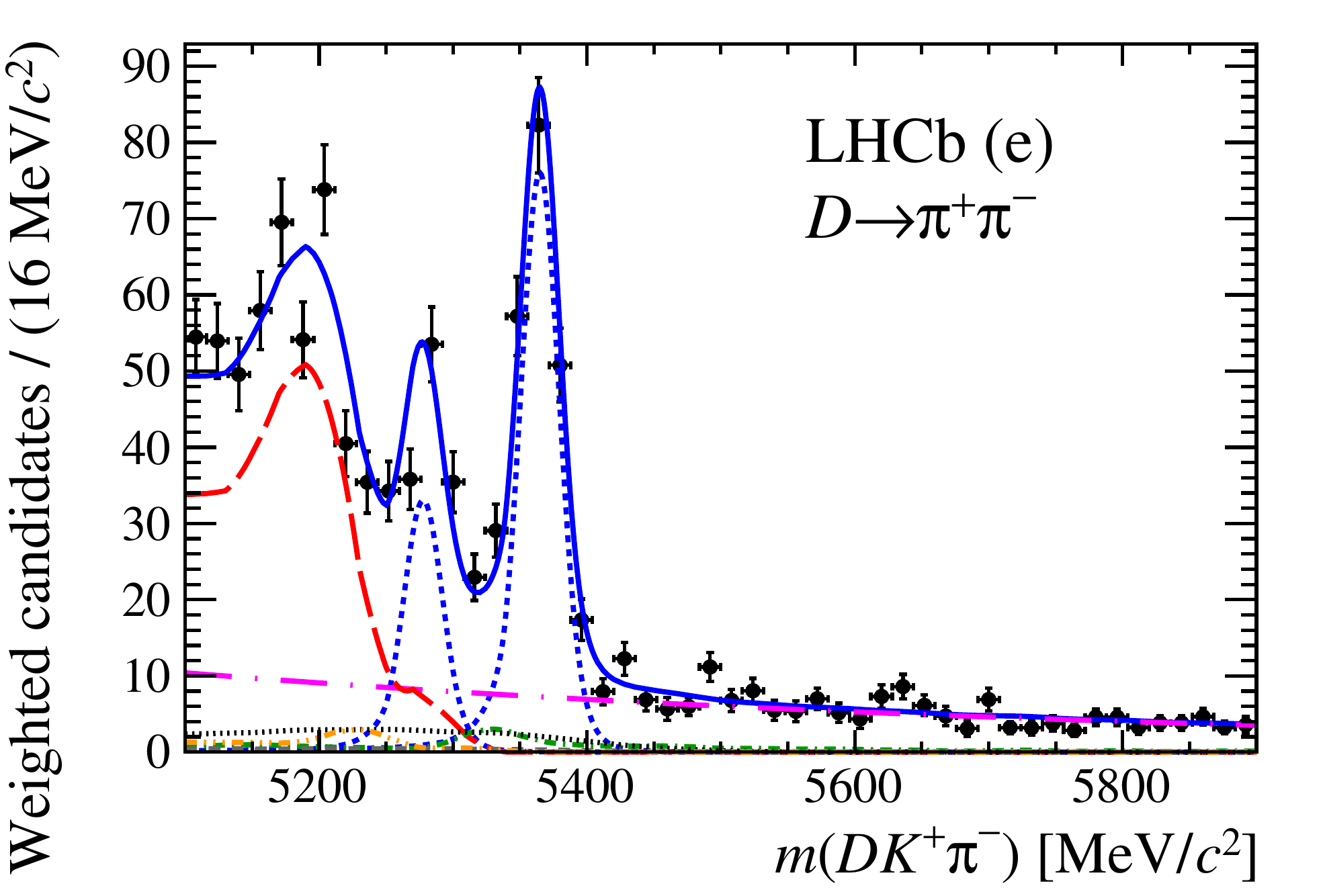}
  \includegraphics[width=0.427\textwidth]{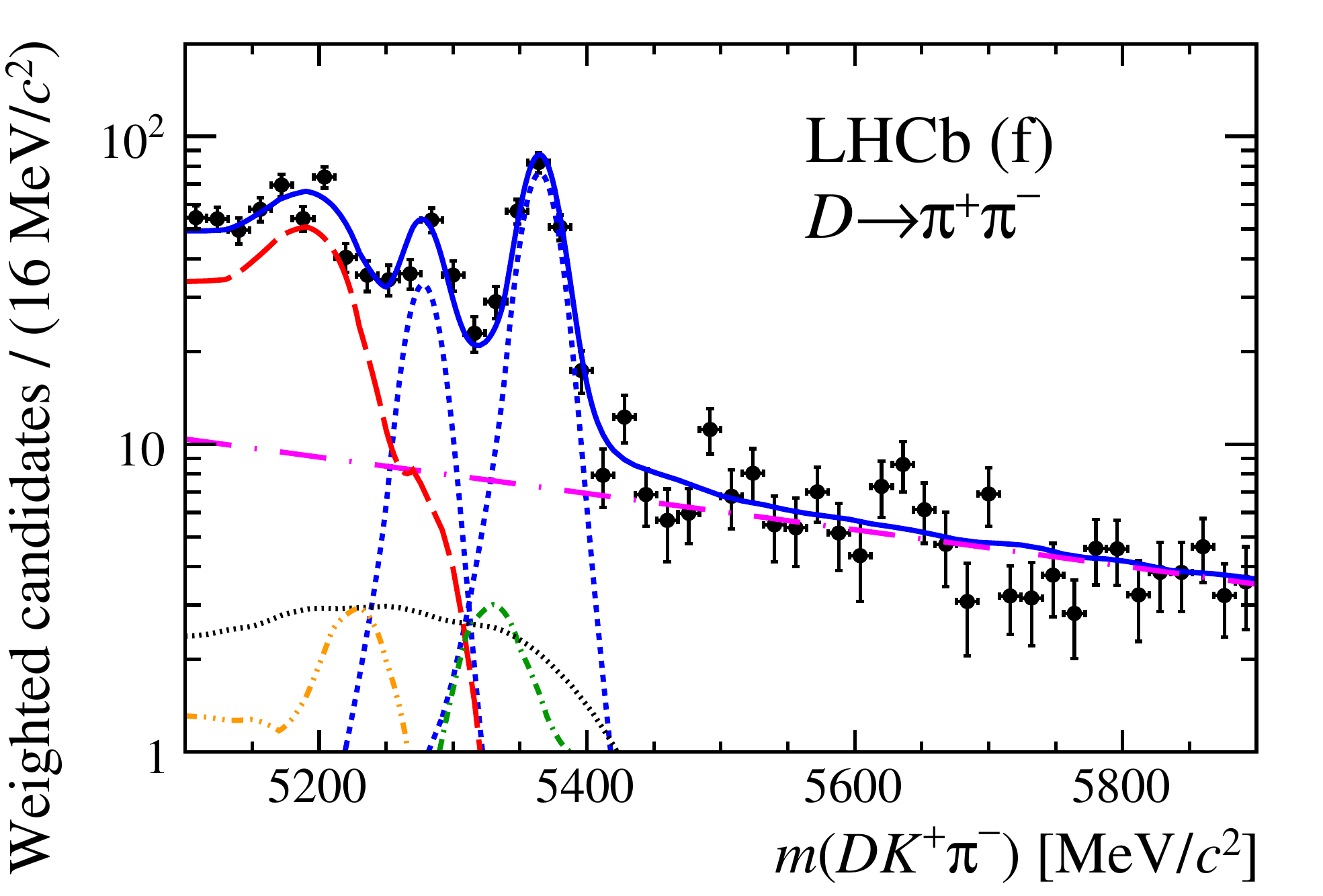}
  \vspace{1.1ex}
  \includegraphics[width=0.86\textwidth]{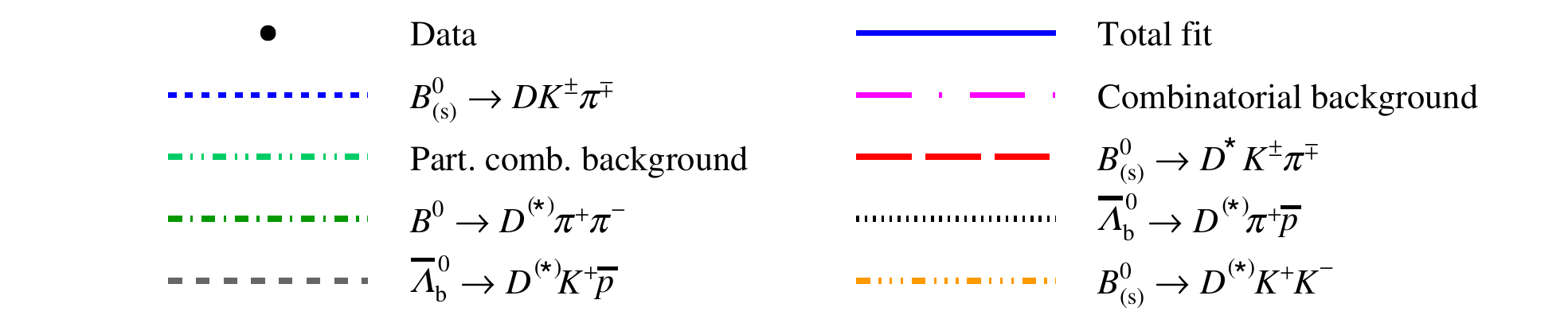}
  \caption{\small
    Results of fits to $D\Kp\pim$ candidates in the (a,b) $D\to \Kp\pim$, (c,d) $D\to \Kp\Km$ and (e,f) $D\to \pip\pim$ samples.
    The data and the fit results in each NN output bin have been weighted according to ${\cal S}/({\cal S}+{\cal B})$ as described in the text.
    The left and right plots are identical but with (left) linear and (right) logarithmic $y$-axis scales.
    The components are as described in the legend.
  }
  \label{fig:mass-fits}
\end{figure}

The results of the fits are shown in Fig.~\ref{fig:mass-fits}, in which the NN output bins have been combined by weighting both the data and fit results by ${\cal S}/({\cal S}+{\cal B})$, where ${\cal S}$ (${\cal B}$) is the signal (background) yield in the signal window, defined as $\pm 2.5\,\sigma({\rm core})$ around the \Bd peak in each sample, where $\sigma({\rm core})$ is the core width of the signal shape.
The yields of each category in these regions, which correspond to $5246.6$--$5309.9\mevcc$, $5246.9$--$5310.5\mevcc$ and $5243.1$--$5312.3\mevcc$ in the $\D\to\Kp\pim$, $\Kp\Km$ and $\pip\pim$ samples, are given in Tables~\ref{tab:Bdyields-Kpi},~\ref{tab:Bdyields-KK} and~\ref{tab:Bdyields-pipi}.
In total, there are $2840 \pm 70$ signal decays within the signal window in the $D\to\Kp\pim$ sample, whilst the corresponding values for the $D\to\Kp\Km$ and $D\to\pip\pim$ samples are $339 \pm 22$ and $168 \pm 19$.
The $\chisq/\rm{ndf}$ values for the projections of the fits to the $\D\to\Kp\pim$, $\D\to\Kp\Km$ and $\D\to\pip\pim$ datasets are $171.5/223$, $188.2/223$ and $169.1/222$, respectively, giving a total $\chisq/\rm{ndf} = 528.8/668$.
Note that there are some bins with low numbers of entries which may result in this value not following exactly the expected $\chisq$ distribution.

\begin{table}[!tb]
  \centering
  \caption{\small
    Yields in the signal window of the fit components in the five NN output bins for the $\D\to \Kp\pim$ sample.
    The last column indicates whether or not each component is explicitly modelled in the Dalitz plot fit.
     }
  \label{tab:Bdyields-Kpi}
  \vspace{1ex}
  \begin{tabular}{ccccccc}
    \hline
    Component & \multicolumn{5}{c}{Yield} & Included? \\
    & bin 1 & bin 2 & bin 3 & bin 4 & bin 5 \\
    \hline
    $\Bd \to \D \Kp\pim$                & 597 & 546 & 585 & 571 & 540 & Yes \\
    $\Bs \to \D \Kp\pim$                &   1 &   1 &   1 &   1 &   1 & No \\
    comb.~bkgd.                         & 540 &  58 &  16 &   6 &   1 & Yes \\
    $\Bp\to\DorDstar\Kp + X^-$          & 305 &  33 &   9 &   3 &   1 & Yes \\
    $\Bd\to\Dstar \Kp\pim$              &   1 &   1 &   1 &   1 &   1 & No \\
    $\Bd\to\DorDstar \pip\pim$          &  20 &  18 &  20 &  19 &  18 & Yes \\
    $\Lbbar\to\DorDstar \Kp\antiproton$ &  21 &  19 &  21 &  20 &  19 & Yes \\
    $\Bd\to\DorDstar \Kp\Km$            &   8 &   7 &   8 &   7 &   7 & No \\
    $\Bs\to\DorDstar \Kp\Km$            &  10 &   9 &  10 &  10 &   9 & No \\
    \hline
  \end{tabular}
\end{table}

\begin{table}[!tb]
  \centering
  \caption{\small
    Yields in the signal window of the fit components in the five NN output bins for the $\D\to \Kp\Km$ sample.
    The last column indicates whether or not each component is explicitly modelled in the Dalitz plot fit.
     }
  \label{tab:Bdyields-KK}
  \vspace{1ex}
  \begin{tabular}{ccccccc}
    \hline
    Component & \multicolumn{5}{c}{Yield} & Included? \\
    & bin 1 & bin 2 & bin 3 & bin 4 & bin 5 \\
    \hline
    $\Bd \to \D \Kp\pim$                &  70 &  63 &  68 &  73 &  65 & Yes \\
    $\Bsb \to \D \Kp\pim$               &   5 &   5 &   5 &   6 &   5 & Yes \\
    comb.~bkgd.                         & 173 &  19 &   9 &   3 &   0 & Yes \\
    $\Bd\to\Dstar \Kp\pim$              &   0 &   1 &   1 &   1 &   0 & No \\
    $\Bsb\to\Dstar \Kp\pim$             &  19 &  28 &  34 &  28 &  20 & Yes \\
    $\Bd\to\DorDstar \pip\pim$          &   4 &   3 &   4 &   4 &   3 & Yes \\
    $\Lb\to\DorDstar \proton\pim$       &  11 &  10 &  10 &  11 &  10 & Yes \\
    $\Lbbar\to\DorDstar \Kp\antiproton$ &   2 &   1 &   2 &   2 &   2 & No \\
    $\Bd\to\DorDstar \Kp\Km$            &   2 &   1 &   2 &   2 &   1 & No \\
    $\Bs\to\DorDstar \Kp\Km$            &   1 &   1 &   1 &   2 &   1 & No \\
    \hline
  \end{tabular}
\end{table}

\begin{table}[!tb]
  \centering
  \caption{\small
    Yields in the signal window of the fit components in the five NN output bins for the $\D\to \pip\pim$ sample.
    The last column indicates whether or not each component is explicitly modelled in the Dalitz plot fit.
     }
  \label{tab:Bdyields-pipi}
  \vspace{1ex}
  \begin{tabular}{ccccccc}
    \hline
    Component & \multicolumn{5}{c}{Yield} & Included? \\
    & bin 1 & bin 2 & bin 3 & bin 4 & bin 5 \\
    \hline
    $\Bd \to \D \Kp\pim$                &  36 &  31 &  38 &  32 &  31 & Yes \\
    $\Bsb \to \D \Kp\pim$               &   3 &   2 &   3 &   3 &   2 & Yes \\
    comb.~bkgd.                         & 119 &  17 &   4 &   3 &   2 & Yes \\
    $\Bd\to\Dstar \Kp\pim$              &   0 &   0 &   0 &   0 &   0 & No \\
    $\Bsb\to\Dstar \Kp\pim$             &   9 &  16 &  15 &  12 &  10 & Yes \\
    $\Bd\to\DorDstar \pip\pim$          &   2 &   2 &   2 &   2 &   2 & Yes \\
    $\Lb\to\DorDstar \proton\pim$       &   6 &   5 &   6 &   5 &   5 & Yes \\
    $\Lbbar\to\DorDstar \Kp\antiproton$ &   1 &   1 &   1 &   1 &   1 & No \\
    $\Bd\to\DorDstar \Kp\Km$            &   1 &   1 &   1 &   1 &   1 & No \\
    $\Bs\to\DorDstar \Kp\Km$            &   1 &   1 &   1 &   1 &   1 & No \\
    \hline
  \end{tabular}
\end{table}

\begin{figure}[!tb]
\centering
  \includegraphics[scale=0.36]{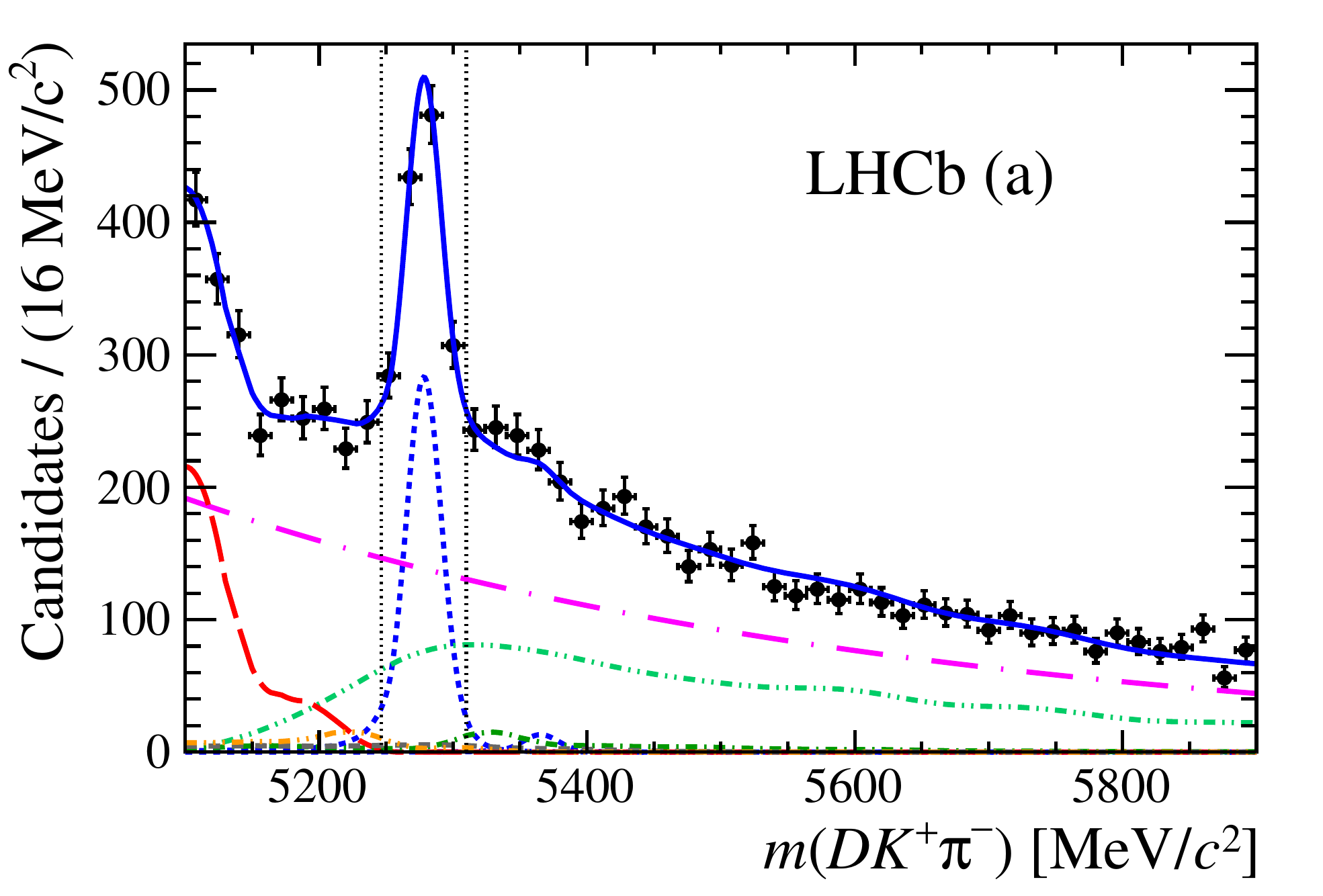}
  \includegraphics[scale=0.36]{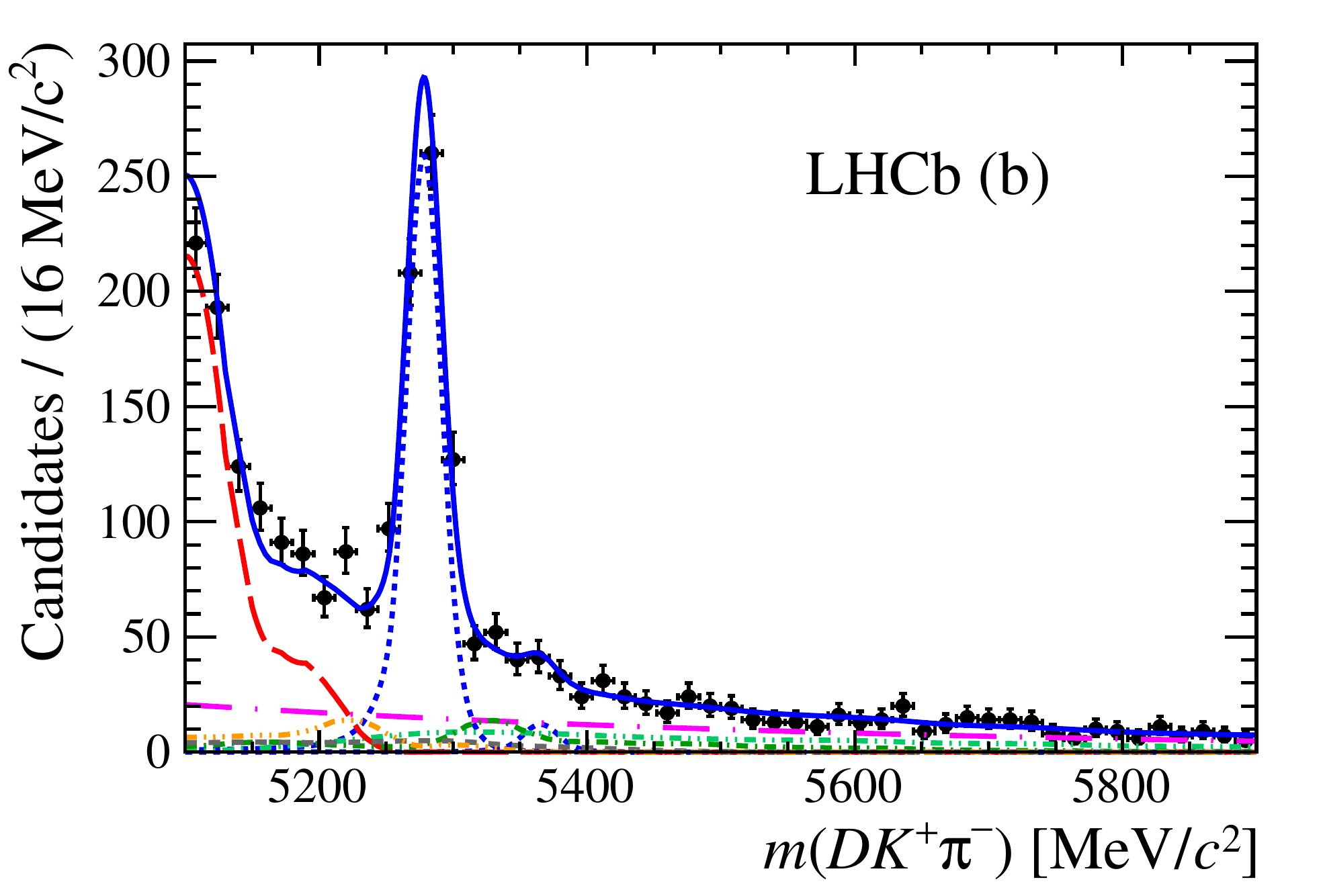}
  \includegraphics[scale=0.36]{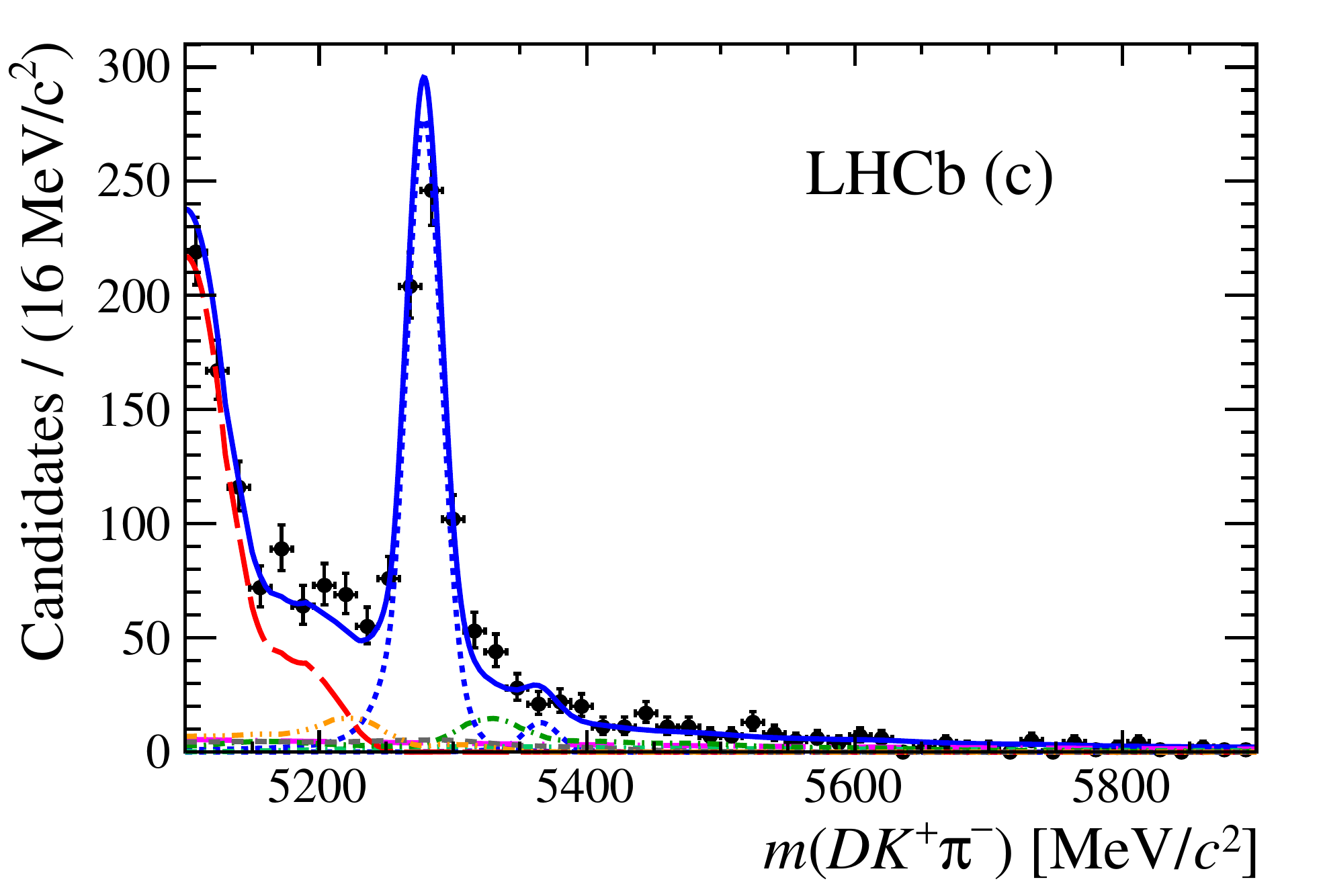}
  \includegraphics[scale=0.36]{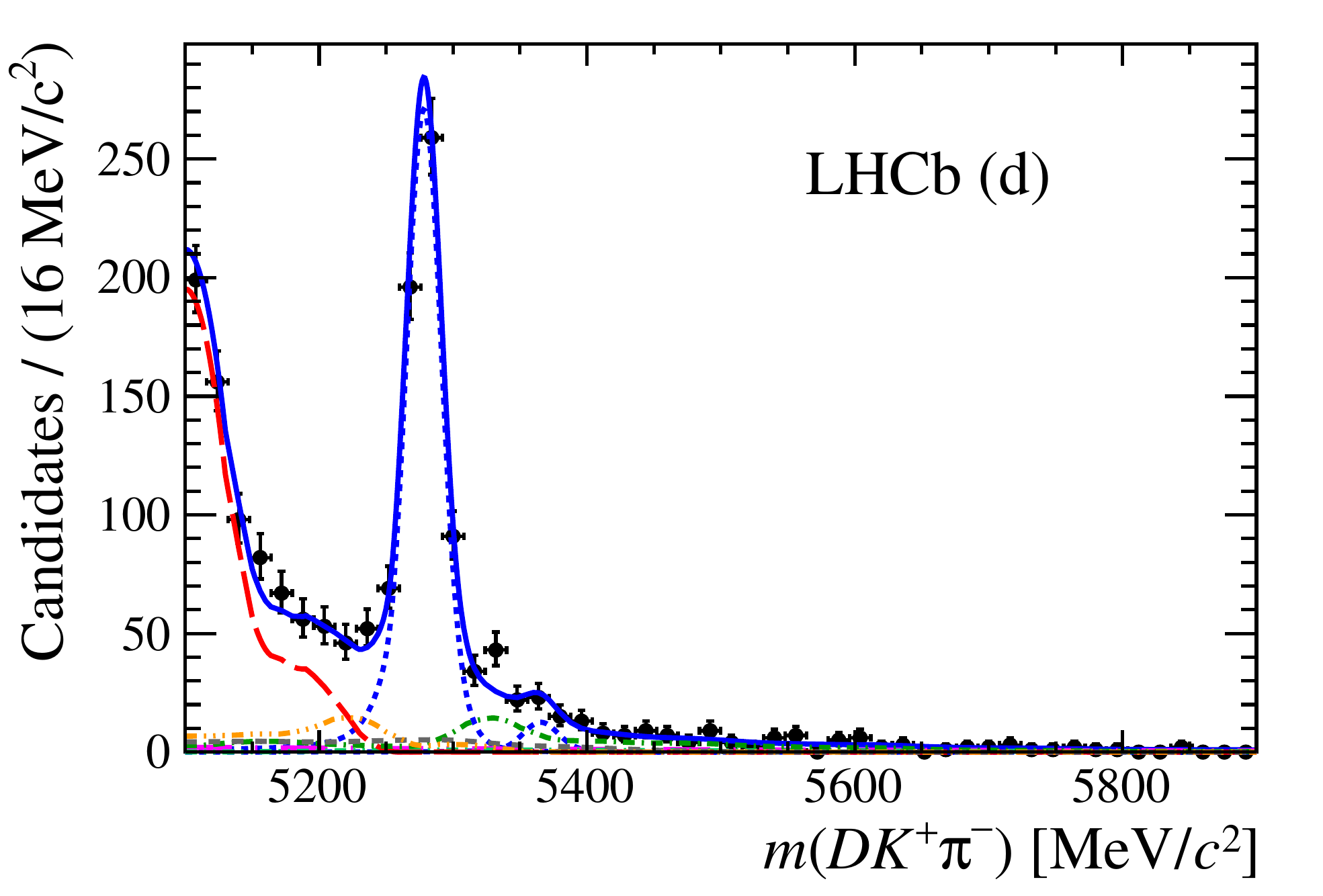}
  \includegraphics[scale=0.36]{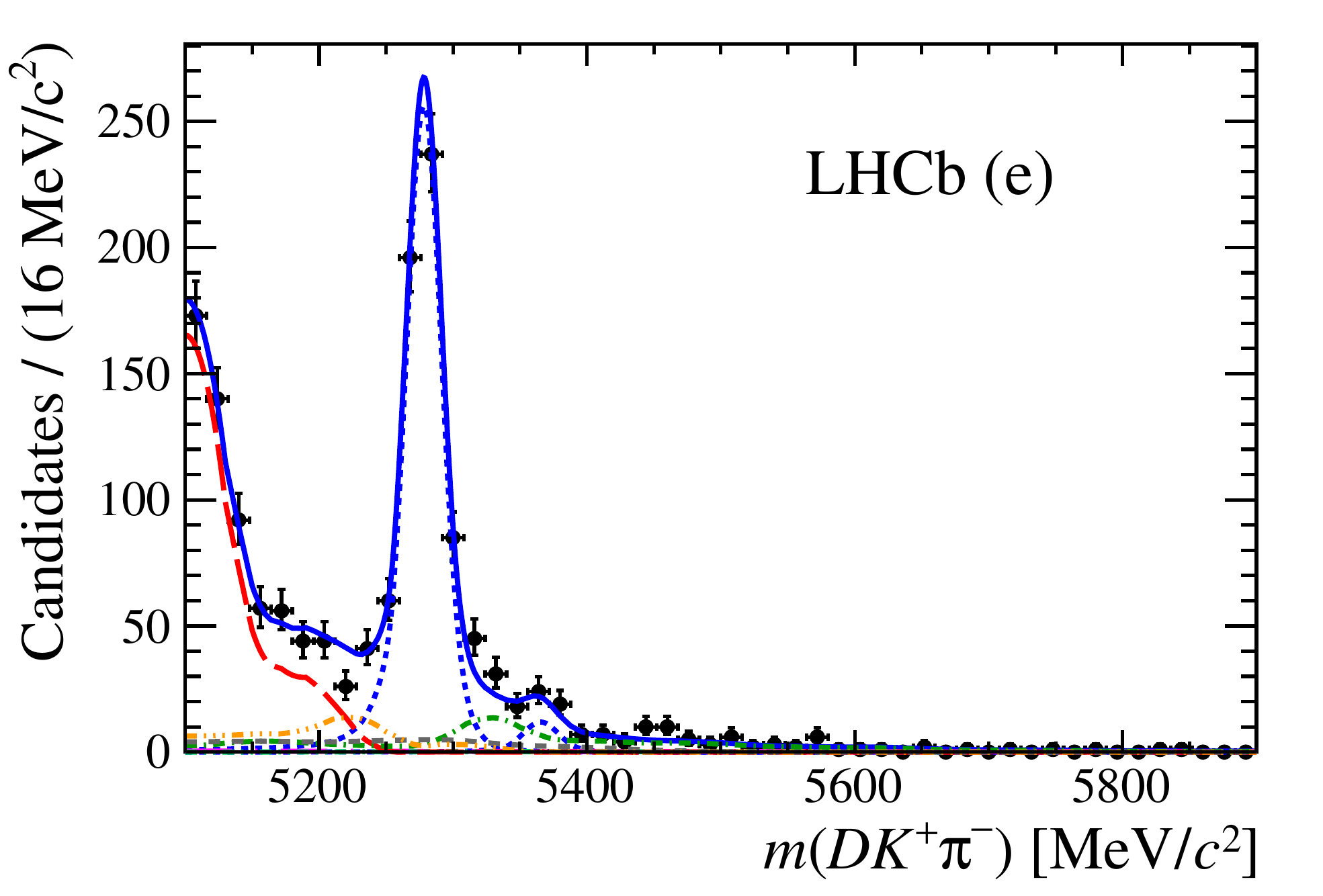}
  \includegraphics[scale=0.36]{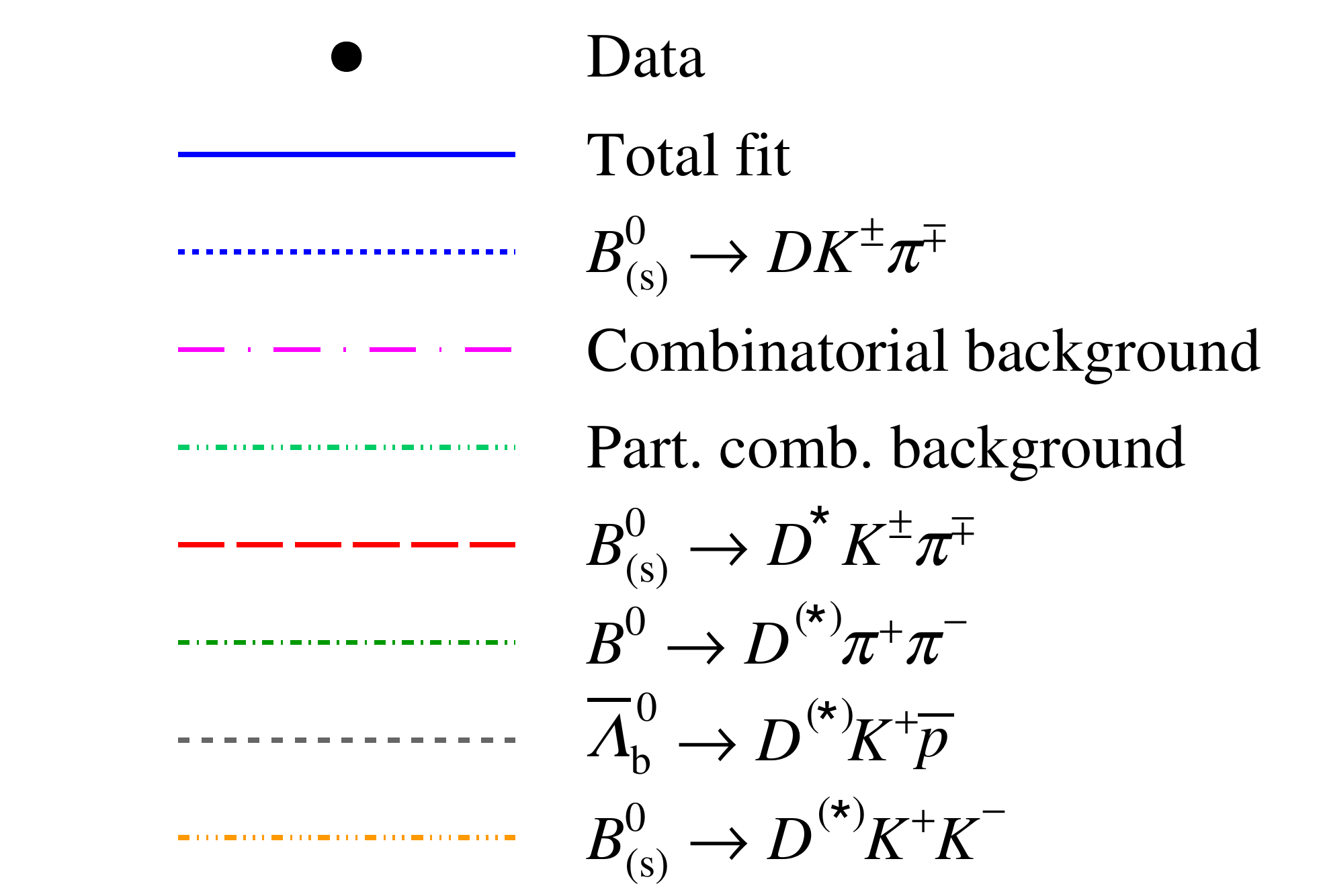}
\caption{\small
  Results of the fit to $\D\Kp\pim$, $\D\to\Kp\pim$ candidates shown separately in the five bins of the neural network output variable.
  The bins are shown, from (a)--(e), in order of increasing ${\cal S}/{\cal B}$.
  The components are as indicated in the legend.
  The vertical dotted lines in (a) show the signal window used for the fit to the Dalitz plot.}
\label{fig:fitKpi}
\end{figure}

\begin{figure}[!tb]
\centering
  \includegraphics[scale=0.36]{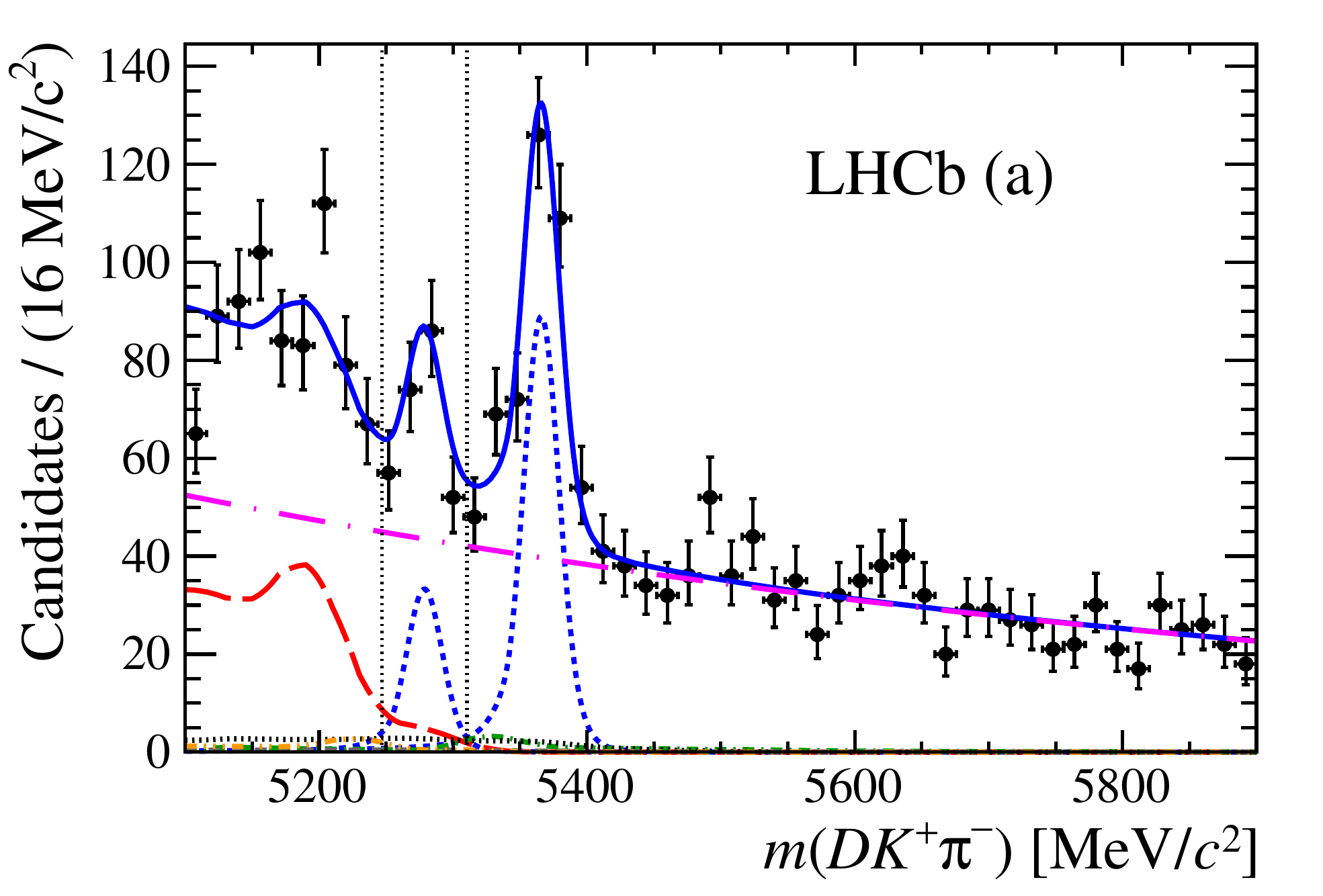}
  \includegraphics[scale=0.36]{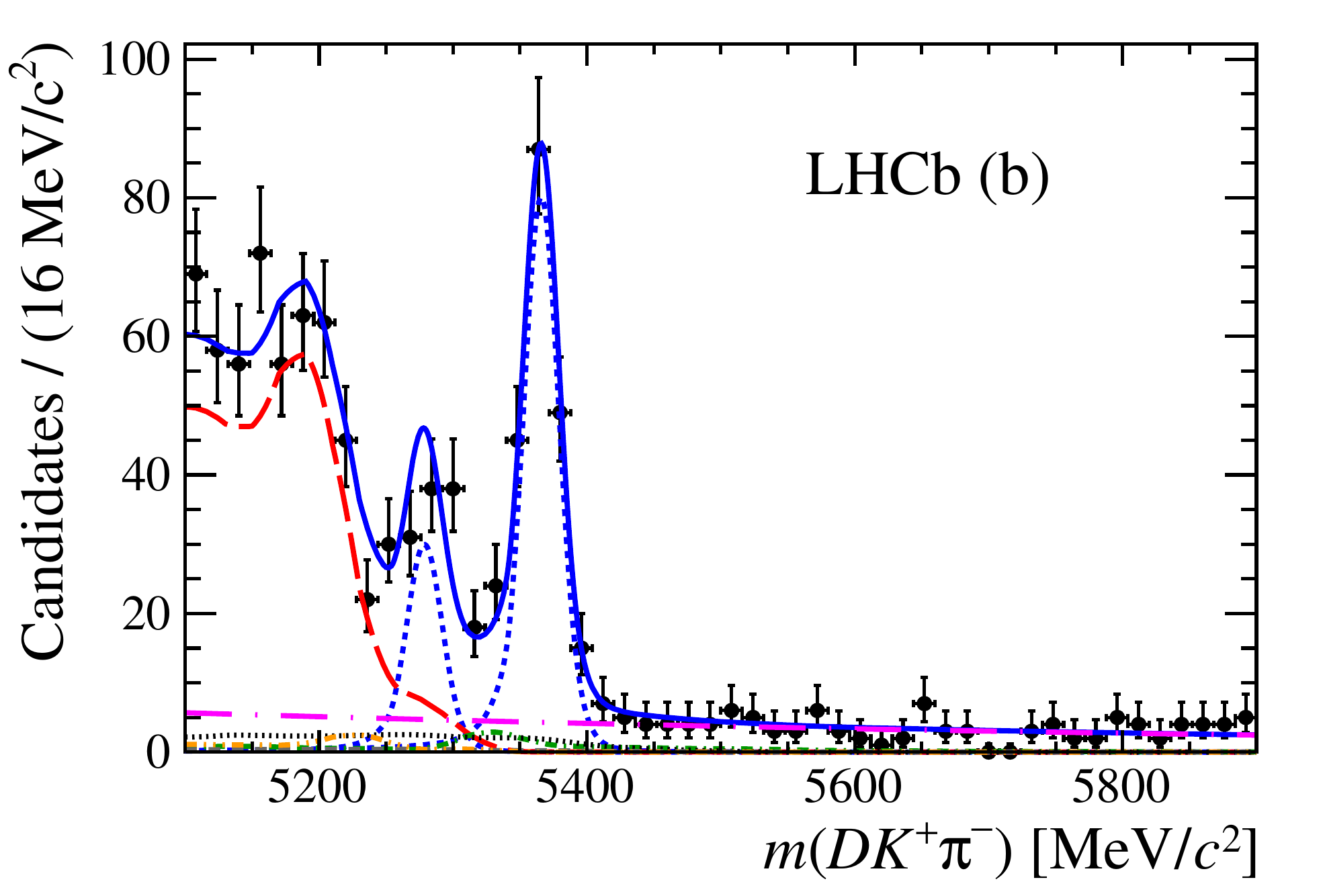}
  \includegraphics[scale=0.36]{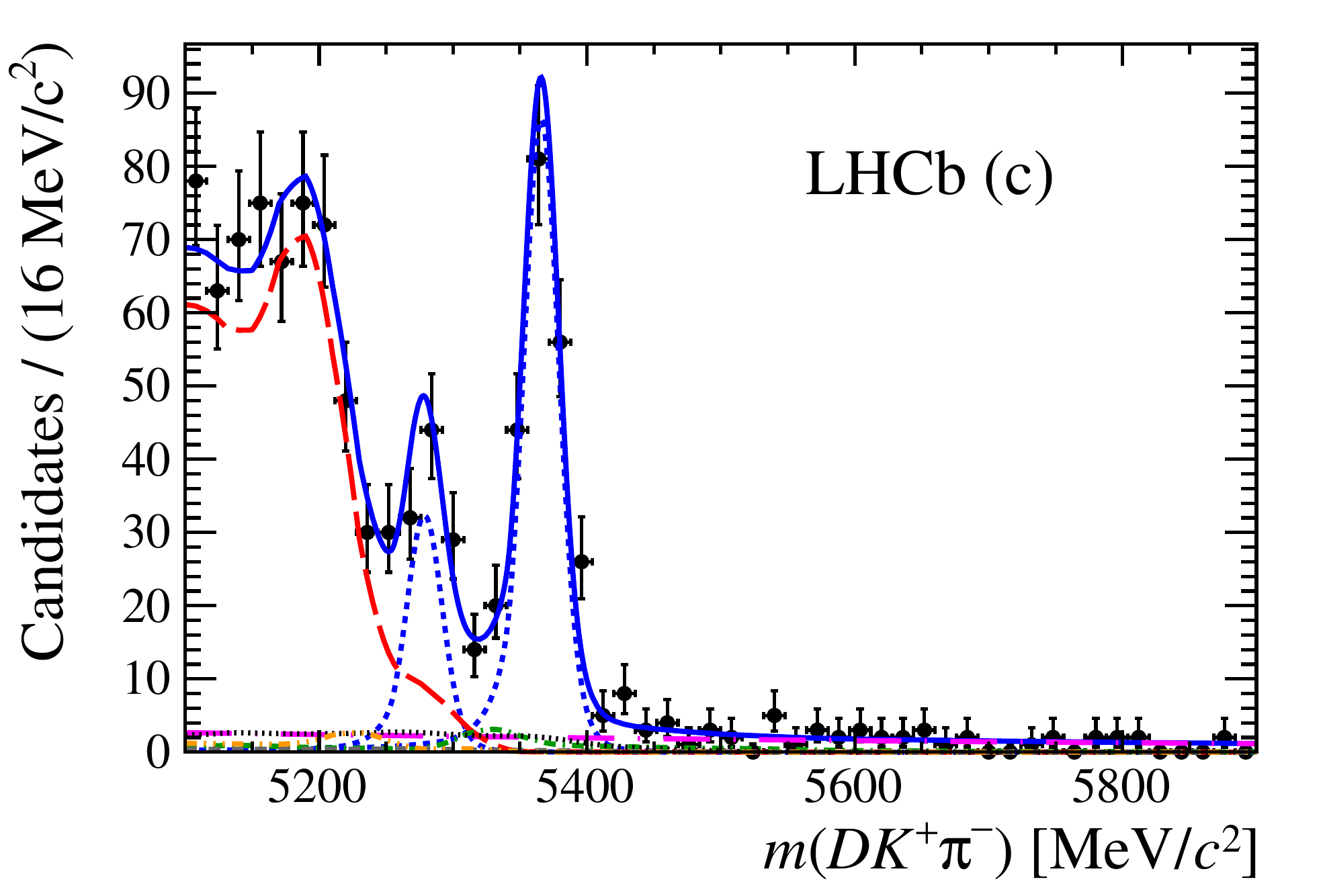}
  \includegraphics[scale=0.36]{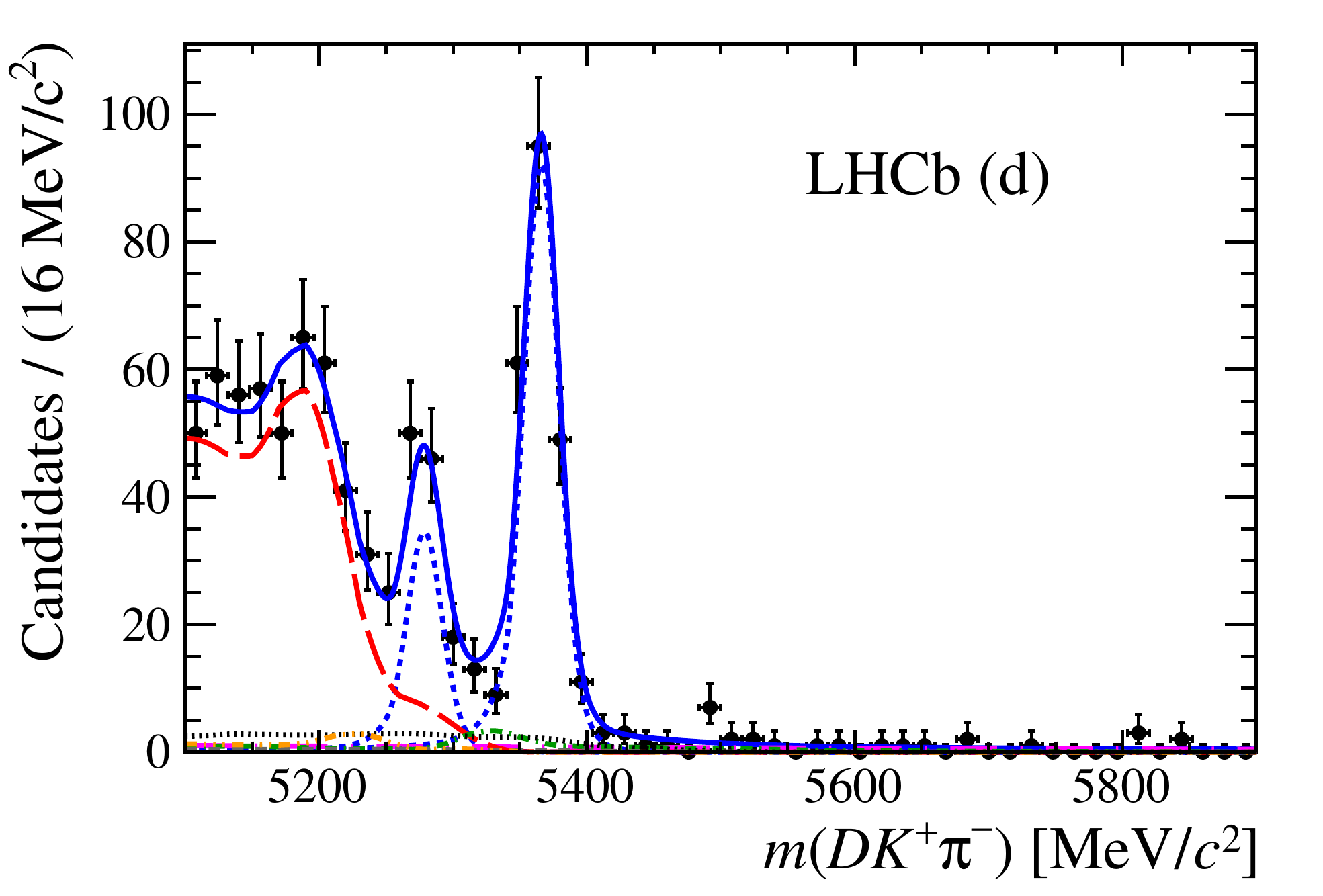}
  \includegraphics[scale=0.36]{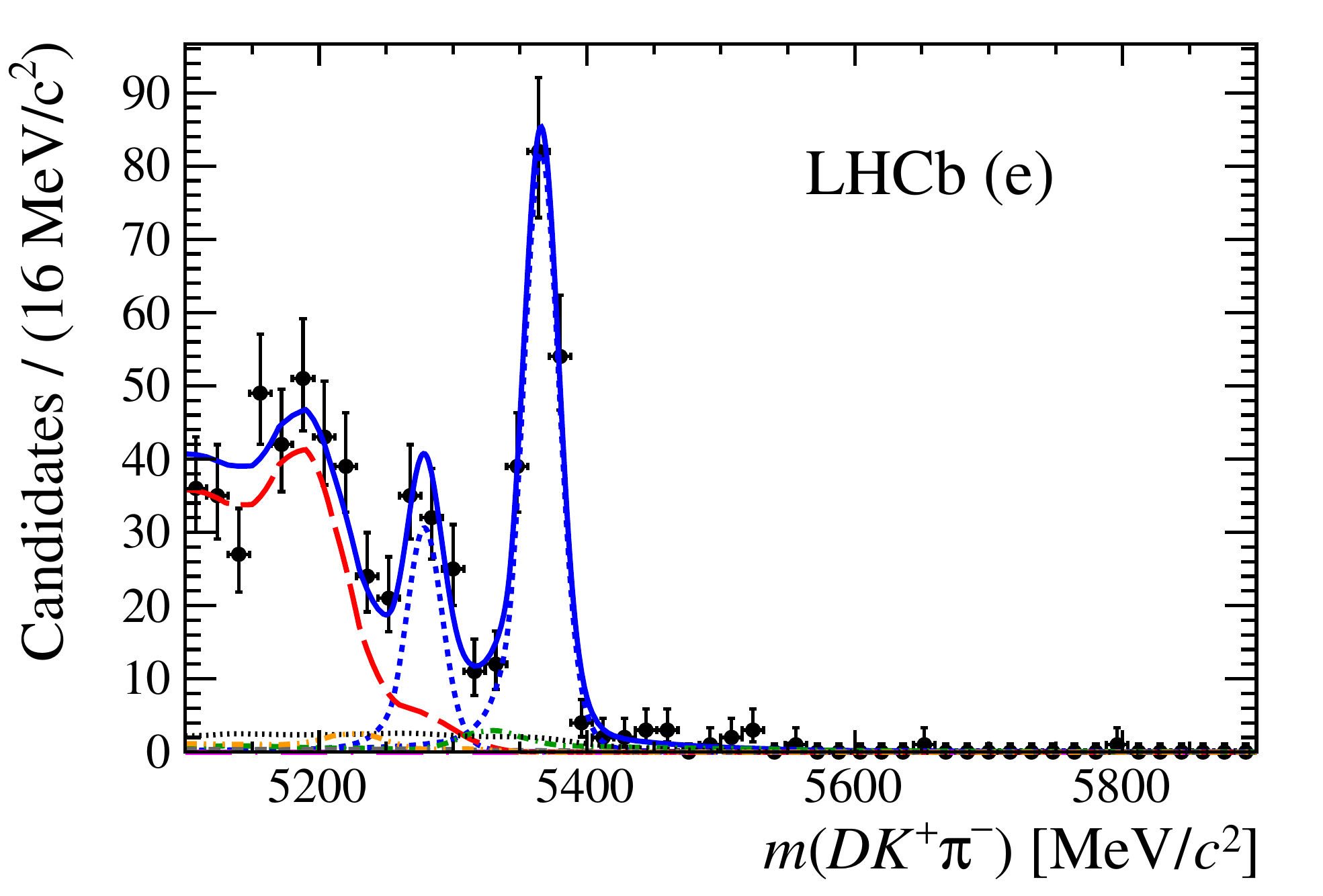}
  \includegraphics[scale=0.36]{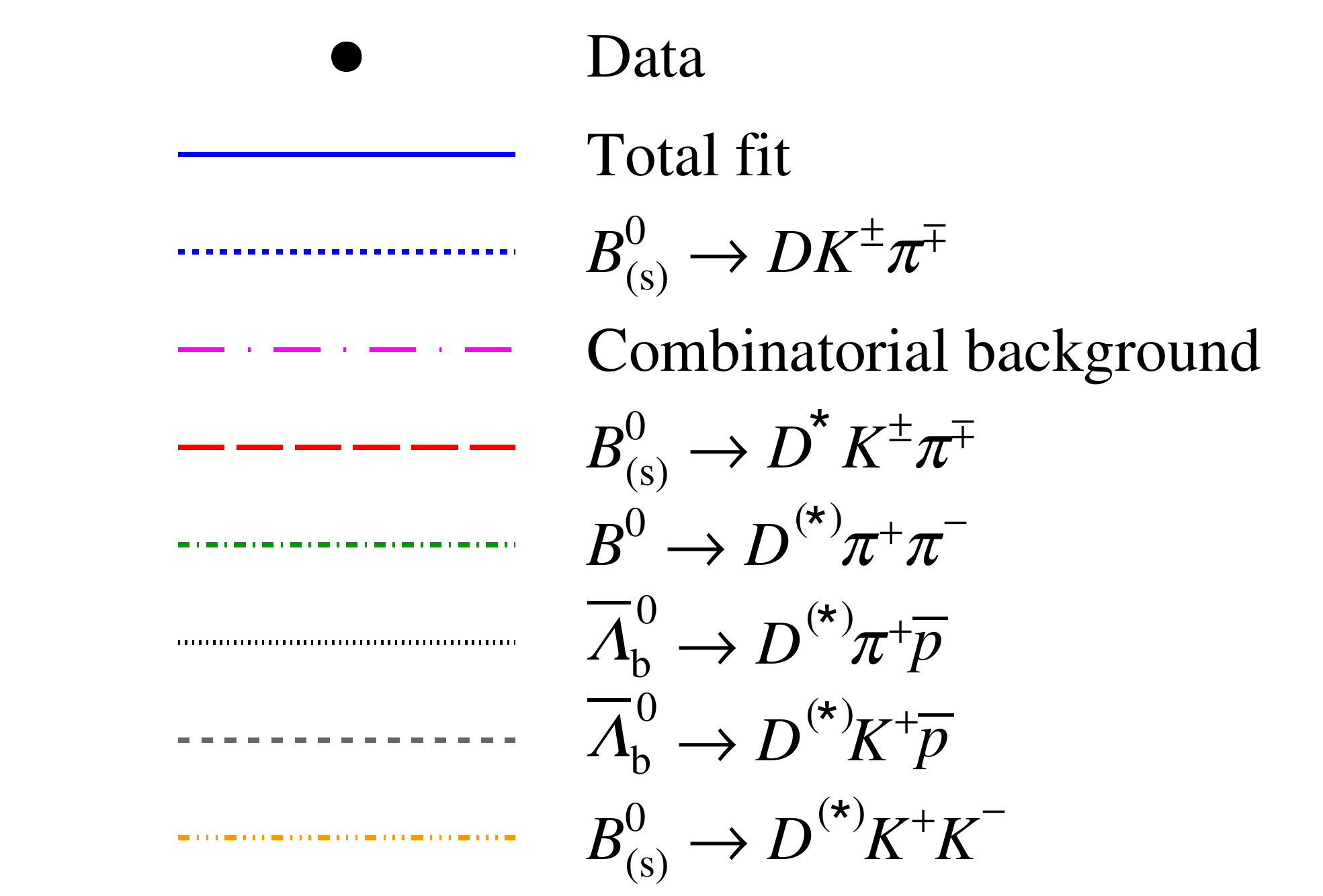}
\caption{\small
  Results of the fit to $\D\Kp\pim$, $\D\to\Kp\Km$ candidates shown separately in the five bins of the neural network output variable.
  The bins are shown, from (a)--(e), in order of increasing ${\cal S}/{\cal B}$.
  The components are as indicated in the legend.
  The vertical dotted lines in (a) show the signal window used for the fit to the Dalitz plot.}
\label{fig:fitKK}
\end{figure}

\begin{figure}[!tb]
\centering
  \includegraphics[scale=0.36]{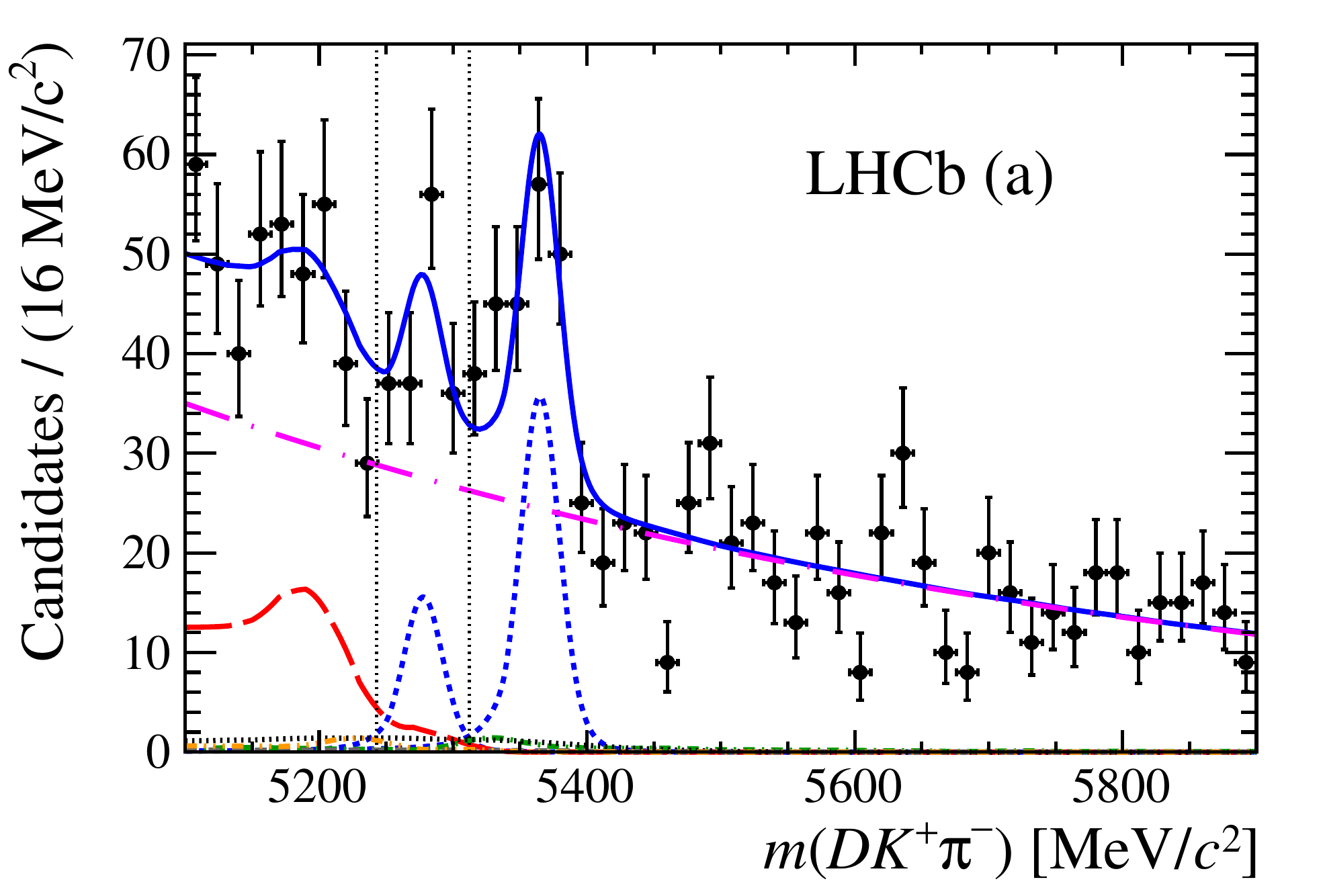}
  \includegraphics[scale=0.36]{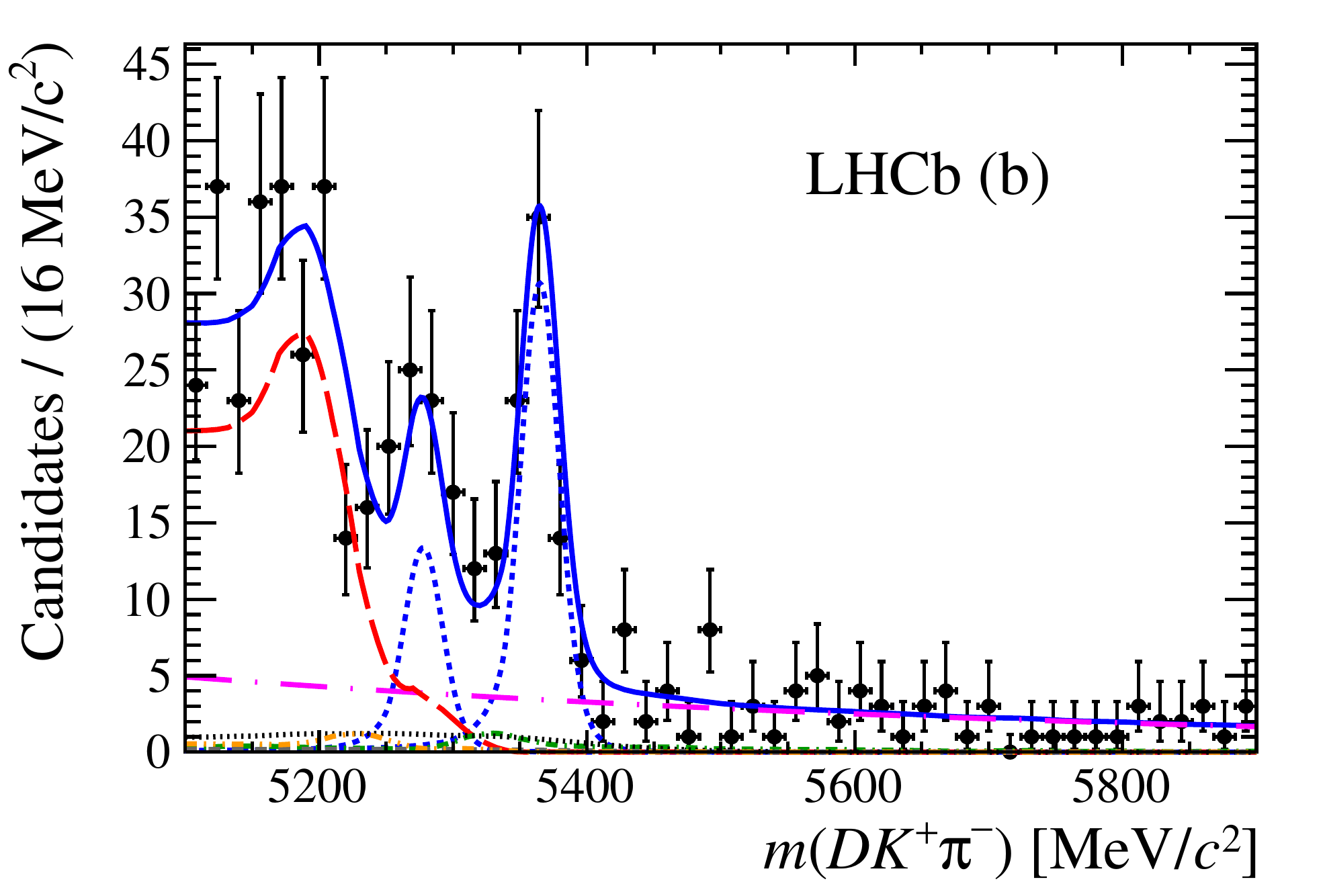}
  \includegraphics[scale=0.36]{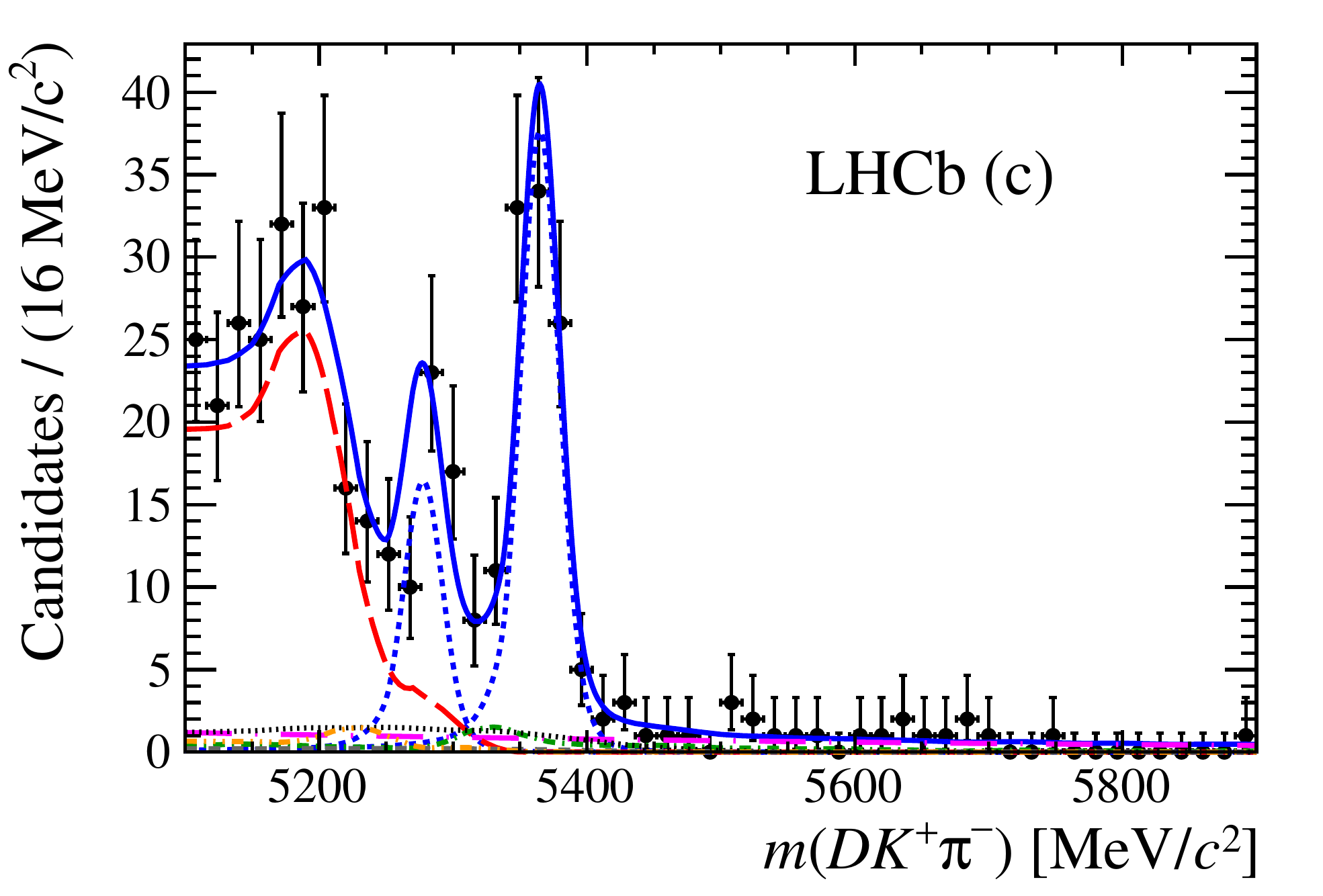}
  \includegraphics[scale=0.36]{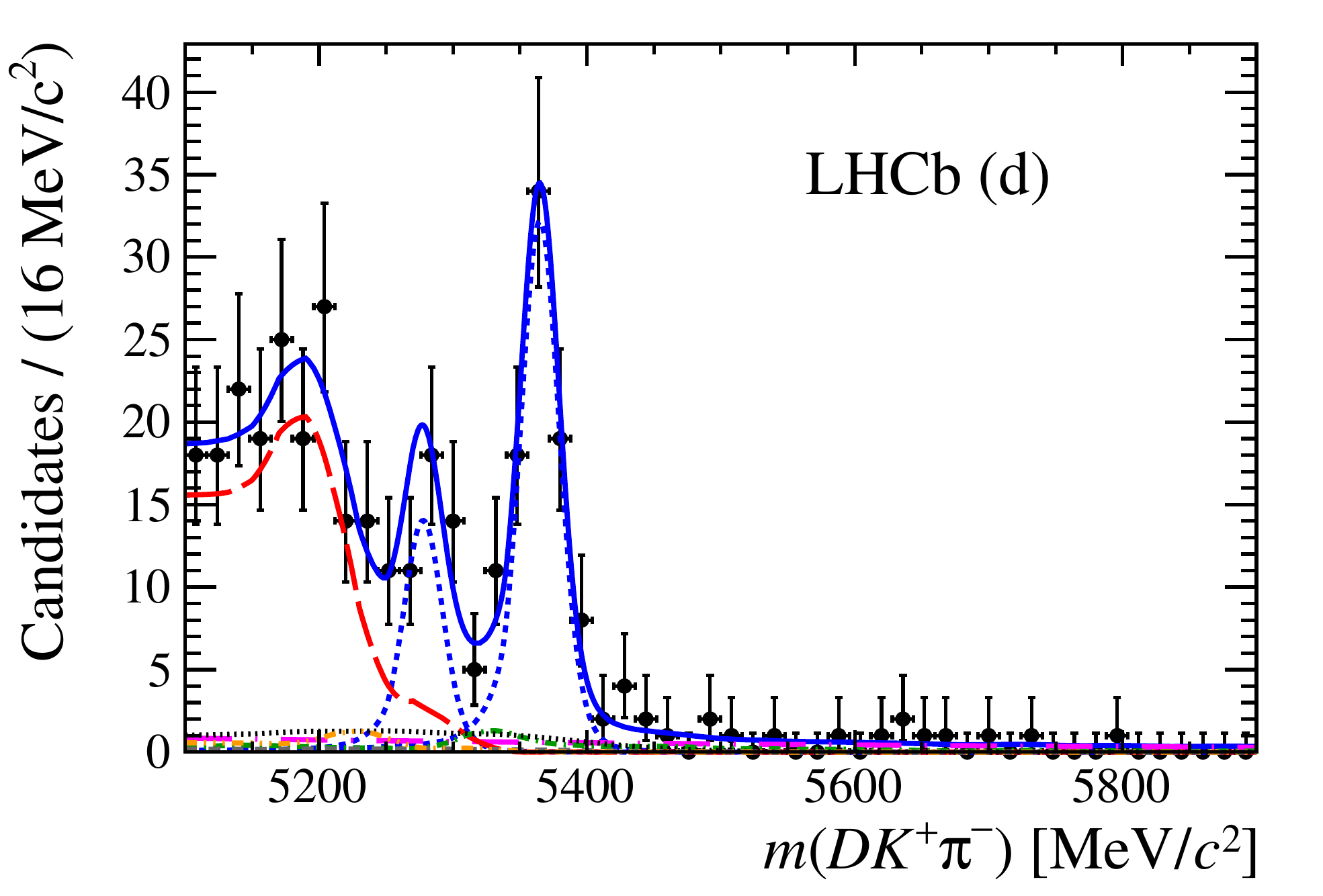}
  \includegraphics[scale=0.36]{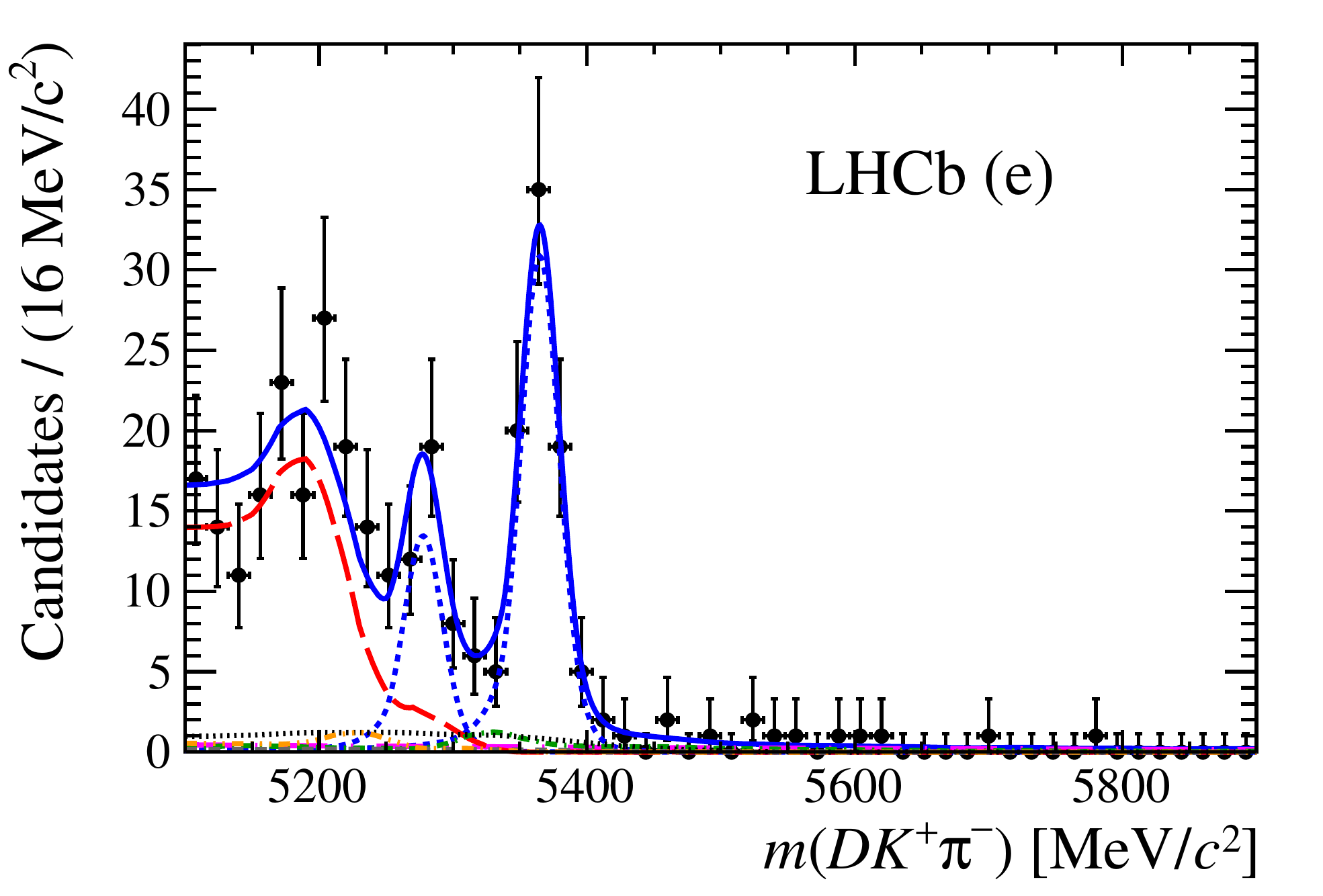}
  \includegraphics[scale=0.36]{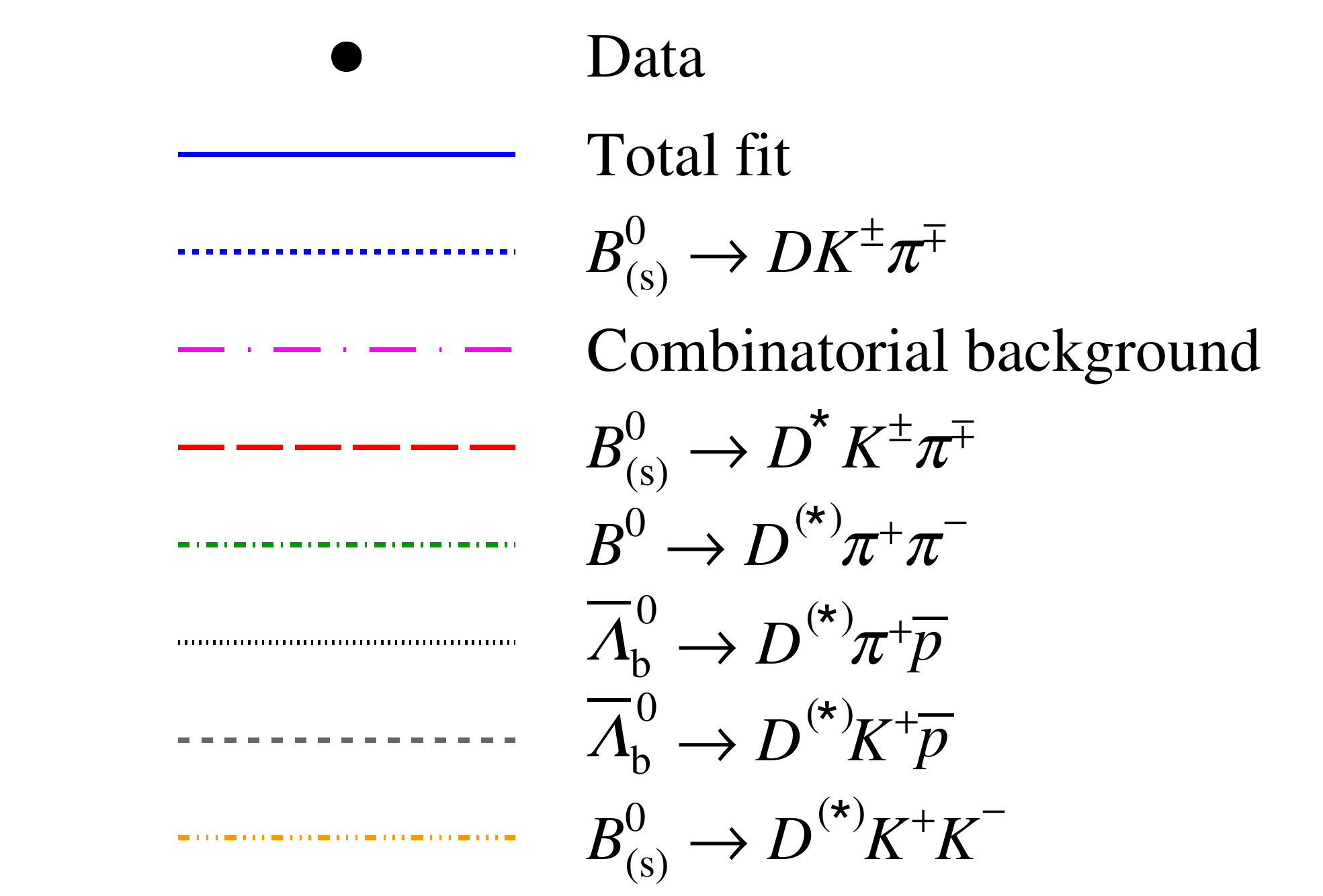}
\caption{\small
  Results of the fit to $\D\Kp\pim$, $\D\to\pip\pim$ candidates shown separately in the five bins of the neural network output variable.
  The bins are shown, from (a)--(e), in order of increasing ${\cal S}/{\cal B}$.
  The components are as indicated in the legend.
  The vertical dotted lines in (a) show the signal window used for the fit to the Dalitz plot.}
\label{fig:fitpipi}
\end{figure}

Projections of the fits separated by NN output bin in each sample are shown in Figs.~\ref{fig:fitKpi},~\ref{fig:fitKK} and~\ref{fig:fitpipi}.
The fitted parameters obtained from all three data samples are reported in Table~\ref{tab:fit}.
The parameters $\mu(B)$, $N({\rm core})/N({\rm total})$, $\sigma({\rm wide})/\sigma({\rm core})$ are, respectively, the peak position, the fraction of the signal function contained in the core and the relative width of the two components of the \Bz signal shape.
Quantities denoted $N$ are total yields of each fit component, while those denoted $f^{i}_{\rm signal}$ are fractions of the signal in NN output bin $i$ (with similar notation for the fractions of the partially reconstructed and combinatorial backgrounds).
The NN output bin labels 1--5 range from the bin with the lowest to highest
value of ${\cal S}/{\cal B}$.

\begin{table}[!tb]
  \centering
  \vspace{3ex}
  \caption{\small
    Results for the unconstrained parameters obtained from the fits to the three data samples.
    Entries where no number is given are fixed to zero.
    Fractions marked * are not varied in the fit, and give the difference between unity and the sum of the other fractions.
  }
  \label{tab:fit}
  \begin{tabular}{cr@{$\,\pm\,$}lr@{$\,\pm\,$}lr@{$\,\pm\,$}l}
    \hline
    & \multicolumn{2}{c}{$\D\to\Kp\pim$} & \multicolumn{2}{c}{$\D\to\Kp\Km$} & \multicolumn{2}{c}{$\D\to\pip\pim$} \\
    Parameter & \multicolumn{6}{c}{Value}\\
    \hline
$\mu(B) \ (\nspmevcc)$                     & $ 5278.3$ & $ 0.4  $    & $ 5278.7$ & $ 0.5  $    & $ 5277.7$ & $ 1.0  $    \\
$\sigma({\rm core}) \ (\nspmevcc)$         & $   12.7$ & $ 0.4  $    & $   12.7$ & $ 0.5  $    & $   13.9$ & $ 0.8  $    \\
$N({\rm core})/N({\rm total})$          & $  0.787$ & $ 0.017$    & $  0.798$ & $ 0.018$    & $  0.797$ & $ 0.018$    \\
$\sigma({\rm wide})/\sigma({\rm core})$ & $   1.80$ & $ 0.05 $    & $   1.75$ & $ 0.05 $    & $   1.76$ & $ 0.05 $    \\
Exp. slope $(c^2/\nspgev)$                  & $  -1.84$ & $ 0.13 $    & $  -1.05$ & $ 0.19 $    & $  -1.35$ & $ 0.26 $    \\
[1ex]
$N(\Bd\to\D K\pi)$                      & $   3125$ & $ 79   $    & $    418$ & $ 27   $    & $    185$ & $ 21   $    \\
$N(\Bs\to\D K\pi)$                      & $    146$ & $ 27   $    & $   1014$ & $ 41   $    & $    429$ & $ 28   $    \\
$N({\rm comb.~bkgd.})$                  & $   5694$ & $ 529  $    & $   2092$ & $ 95   $    & $   1288$ & $ 86   $    \\
$N(\B \to\DorDstar K + X)$              & $   2648$ & $ 454  $    & \multicolumn{2}{c}{---} & \multicolumn{2}{c}{---} \\
$N(\Bd\to\Dstar K\pi)$                  & $   3028$ & $ 115  $    & $    543$ & $ 48   $    & $    183$ & $ 33   $    \\
$N(\Bs\to\Dstar K\pi)$                  & \multicolumn{2}{c}{---} & $   1493$ & $ 77   $    & $    639$ & $ 52   $    \\
$N(\Bd\to\DorDstar \pi\pi)$             & $    783$ & $ 67   $    & $    146$ & $ 17   $    & $     72$ & $ 11   $    \\
$N(\Lb\to\DorDstar p\pi)$               & \multicolumn{2}{c}{---} & $    241$ & $ 47   $    & $    118$ & $ 26   $    \\
$N(\Lb\to\DorDstar pK)$                 & $    416$ & $ 64   $    & $     34$ & $ 9    $    & $     17$ & $ 5    $    \\
$N(\Bd\to\DorDstar KK)$                 & $    371$ & $ 51   $    & $     64$ & $ 15   $    & $     33$ & $ 8    $    \\
$N(\Bs\to\DorDstar KK)$                 & $    171$ & $ 47   $    & $     25$ & $ 11   $    & $     14$ & $ 6    $    \\
[1ex]
$f^{1}_{\rm signal}$                    & $  0.210$ & $ 0.012$    & $  0.187$ & $ 0.017$    & $  0.214$ & $ 0.029$    \\
$f^{2}_{\rm signal}$                    & $  0.192$ & $ 0.008$    & $  0.186$ & $ 0.011$    & $  0.184$ & $ 0.019$    \\
$f^{3}_{\rm signal}$                    & $  0.206$ & $ 0.008$    & $  0.201$ & $ 0.012$    & $  0.225$ & $ 0.019$    \\
$f^{4}_{\rm signal}$                    & $  0.201$ & $ 0.007$    & $  0.215$ & $ 0.012$    & $  0.193$ & $ 0.018$    \\
$f^{5}_{\rm signal}$*                   & $  0.190$ & $ 0.007$    & $  0.211$ & $ 0.011$    & $  0.184$ & $ 0.017$    \\ [1ex]
$f^{1}_{\rm part.~rec.~bkgd.}$          & $  0.214$ & $ 0.023$    & $  0.145$ & $ 0.020$    & $  0.152$ & $ 0.042$    \\
$f^{2}_{\rm part.~rec.~bkgd.}$          & $  0.214$ & $ 0.010$    & $  0.217$ & $ 0.011$    & $  0.254$ & $ 0.021$    \\
$f^{3}_{\rm part.~rec.~bkgd.}$          & $  0.215$ & $ 0.011$    & $  0.267$ & $ 0.013$    & $  0.237$ & $ 0.021$    \\
$f^{4}_{\rm part.~rec.~bkgd.}$          & $  0.193$ & $ 0.010$    & $  0.215$ & $ 0.012$    & $  0.189$ & $ 0.019$    \\
$f^{5}_{\rm part.~rec.~bkgd.}$*         & $  0.164$ & $ 0.009$    & $  0.156$ & $ 0.010$    & $  0.169$ & $ 0.018$    \\
[1ex]
$f^{1}_{\rm comb.~bkgd.}$               & $  0.870$ & $ 0.013$    & $  0.849$ & $ 0.012$    & $  0.828$ & $ 0.018$    \\
$f^{2}_{\rm comb.~bkgd.}$               & $  0.094$ & $ 0.008$    & $  0.092$ & $ 0.009$    & $  0.116$ & $ 0.014$    \\
$f^{3}_{\rm comb.~bkgd.}$               & $  0.025$ & $ 0.004$    & $  0.043$ & $ 0.007$    & $  0.027$ & $ 0.008$    \\
$f^{4}_{\rm comb.~bkgd.}$               & $  0.009$ & $ 0.003$    & $  0.017$ & $ 0.005$    & $  0.019$ & $ 0.007$    \\
$f^{5}_{\rm comb.~bkgd.}$*              & $  0.002$ & $ 0.002$    & $  0.000$ & $ 0.000$    & $  0.010$ & $ 0.006$    \\
\hline
\end{tabular}
\vspace{3ex}
\end{table}

\section{Dalitz plot analysis}
\label{sec:gamma-fit}

Candidates within the signal region are used in the DP analysis.
A simultaneous fit is performed to the samples with different $D$ decays by using the {\it J}{\sc fit} method~\cite{Ben-Haim:2014afa} as implemented in the {\tt Laura++} package~\cite{Laura++}.
The likelihood function contains signal and background terms, with yields in each NN output bin fixed according to the results obtained previously.
The NN output bin with the lowest ${\cal S}/{\cal B}$ value in the $D \to \Kp\pim$ sample only is found not to contribute significantly to the sensitivity and is susceptible to mismodelling of the combinatorial background; it is therefore excluded from the subsequent analysis.

The signal probability function is derived from the isobar model obtained in Ref.~\cite{LHCb-PAPER-2015-017}, with amplitude
\begin{equation}
  \label{eqn:amp}
  {\cal A}\left(m^2(D\pim), m^2(\Kp\pim)\right) = \sum_{j=1}^{N} c_j F_j\left(m^2(D\pim), m^2(\Kp\pim)\right) \,,
\end{equation}
where $c_j$ are complex coefficients describing the relative contribution for each intermediate process, and the $F_j\left(m^2(D\pim),m^2(\Kp\pim)\right)$ terms describe the resonant dynamics through the lineshape, angular distribution and barrier factors.
The sum is over amplitudes from the $D_0^*(2400)^-$, $D_2^*(2460)^-$, $\Kstar(892)^0$, $\Kstar(1410)^0$ and $K_2^*(1430)^0$ resonances as well as a $\Kp\pim$ S-wave component and both S-wave and P-wave nonresonant $D\pim$ amplitudes~\cite{LHCb-PAPER-2015-017}.
The masses and widths of $\Kp\pim$ resonances are fixed, and those of $D\pim$ resonances are constrained within uncertainties to known values~\cite{PDG2014,LHCb-PAPER-2014-070,LHCb-PAPER-2013-026,LHCb-PAPER-2015-017}.
The values of the $c_j$ coefficients are allowed to vary in the fit, as are the shape parameters of the nonresonant amplitudes.

For the $D \to \Kp\pim$ sample, the contribution from the $V_{ub}$ amplitude followed by doubly-Cabibbo-suppressed \D decay is negligible.
This sample can therefore be treated as if only the $V_{cb}$ amplitude contributes, and the signal probability function is given by Eq.~(\ref{eqn:amp}).
For the samples with $D \to \Kp\Km$ and $\pip\pim$ decays, the $c_j$ terms are modified,
\begin{equation}
  \label{eq:param1}
  c_j \longrightarrow
  \left\{
    \begin{array}{cl}
      c_j & {\rm for \ a} \ \D\pim \ {\rm resonance} \, ,\\
      c_j \left[ 1 + x_{\pm,\,j} + i y_{\pm,\,j} \right] & {\rm for \ a} \ \Kp\pim \ {\rm resonance} \, ,
    \end{array}
    \right.
\end{equation}
with $x_{\pm,\,j} = r_{B,\,j} \cos\left(\delta_{B,\,j} \pm \gamma\right)$ and $y_{\pm,\,j} = r_{B,\,j} \sin\left(\delta_{B,\,j} \pm \gamma\right)$, where the $+$ and $-$ signs correspond to $\Bz$ and $\Bzb$ DPs, respectively.
Here $r_{B,\,j}$ and $\delta_{B,\,j}$ are the relative magnitude and strong phase of the $V_{ub}$ and $V_{cb}$ amplitudes for each $\Kp\pim$ resonance $j$.
In this analysis the $x_{\pm,\,j}$ and $y_{\pm,\,j}$ parameters are measured only for the $\Kstar(892)^0$ resonance, which has a large enough yield and a sufficiently well-understood lineshape to allow reliable determinations of these parameters; therefore the $j$ subscript is omitted hereafter.
In addition, a component corresponding to the $\Bz \to D_{s1}^*(2700)^+\pim$ decay, which is mediated by the $V_{ub}$ amplitude alone, is included in the fit with mass and width parameters fixed to their known values~\cite{PDG2014,Lees:2014abp} and magnitude constrained according to expectation based on the $\Bz \to D_{s1}^*(2700)^+\Dm$ decay rate~\cite{Lees:2014abp}.

The signal efficiency and backgrounds are modelled in the likelihood function, separately for each of the samples, following Refs.~\cite{LHCb-PAPER-2015-017,LHCb-PAPER-2014-035,LHCb-PAPER-2014-036}.
The DP distribution of combinatorial background is obtained from a sideband in \B candidate mass, defined as $5400 \ (5450) < m(D\Kp\pim) < 5900 \mevcc$ for the samples with $D \to \Kp\pim$ ($D \to \Kp\Km$ or $\pip\pim$).
The shapes of partially reconstructed and misidentified backgrounds are obtained from simulated samples based on known DP distributions~\cite{LHCb-PAPER-2014-035,LHCb-PAPER-2014-036,LHCb-PAPER-2014-070,Kuzmin:2006mw,Abe:2004cw,LHCb-PAPER-2012-018,LHCb-PAPER-2013-056}.
Combinatorial background is the largest component in the NN output bins with the lowest ${\cal S}/{\cal B}$ values, while in the purest bins in the $D \to \Kp\Km$ and $\pip\pim$ samples the $\Bs \to \Dstar\Km\pip$ background makes an important contribution.
Background sources with yields below $2\,\%$ relative to the signal in all NN bins are neglected, as indicated in Tables~\ref{tab:Bdyields-Kpi},~\ref{tab:Bdyields-KK} and~\ref{tab:Bdyields-pipi}.

The fit procedure is validated with ensembles of pseudoexperiments.
In addition, samples of $\Bs \to D\Km\pip$ decays are selected for each of the $D$ decays.
These are used to test the fit with a model based on that of Refs.~\cite{LHCb-PAPER-2014-035,LHCb-PAPER-2014-036} and where $D\Km$ resonances have contributions only from $V_{cb}$ amplitudes, while the coefficients for $\Km\pip$ resonances are parametrised by Eq.~(\ref{eq:param1}).
The results are
\begin{equation*}
  \begin{array}{c}
    x_{+}(\Bs \to D\Kstarb(892)^0) = \phantom{-}0.05 \pm 0.05 \, , \qquad
    y_{+}(\Bs \to D\Kstarb(892)^0) = -0.08 \pm 0.11 \, ,\\
    x_{-}(\Bs \to D\Kstarb(892)^0) = \phantom{-}0.01 \pm 0.05 \, , \qquad
    y_{-}(\Bs \to D\Kstarb(892)^0) = -0.08 \pm 0.12 \, ,
  \end{array}
\end{equation*}
where the uncertainties are statistical only.
No significant \CP violation effect is observed, consistent with the expectation that $V_{ub}$ amplitudes are highly suppressed in this control channel.

\section{Systematic uncertainties}
\label{sec:systematic}

Sources of systematic uncertainty on the $x_\pm$ and $y_\pm$ parameters can be divided into two categories: experimental and model uncertainties.
These are summarised in Tables~\ref{tab:expt-syst} and~\ref{tab:model-syst}.
The former category includes effects due to knowledge of the signal and background yields in the signal region (denoted ``${\cal S}/{\cal B}$'' in Table~\ref{tab:expt-syst}), the variation of the efficiency ($\epsilon$) across the Dalitz plot, the background Dalitz plot distributions (${\cal B}$ DP) and fit bias, all of which are evaluated in similar ways to those described in Ref.~\cite{LHCb-PAPER-2015-017}.
Additionally, effects that may induce fake asymmetries, including asymmetry between \Bzb and \Bz candidates in the background yields (${\cal B}$ asym.) as well as asymmetries in the background Dalitz plot distributions (${\cal B}$ DP asym.) and in the efficiency variation ($\epsilon$ asym.) are accounted for.
The largest source of uncertainty in this category arises from lack of knowledge of the DP distribution for the $\Bs \to \Dstar\Km\pip$ background.

Model uncertainties arise due to fixing parameters in the amplitude model (denoted ``fixed pars.'' in Table~\ref{tab:model-syst}), the addition or removal of marginal components, namely the $\Kstar(1410)^0$, $\Kstar(1680)^0$, $D_1^*(2760)^-$, $D_3^*(2760)^-$ and $D_{s2}^*(2573)^+$ resonances, in the Dalitz plot fit (add/rem.), and the use of alternative models for the $\Kp\pim$ S-wave and $D\pim$ nonresonant amplitudes (alt.\ mod.); all of these are evaluated as in Ref.~\cite{LHCb-PAPER-2015-017}.
The possibilities of \CP violation associated to the $D_{s1}^*(2700)^+$ amplitude ($D_s^{**}$ $\CP$V), and of independent \CP violation parameters in the two components of the $\Kp\pim$ S-wave amplitude~\cite{lass} ($K\pi_{\rm S-wave}$ $\CP$V), are also accounted for.
The largest source of uncertainty in this category arises from changing the description of the $\Kp\pim$ S-wave.
Other possible sources of systematic uncertainty, such as production asymmetry~\cite{LHCb-PAPER-2014-042} or \CP violation in the $D\to\Kp\Km$ and $\pip\pim$ decays~\cite{HFAG,LHCb-PAPER-2014-069,LHCb-PAPER-2015-055}, are found to be negligible.

The total uncertainties are obtained by combining all sources in quadrature.
The leading sources of systematic uncertainty are expected to be reducible with larger data samples.

\begin{table}[!tb]
  \centering
  \caption{\small
   Experimental systematic uncertainties.}
  \label{tab:expt-syst}
  \begin{tabular}{ccccccccc}
    \hline
    Parameter & \multicolumn{8}{c}{Uncertainty}\\
    & ${\cal S}/{\cal B}$ & $\epsilon$ & ${\cal B}$ DP & fit bias & ${\cal B}$ asym. & ${\cal B}$ DP asym. & $\epsilon$ asym. & total \\
    \hline
    $x_{+}$ & 0.010 & 0.035 & 0.046 & 0.021 & 0.007 & 0.049 & 0.000 & 0.079 \\
    $x_{-}$ & 0.026 & 0.028 & 0.063 & 0.019 & 0.010 & 0.045 & 0.001 & 0.089 \\
    $y_{+}$ & 0.019 & 0.042 & 0.122 & 0.066 & 0.017 & 0.027 & 0.000 & 0.149 \\
    $y_{-}$ & 0.024 & 0.022 & 0.054 & 0.035 & 0.018 & 0.071 & 0.000 & 0.103 \\
    \hline
  \end{tabular}
\end{table}

\begin{table}[!tb]
 \centering
 \caption{\small
    Model uncertainties.}
  \label{tab:model-syst}
  \begin{tabular}{ccccccc}
    \hline
    Parameter & \multicolumn{6}{c}{Uncertainty}\\
    & fixed pars. & add/rem. & alt.\ mod. & $D_s^{**}$ $\CP$V & $K\pi_{\rm S-wave}$ $\CP$V & total \\
    \hline
    $x_{+}$ & 0.027 & 0.028 & 0.068 & 0.008 & 0.003 & 0.079\\
    $x_{-}$ & 0.030 & 0.034 & 0.076 & 0.056 & 0.022 & 0.107\\
    $y_{+}$ & 0.075 & 0.061 & 0.131 & 0.012 & 0.047 & 0.170\\
    $y_{-}$ & 0.040 & 0.066 & 0.255 & 0.286 & 0.064 & 0.396\\
    \hline
  \end{tabular}
\end{table}

\section{Results and summary}
\label{sec:results}

\begin{figure}[!tb]
  \centering
  \includegraphics[width=0.45\textwidth]{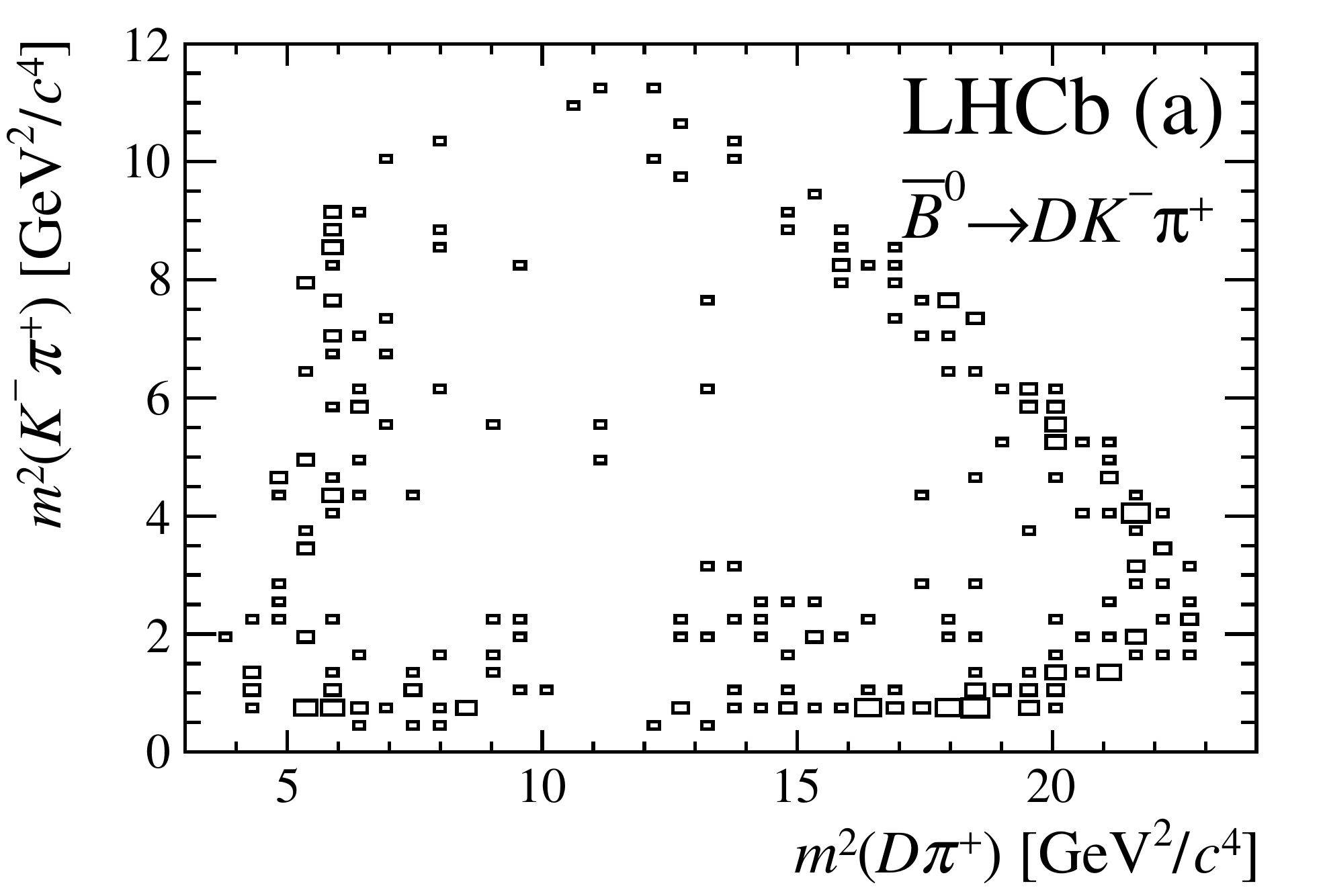}
  \includegraphics[width=0.45\textwidth]{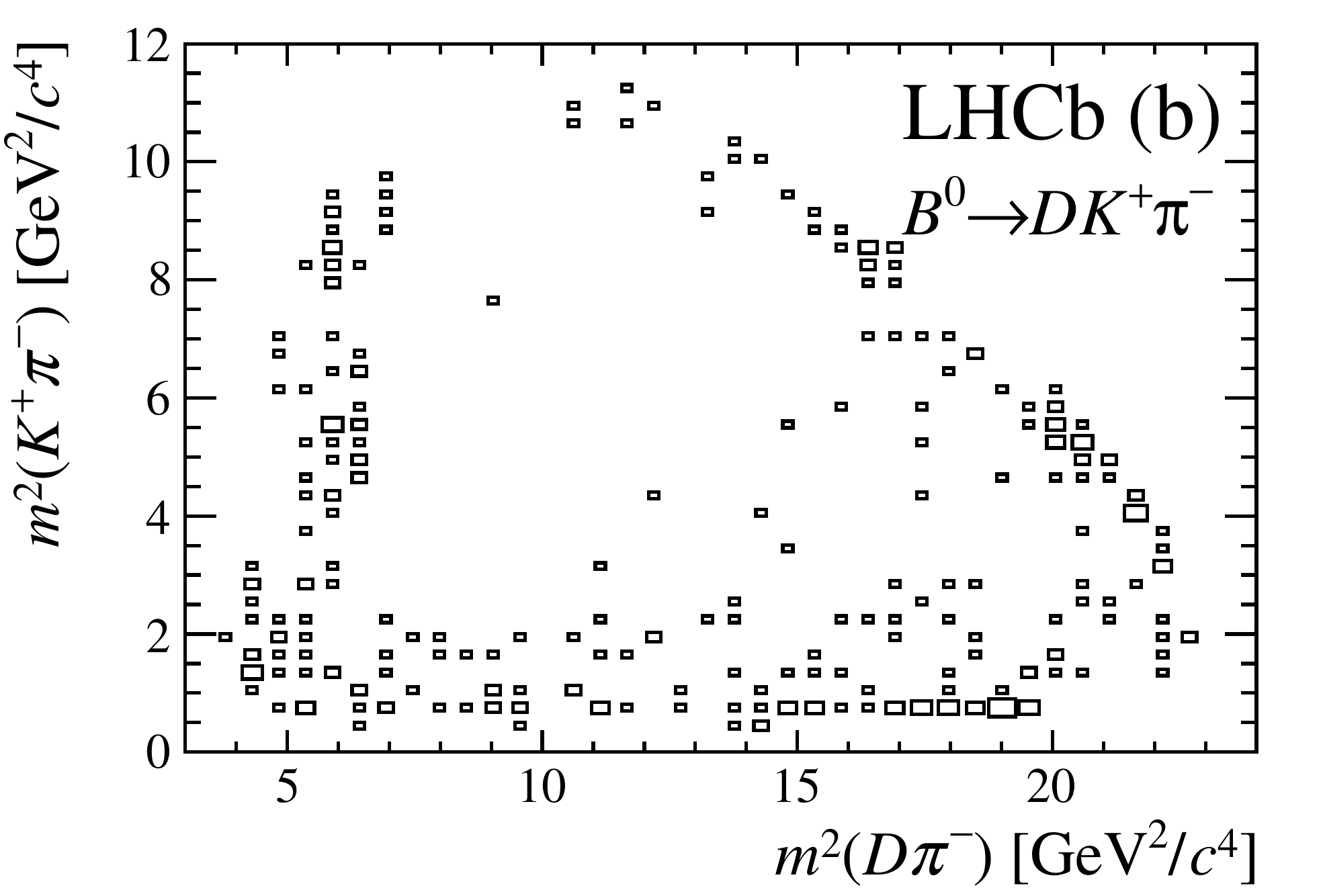}
  \caption{\small
    Dalitz plots for candidates in the \B candidate mass signal region in the $\D\to\Kp\Km$ and $\pip\pim$ samples for (a) \Bzb and (b) \Bz candidates.
    Only candidates in the three purest NN bins are included.
    Background has not been subtracted, and therefore some contribution from $\Bsb \to \Dstarz\Kp\pim$ decays is expected at low $m(D\Kp)$ (\ie\ along the top right diagonal).
  }
  \label{fig:CPDPs}
\end{figure}

\begin{figure}[!tb]
  \centering
  \includegraphics[width=0.47\textwidth]{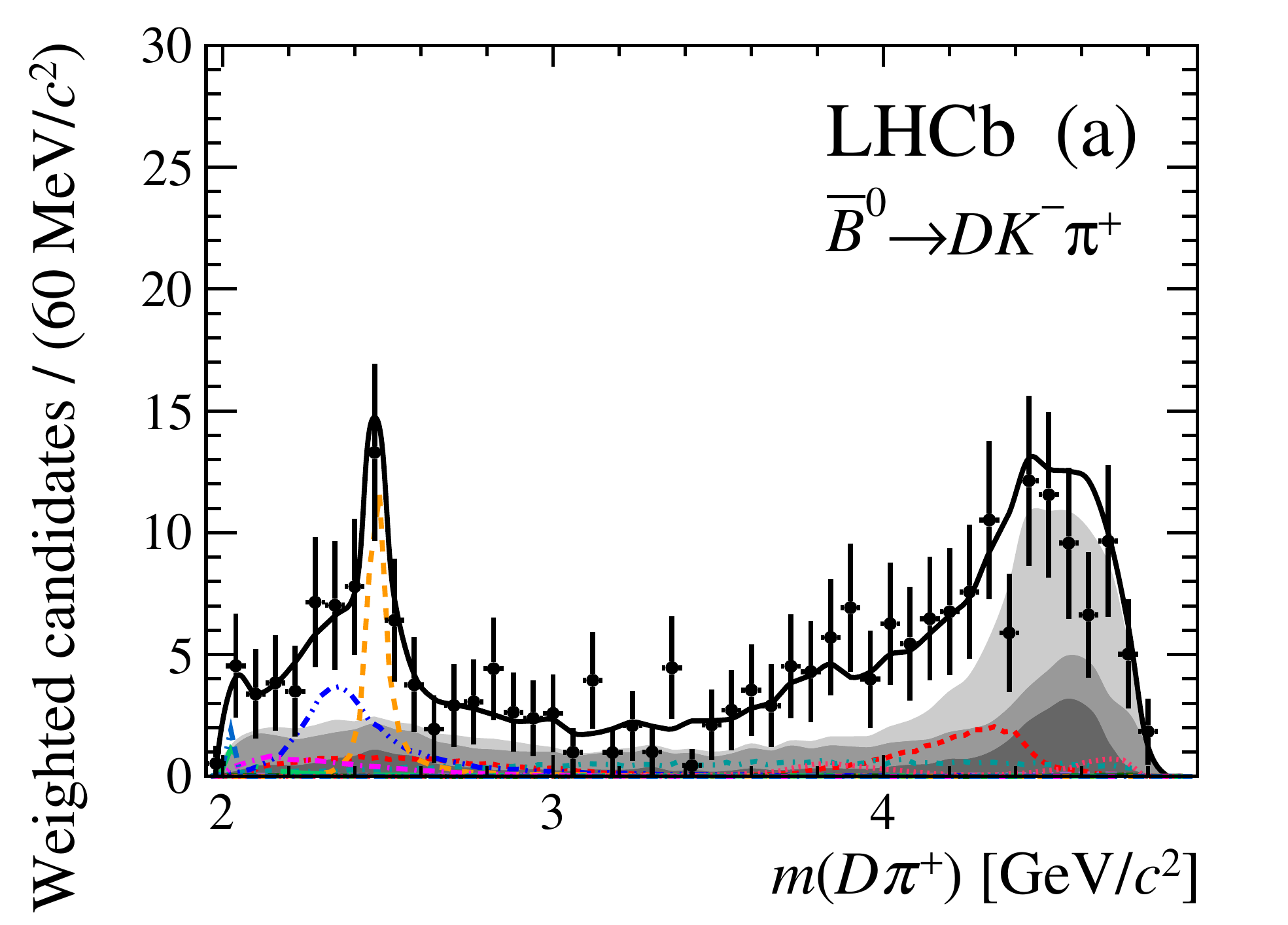}
  \includegraphics[width=0.47\textwidth]{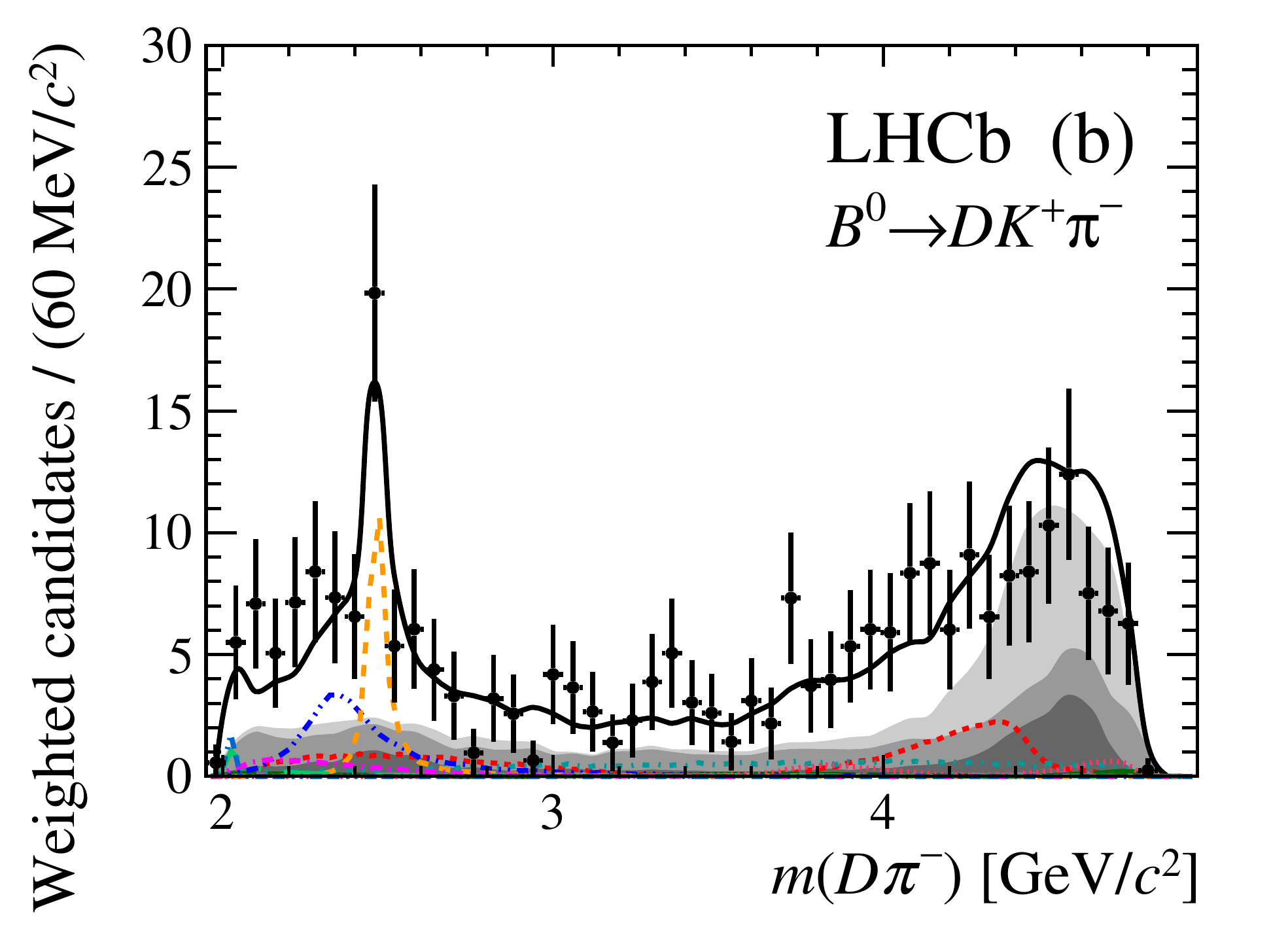}
  \includegraphics[width=0.47\textwidth]{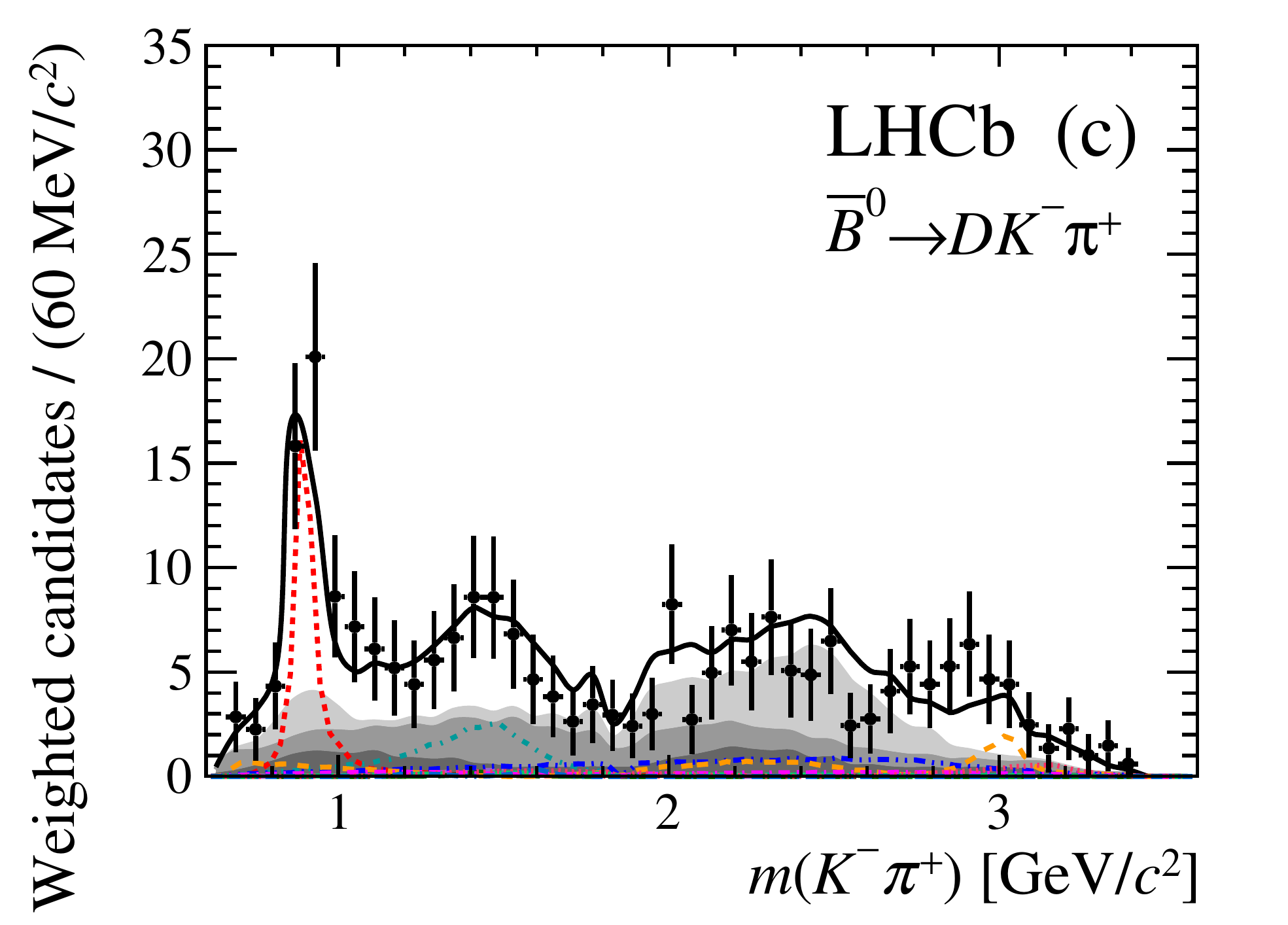}
  \includegraphics[width=0.47\textwidth]{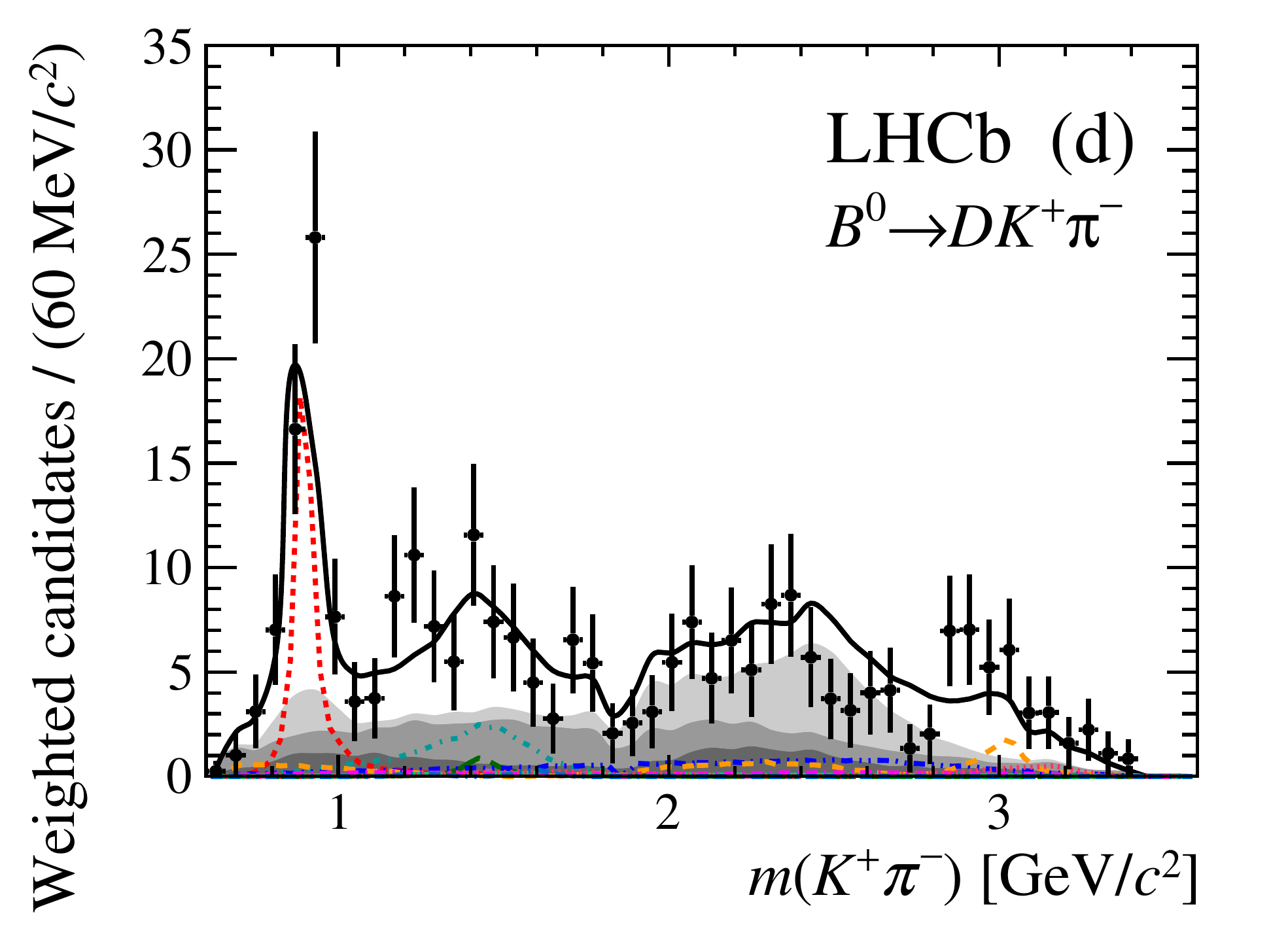}
  \includegraphics[width=0.47\textwidth]{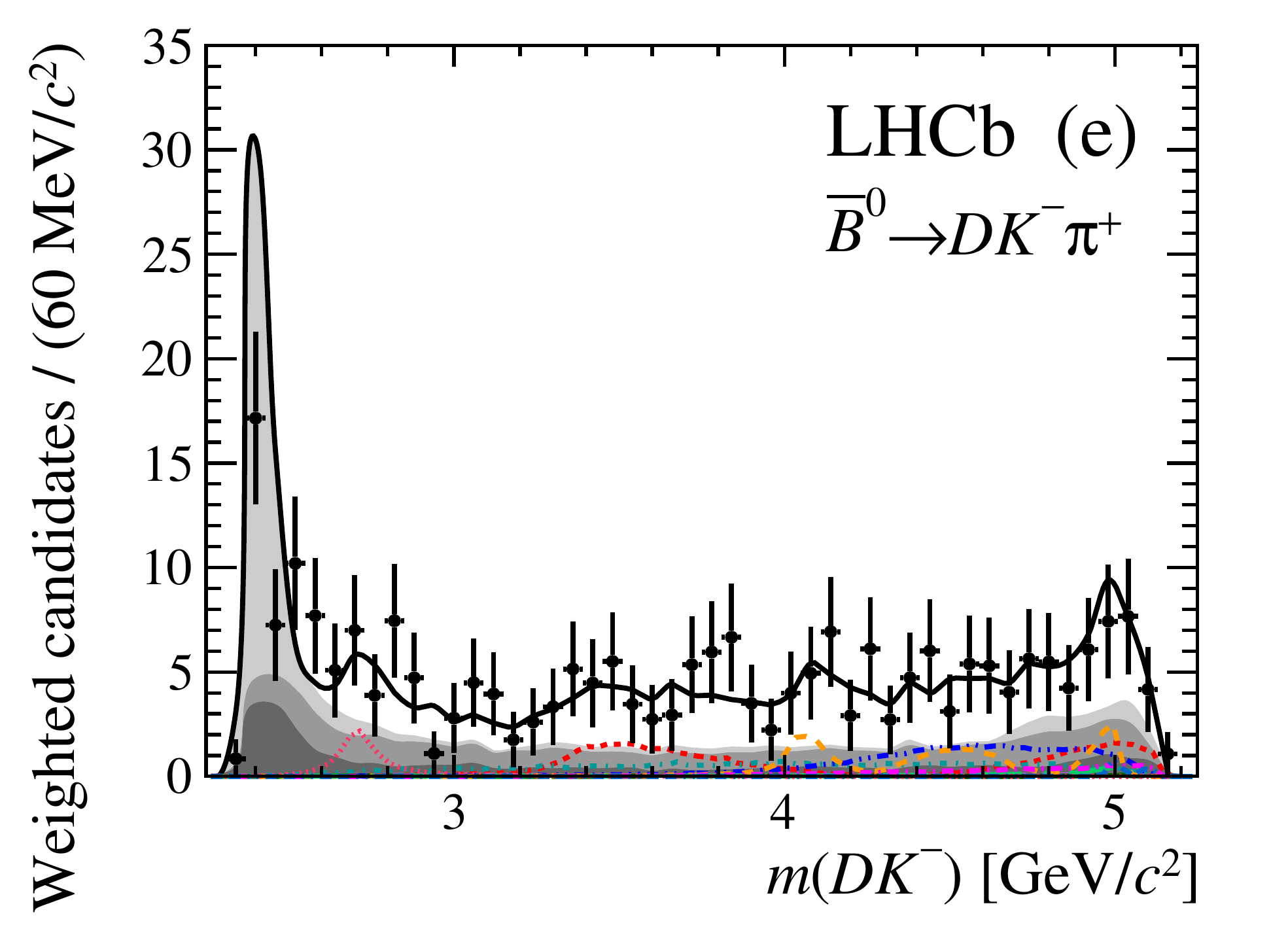}
  \includegraphics[width=0.47\textwidth]{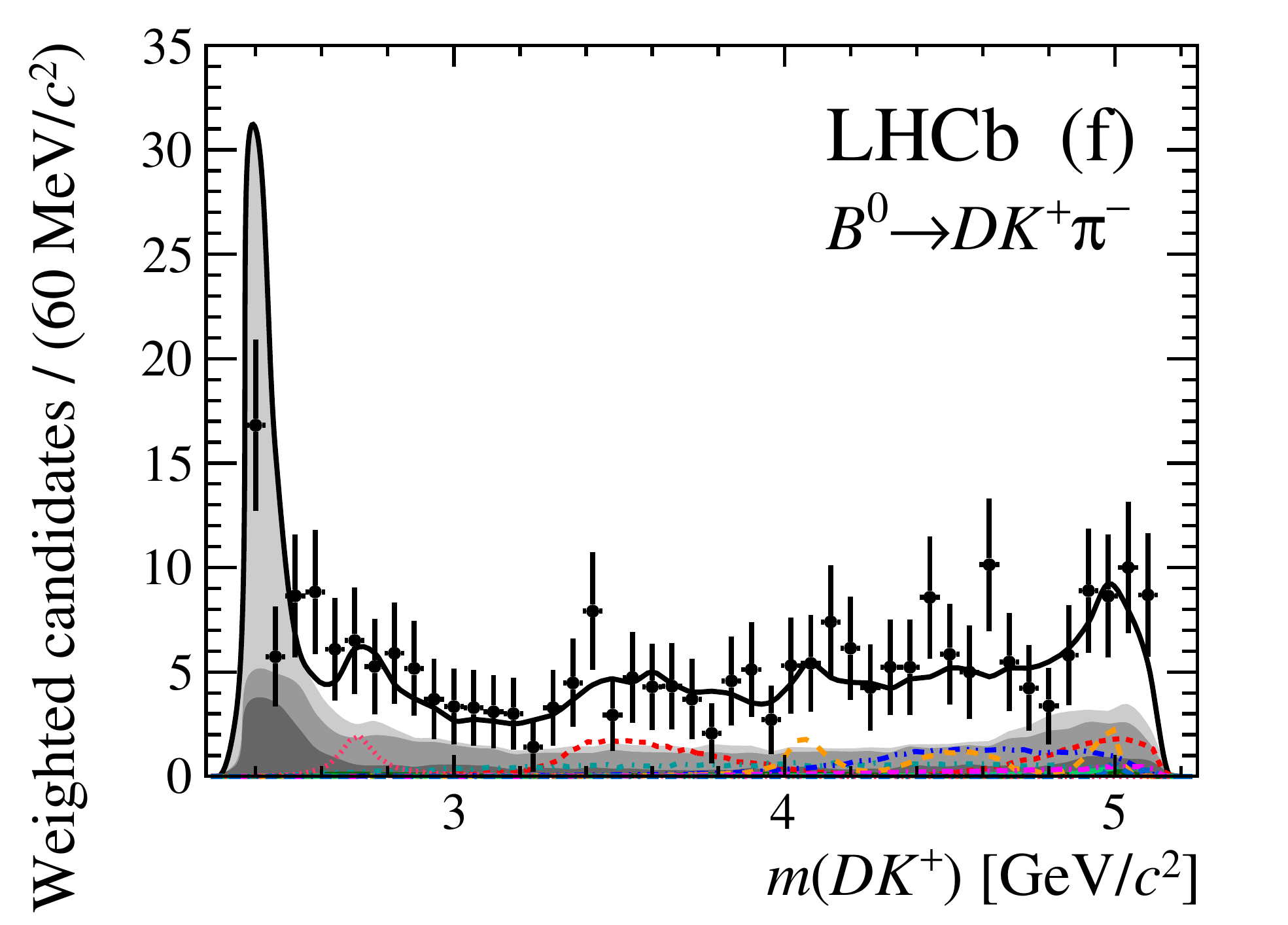}
  \includegraphics[width=0.80\textwidth]{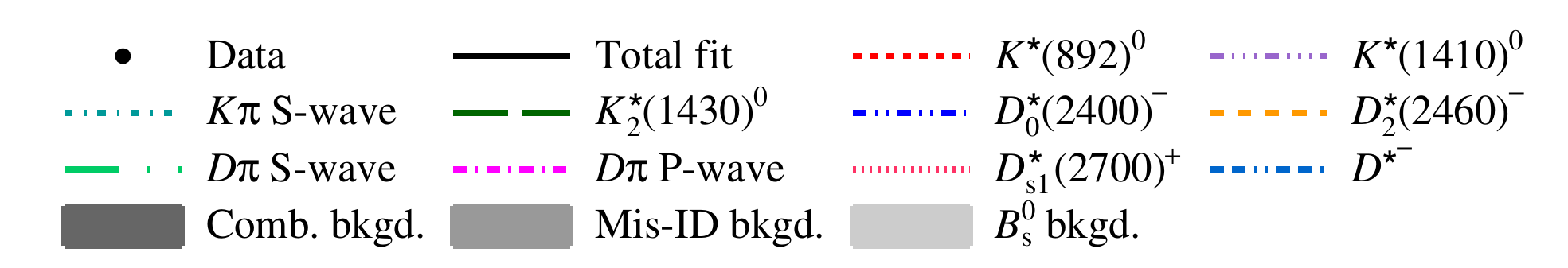}
  \caption{\small
    Projections of the $D\to\Kp\Km$ and $\pip\pim$ samples and the fit result onto (a,b)~$m(\D\pimp)$, (c,d)~$m(\Kpm\pimp)$ and (e,f)~$m(\D\Kpm)$ for (a,c,e)~\Bzb and (b,d,f)~\Bz candidates.
    The data and the fit results in each NN output bin have been weighted according to ${\cal S}/({\cal S}+{\cal B})$ and combined.
    The components are described in the legend.
  }
  \label{fig:CPDP-proj}
\end{figure}

The DPs for candidates in the \B candidate mass signal region in the $\D\to\Kp\Km$ and $\pip\pim$ samples are shown separately for \Bzb and \Bz candidates in Fig.~\ref{fig:CPDPs}.
Projections of the fit results onto $m(D\pi)$, $m(K\pi)$ and $m(DK)$ for the $D\to\Kp\Km$ and $\pip\pim$ samples are shown separately for $\Bzb$ and $\Bz$ candidates in Fig.~\ref{fig:CPDP-proj}.
No significant \CP violation effect is seen.

The results, with statistical uncertainties only, for the complex coefficients $c_j$ are given in Table~\ref{tab:BdFit1}.
Due to the changes in the selection requirements, the overlap between the $D\to\Kp\pim$ sample and the dataset used in Ref.~\cite{LHCb-PAPER-2015-017} is only around $60\,\%$, and the results are found to be consistent.

\begin{table}[!tb]
  \centering
  \caption{\small
      Results for the complex coefficients $c_j$ from the fit to data.
      Uncertainties are statistical only.
      All reported quantities are unconstrained in the fit, except that the $D^{*}_{2}(2460)^{-}$ component is fixed as a reference amplitude, and the magnitude of the $D^{*}_{s1}(2700)^{+}$ component is constrained.
      The $\Kp\pim$ S-wave is the coherent sum of the $\Kstarzero(1430)^{0}$ and the nonresonant $K\pi$ S-wave component~\cite{lass}.
  }
  \label{tab:BdFit1}
  \begin{tabular}{lcc}
    \hline
    Resonance & Real part & Imaginary part \\
    \hline \\ [-2.3ex]
    $\Kstar(892)^{0}$        & $          -0.07 \pm 0.10$ & $          -1.19 \pm 0.04$\\
    $\Kstar(1410)^{0}$       & $\phantom{-}0.16 \pm 0.04$ & $\phantom{-}0.21 \pm 0.06$\\
    $\Kstarzero(1430)^{0}$   & $\phantom{-}0.40 \pm 0.08$ & $\phantom{-}0.67 \pm 0.06$\\
    Nonresonant $K\pi$ S-wave & $\phantom{-}0.37 \pm 0.07$ & $\phantom{-}0.69 \pm 0.07$\\
    $\Kstartwo(1430)^{0}$   & $          -0.01 \pm 0.06$ & $          -0.48 \pm 0.04$\\
    $D^{*}_{0}(2400)^{-}$     & $          -1.10 \pm 0.05$ & $          -0.18 \pm 0.07$\\
    $D^{*}_{2}(2460)^{-}$     & $\phantom{-}1.00$          & $\phantom{-}0.00$         \\
    \hline
    Nonresonant $D\pi$ S-wave & $          -0.44 \pm 0.06$ & $\phantom{-}0.02 \pm 0.07$\\
    Nonresonant $D\pi$ P-wave & $          -0.61 \pm 0.05$ & $          -0.08 \pm 0.06$\\
    \hline
    $D^{*}_{s1}(2700)^{+}$    & $\phantom{-}0.57 \pm 0.05$ & $          -0.09 \pm 0.19$\\
    \hline
  \end{tabular}
\end{table}

\begin{figure}[!tb]
  \centering
  \includegraphics[width=0.55\textwidth]{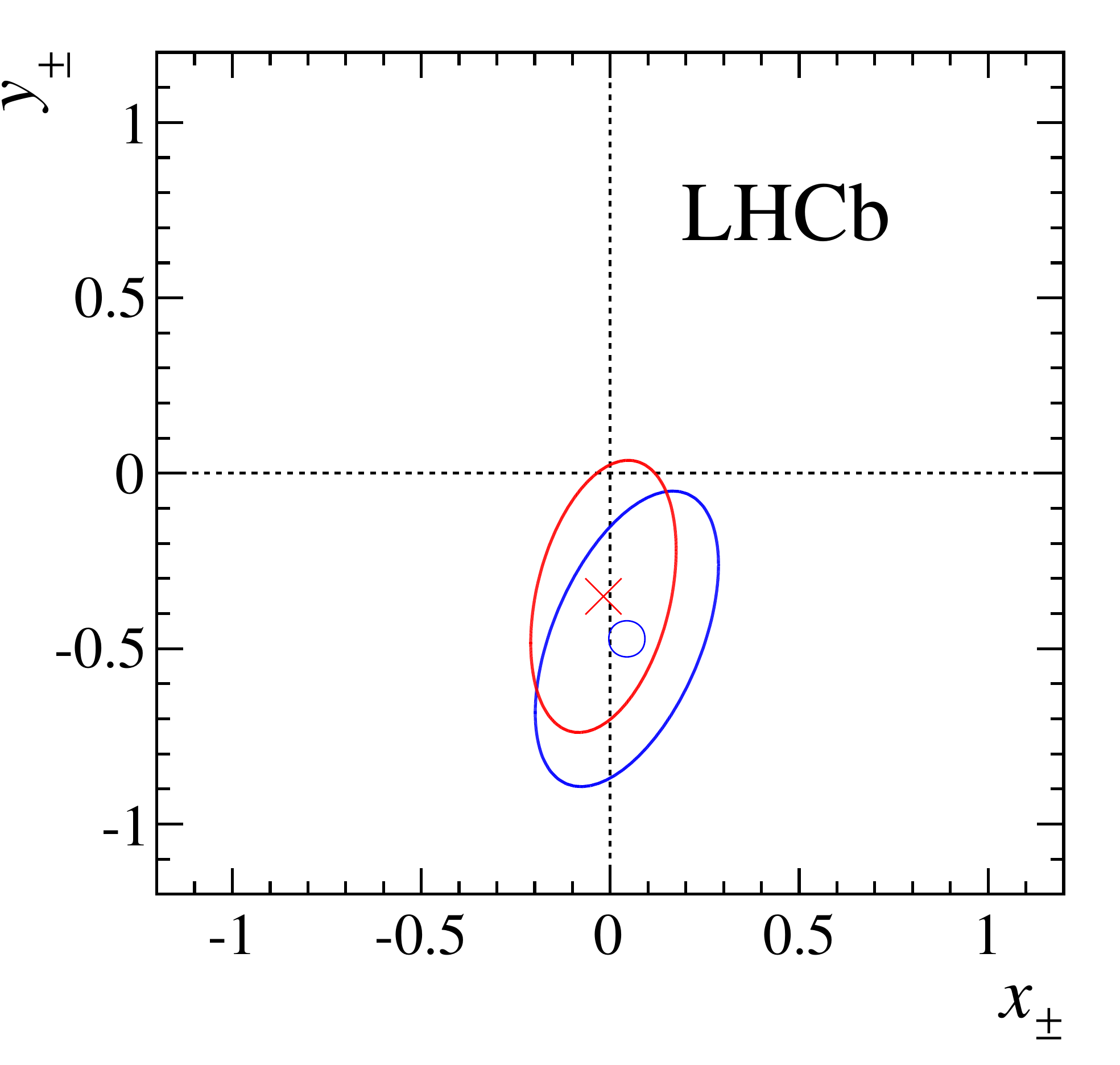}
  \caption{\small
    Contours at $68\,\%$ CL for the (blue) $(x_+,y_+)$ and (red) $(x_-,y_-)$ parameters associated with the $\Bz \to D\Kstar(892)^0$ decay, with statistical uncertainties only. The central values are marked by a circle and a cross, respectively.
  }
  \label{fig:xy-contours}
\end{figure}

The results for the \CP violation parameters associated with the $\Bz \to D\Kstar(892)^0$ decay are
\begin{equation*}
  \begin{array}{c}
    x_{+} = \phantom{-}0.04 \pm 0.16 \pm 0.11 \, , \qquad
    y_{+} = -0.47 \pm 0.28 \pm 0.22 \, ,\\
    x_{-} = -0.02 \pm 0.13 \pm 0.14 \, , \qquad
    y_{-} = -0.35 \pm 0.26 \pm 0.41 \, ,
  \end{array}
\end{equation*}
where the uncertainties are statistical and systematic.
The statistical and systematic correlation matrices are given in Table~\ref{tab:correlations}.
The results for $(x_+,y_+)$ and $(x_-,y_-)$ are shown as contours in Fig.~\ref{fig:xy-contours}.

\begin{table}[!tb]
  \centering
  \caption{\small
    Correlation matrices associated to the (left) statistical and (right) systematic uncertainties of the \CP violation parameters associated with the $\Bz \to D\Kstar(892)^0$ decay.
  }
  \label{tab:correlations}
  \begin{tabular}{c|cccc}
    \hline
    & $x_-$ & $y_-$ & $x_+$ & $y_+$ \\
    \hline
    $x_-$ & 1.00 &      &      &      \\
    $y_-$ & 0.34 & 1.00 &      &      \\
    $x_+$ & 0.10 & 0.05 & 1.00 &      \\
    $y_+$ & 0.13 & 0.15 & 0.50 & 1.00 \\
    \hline
  \end{tabular}
  \qquad
  \begin{tabular}{c|cccc}
  \hline
  & $x_-$ & $y_-$ & $x_+$ & $y_+$ \\
  \hline
  $x_-$ & 1.00 &      &      &      \\
  $y_-$ & 0.87 & 1.00 &      &      \\
  $x_+$ & 0.25 & 0.29 & 1.00 &      \\
  $y_+$ & 0.37 & 0.41 & 0.73 & 1.00 \\
  \hline
  \end{tabular}
\end{table}

\begin{figure}[!htb]
\setlength{\unitlength}{\textwidth}
  \centering
  \begin{picture}(0.9,0.7)
    \put(0.000,0.35){
      \includegraphics[width=0.45\textwidth]{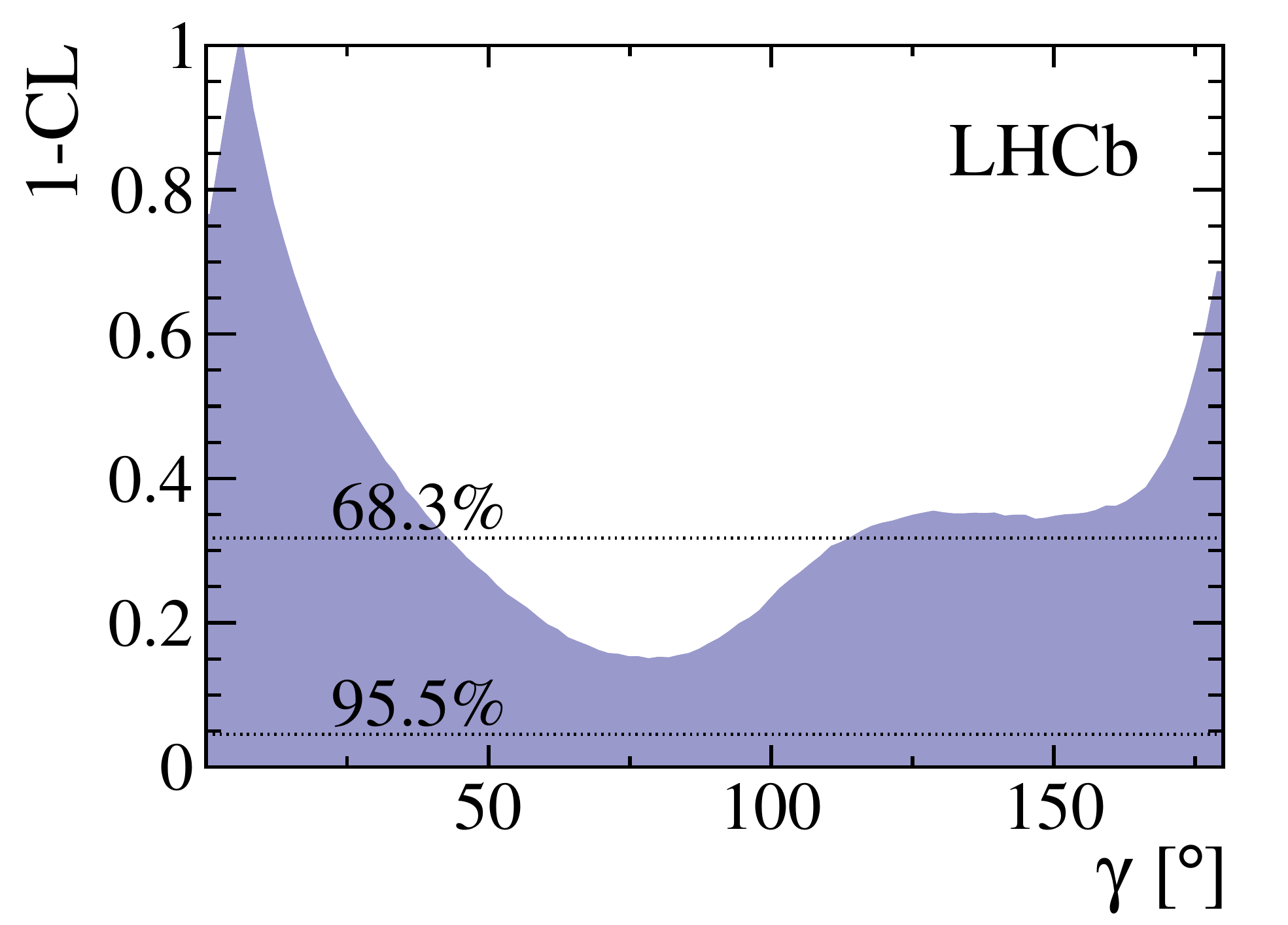}
    }
    \put(0.450,0.35){
      \includegraphics[width=0.45\textwidth]{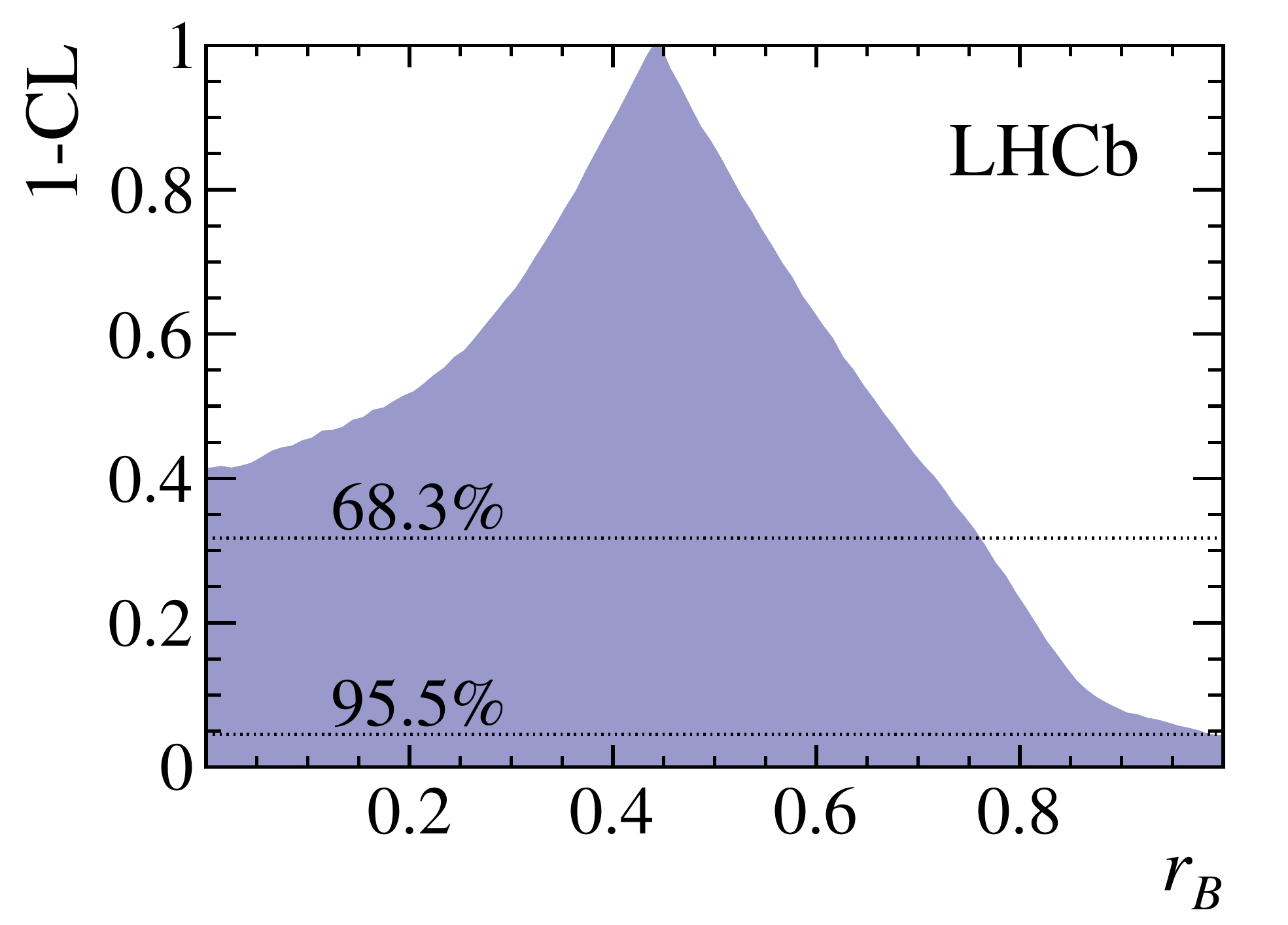}
    }
    \put(0.225,0.0){
      \includegraphics[width=0.45\textwidth]{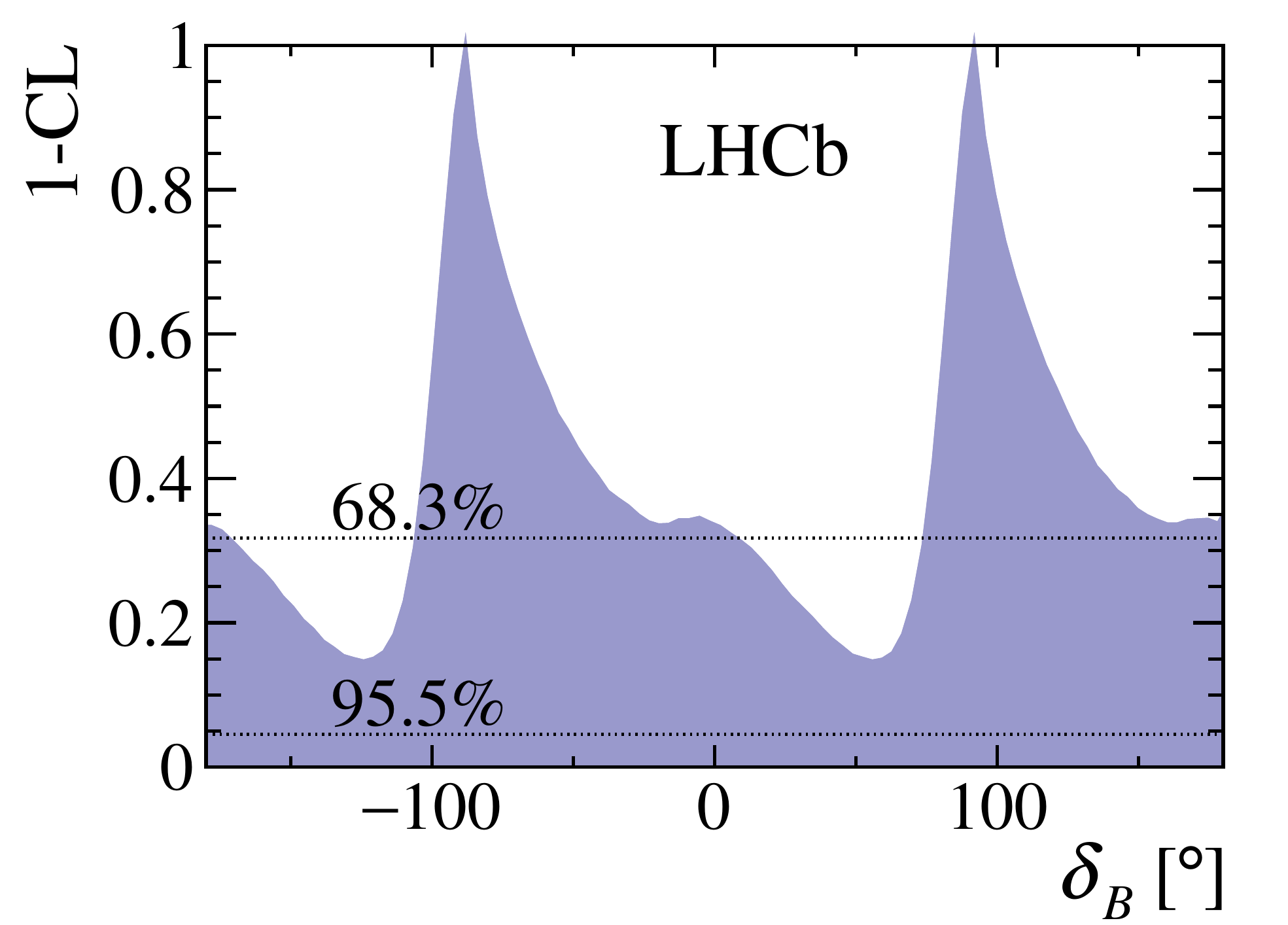}
    }
    \put( 0.375,0.57){(a)}
    \put( 0.825,0.57){(b)}
    \put( 0.505,0.225){(c)}
  \end{picture}
  \caption{\small
    Results of likelihood scans for (a) $\gamma$, (b) $r_B$ and (c) $\delta_B$.
  }
  \label{fig:1dgcscans}
\end{figure}

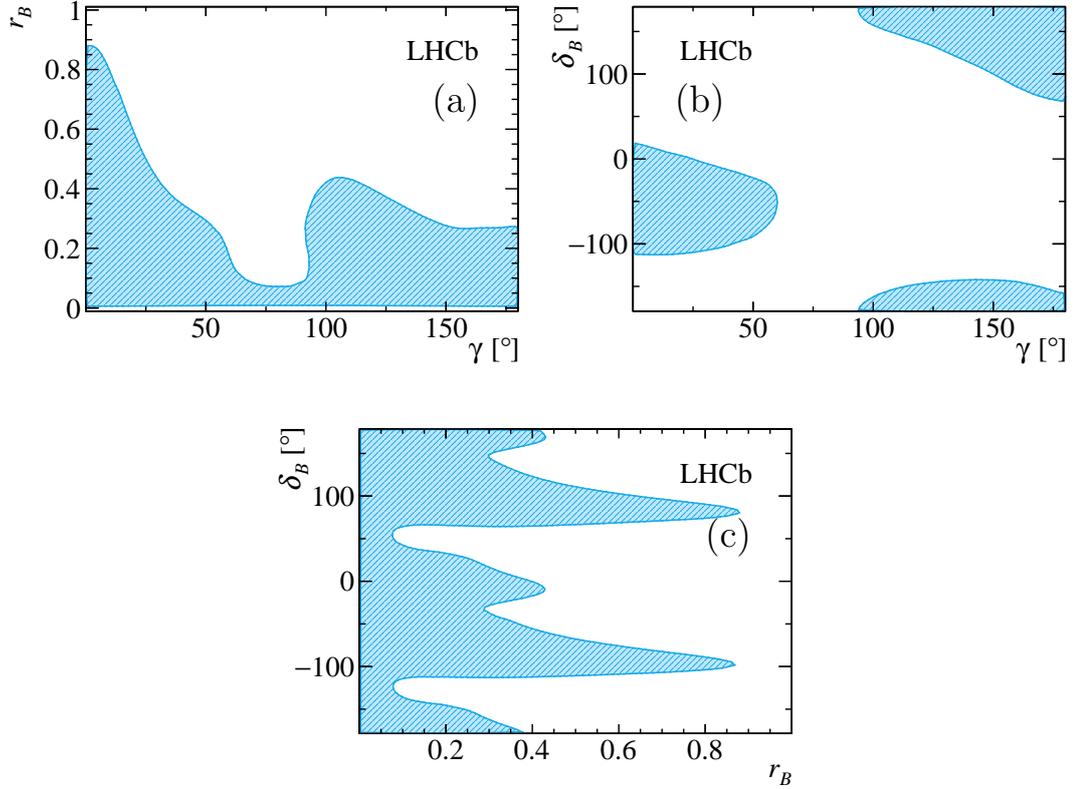
\begin{figure}[!tb]
\setlength{\unitlength}{\textwidth}
  \centering
  \begin{picture}(0.9,0.7)
    \put(0.000,0.35){
   	\includegraphics[width=0.45\textwidth]{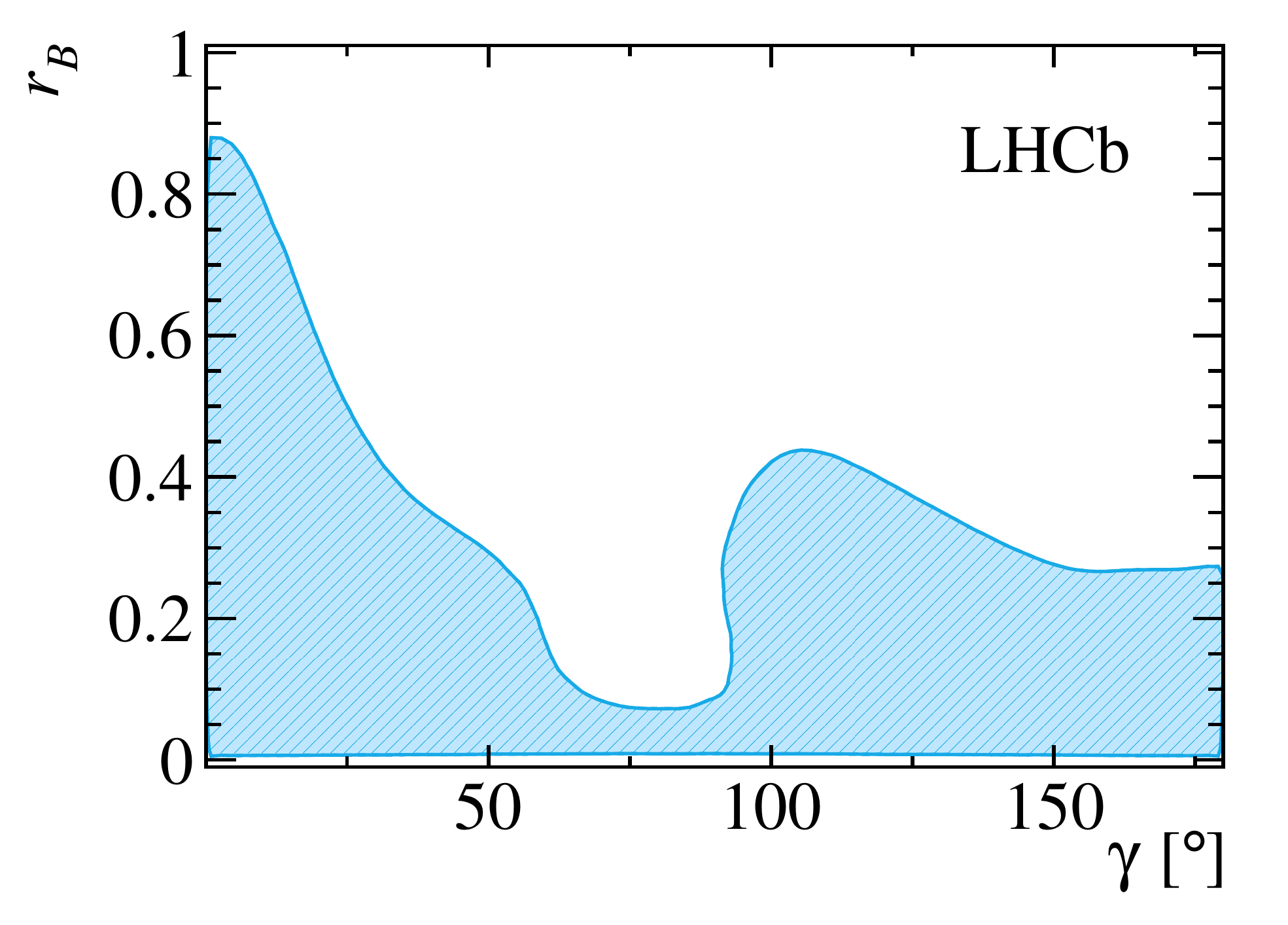}
    }
    \put(0.450,0.35){
    	\includegraphics[width=0.45\textwidth]{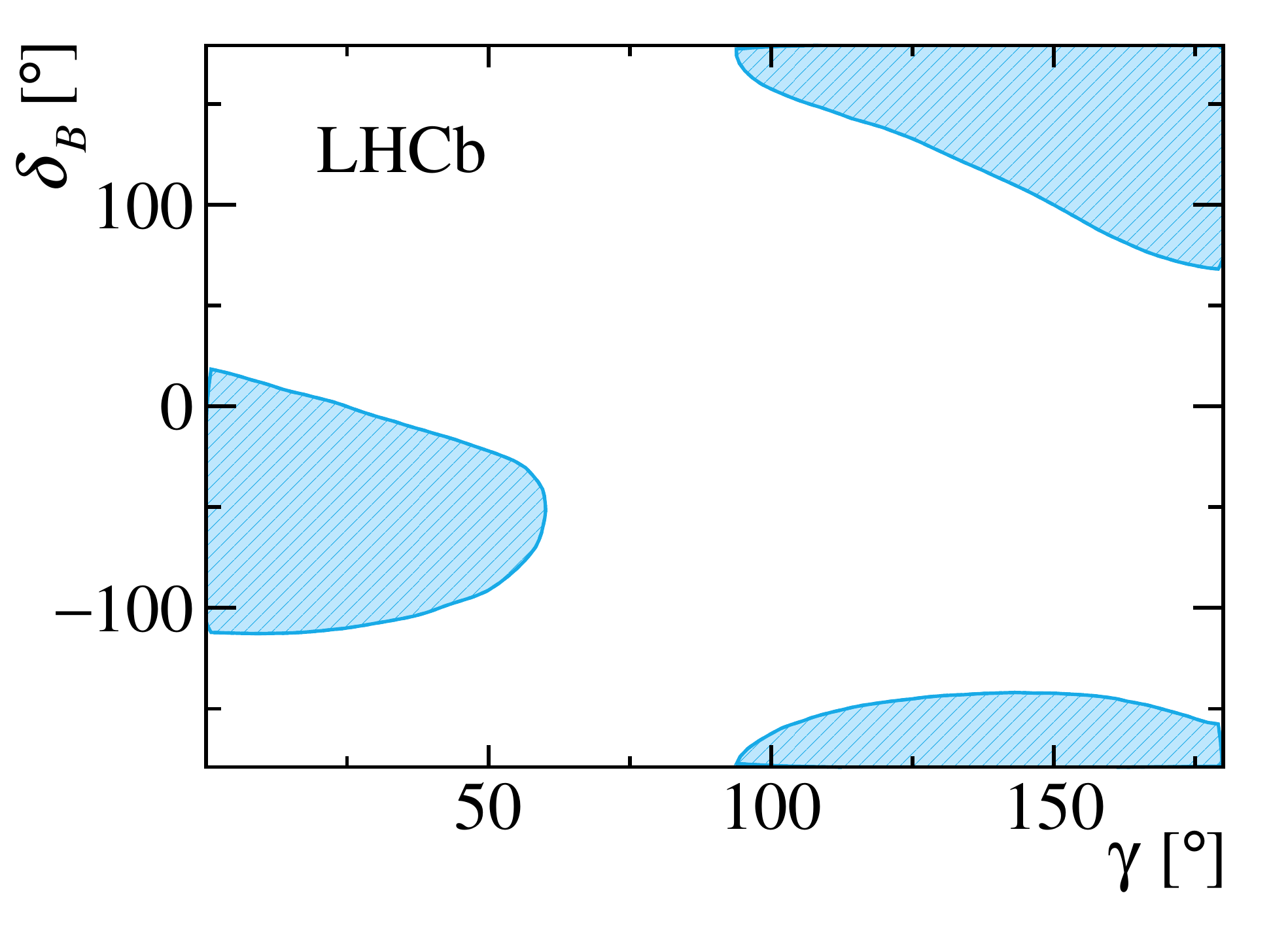}
    }
    \put(0.225,0.00){
    	\includegraphics[width=0.45\textwidth]{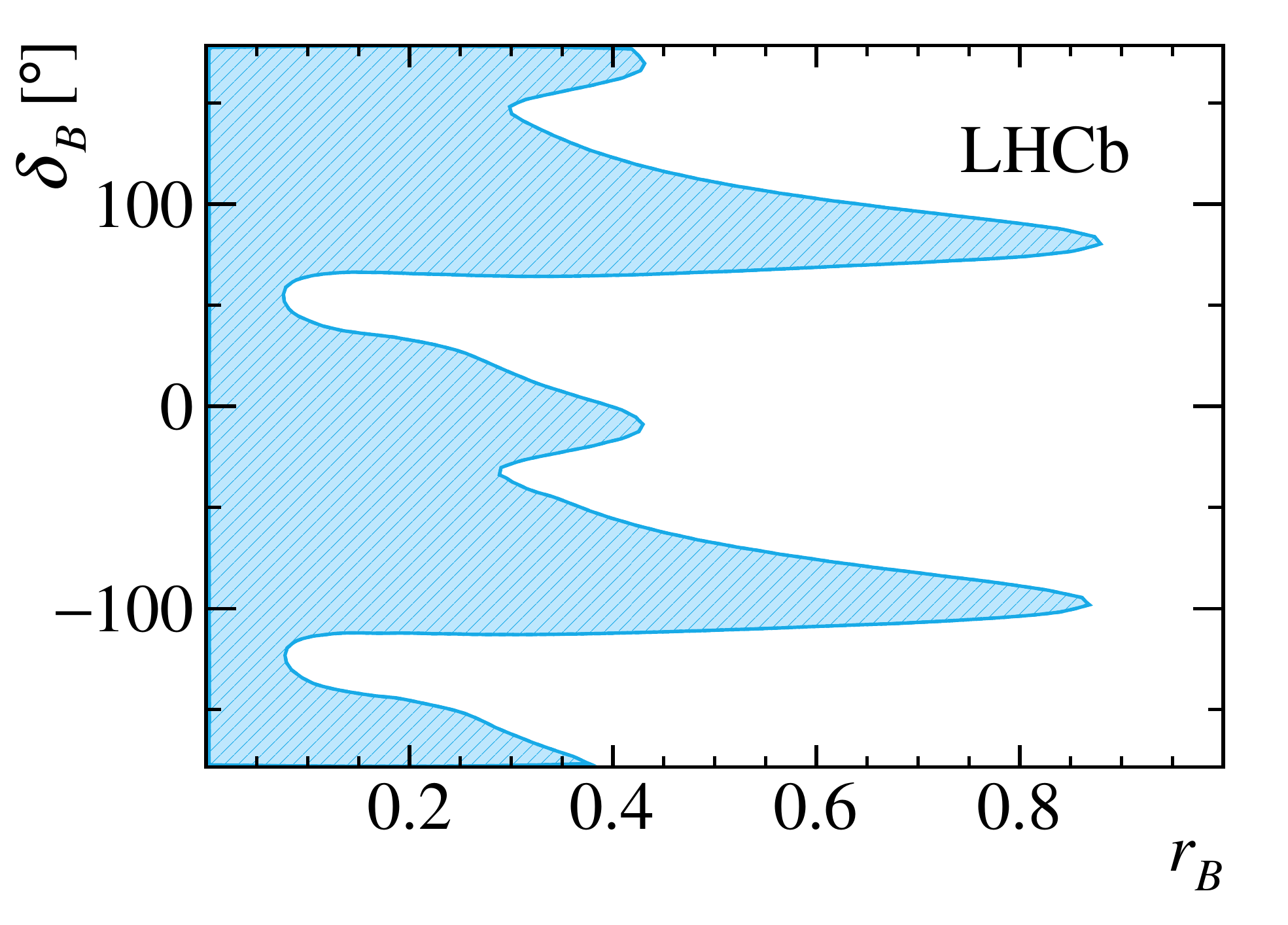}
    }
    \put( 0.365,0.57){\large (a)}
    \put( 0.565,0.57){\large (b)}
    \put( 0.590,0.21){\large (c)}
  \end{picture}
  \caption{\small
    Confidence level contours for (a) $\gamma$ and $r_B$, (b) $\gamma$ and $\delta_B$ and (c) $r_B$ and $\delta_B$.
    The shaded regions are allowed at $68\,\%$ CL.
  }
  \label{fig:2dgcscans}
\end{figure}

The {\tt GammaCombo} package~\cite{GammaCombo} is used to evaluate constraints from these results on $\gamma$ and the hadronic parameters $r_B$ and $\delta_B$ associated with the $\Bz \to D\Kstar(892)^0$ decay.
A frequentist treatment referred to as the ``plug-in'' method, described in
Refs.~\cite{LHCb-PAPER-2013-020,LHCb-CONF-2014-004,LHCb-CONF-2016-001,plugin}, is used.
Figure~\ref{fig:1dgcscans} shows the results of likelihood scans for $\gamma$, $r_B$ and $\delta_B$.
Figure~\ref{fig:2dgcscans} shows the two-dimensional $68\,\%$ confidence level for each pair of observables from $\gamma$, $r_B$ and $\delta_B$.
No value of $\gamma$ is excluded at 95\,\% confidence level (CL); the world-average value for $\gamma$~\cite{Charles:2004jd,Bona:2005vz} has a CL of 0.85.

The $\Bz \to D\Kstar(892)^0$ decay can also be used to determine parameters sensitive to $\gamma$ with a quasi-two-body approach, as has been done with $D \to \Kp\Km$, $\pip\pim$~\cite{LHCb-PAPER-2014-028}, $\Kpm\pimp$, $\Kpm\pimp\piz$, $\Kpm\pimp\pip\pim$~\cite{LHCb-PAPER-2014-028,Aubert:2009ao,Negishi:2012uxa} and $D \to \KS\pip\pim$ decays~\cite{Aubert:2008yn,Negishi:2015vqa,LHCb-PAPER-2016-006,LHCb-PAPER-2016-007}.
In the quasi-two-body analysis, the results depend on the effective hadronic parameters $\kappa$, $\bar{r}_B$ and $\bar{\delta}_B$, which are, respectively, the coherence factor and the relative magnitude and strong phase of the $V_{ub}$ and $V_{cb}$ amplitudes averaged over the selected region of phase space~\cite{Gronau:2002mu}.
Precise definitions are given in the Appendix.
These parameters are calculated from the models for $V_{cb}$ and $V_{ub}$ amplitudes obtained from the fit for the $\Kstar(892)^0$ selection region $\left| m(\Kp\pim) - m_{\Kstar(892)^0} \right| < 50 \mevcc$ and $\left| \cos \theta_{\Kstarz} \right| > 0.4$, where $m_{\Kstar(892)^0}$ is the known value of the $\Kstar(892)^0$ mass~\cite{PDG2014} and $\theta_{\Kstarz}$ is the \Kstarz helicity angle, \ie\ the angle between the \Kp and $D$ directions in the $\Kp\pim$ rest frame.
To reduce correlations with the values for $r_B$ and $\delta_B$ determined from the DP analysis,
the quantities $\bar{R}_B = \bar{r}_B/r_{B}$ and $\Delta \bar{\delta}_B = \bar{\delta}_B - \delta_{B}$ are calculated.
The results are
\begin{equation*}
  \kappa = 0.958 \,^{+0.005}_{-0.010} \,^{+0.002}_{-0.045} \, , \quad
  \bar{R}_B = 1.02 \,^{+0.03}_{-0.01} \pm 0.06 \, , \quad
  \Delta \bar{\delta}_B = 0.02 \,^{+0.03}_{-0.02} \pm 0.11 \, ,
\end{equation*}
where the uncertainties are statistical and systematic and are evaluated as described in the Appendix.

In summary, a data sample corresponding to $3.0\invfb$ of $pp$ collisions collected with the LHCb detector has been used to measure, for the first time, parameters sensitive to the angle $\gamma$ from a Dalitz plot analysis of $\Bz \to D\Kp\pim$ decays.
No significant \CP violation effect is seen.
The results are consistent with, and supersede, the results for ${\cal A}_d^{KK,\pi\pi}$ and  ${\cal R}_d^{KK,\pi\pi}$ from Ref.~\cite{LHCb-PAPER-2014-028}.
Parameters that are needed to determine $\gamma$ from quasi-two-body analyses of $\Bz \to D\Kstar(892)^0$ decays are measured.
These results can be combined with current and future measurements with the $\Bz \to D\Kstar(892)^0$ channel to obtain stronger constraints on $\gamma$.

\section*{Acknowledgements}

\noindent We express our gratitude to our colleagues in the CERN
accelerator departments for the excellent performance of the LHC. We
thank the technical and administrative staff at the LHCb
institutes. We acknowledge support from CERN and from the national
agencies: CAPES, CNPq, FAPERJ and FINEP (Brazil); NSFC (China);
CNRS/IN2P3 (France); BMBF, DFG and MPG (Germany); INFN (Italy);
FOM and NWO (The Netherlands); MNiSW and NCN (Poland); MEN/IFA (Romania);
MinES and FANO (Russia); MinECo (Spain); SNSF and SER (Switzerland);
NASU (Ukraine); STFC (United Kingdom); NSF (USA).
We acknowledge the computing resources that are provided by CERN, IN2P3 (France), KIT and DESY (Germany), INFN (Italy), SURF (The Netherlands), PIC (Spain), GridPP (United Kingdom), RRCKI and Yandex LLC (Russia), CSCS (Switzerland), IFIN-HH (Romania), CBPF (Brazil), PL-GRID (Poland) and OSC (USA). We are indebted to the communities behind the multiple open
source software packages on which we depend.
Individual groups or members have received support from AvH Foundation (Germany),
EPLANET, Marie Sk\l{}odowska-Curie Actions and ERC (European Union),
Conseil G\'{e}n\'{e}ral de Haute-Savoie, Labex ENIGMASS and OCEVU,
R\'{e}gion Auvergne (France), RFBR and Yandex LLC (Russia), GVA, XuntaGal and GENCAT (Spain), The Royal Society, Royal Commission for the Exhibition of 1851 and the Leverhulme Trust (United Kingdom).

\appendix

\clearpage

\section*{Appendix: Quasi-two-body parameters}
\label{app:q2b}

In the quasi-two-body analyses of $\Bz \to D\Kstar(892)^0$ decays, the following parameters are defined~\cite{Gronau:2002mu}:
\begin{eqnarray}
  \kappa & = & \left| \frac{\int \left| A_{cb}(p)A_{ub}(p) \right| \exp \left[ i\delta(p)\right] {\rm d}p}{\sqrt{\int\left| A_{cb}(p) \right|^2 {\rm d}p \int \left| A_{ub}(p) \right|^2 {\rm d}p}} \right| \, , \label{eq:kappa} \\
  \bar{r}_B & = & \sqrt{\frac{\int \left| A_{ub}(p) \right|^2 {\rm d}p}{\int\left| A_{cb}(p) \right|^2 {\rm d}p}} \, , \label{eq:rbar} \\
  \bar{\delta}_B & = & \arg \left( \frac{\int \left| A_{cb}(p)A_{ub}(p) \right| \exp \left[ i\delta(p)\right] {\rm d}p}{\sqrt{\int\left| A_{cb}(p) \right|^2 {\rm d}p \int \left| A_{ub}(p) \right|^2 {\rm d}p}} \right) \, ,  \label{eq:deltabar}
\end{eqnarray}
where all the integrations are over the part of the phase space $p$ inside the used $\Kstar(892)^0$ selection window.
In these equations, $\left| A_{cb}(p) \right|$ and $\left| A_{ub}(p) \right|$ refer to the magnitudes of the total $V_{cb}$ and $V_{ub}$ amplitudes, and $\delta(p)$ is their relative strong phase.
In terms of the parameters used in this analysis,
\begin{eqnarray}
  \left| A_{cb}(p) \right| & = & \left| \sum_j c_j F_j(p) \right| \, , \label{eq:Acb} \\
  \left| A_{ub}(p) \right| & = & \left| \sum_j c_j r_{B,j} \exp\left[ i\delta_{B,j} \right] F_j(p) \right| \, , \label{eq:Aub} \\
  \delta(p) & = & \arg \left( \frac{\sum_j c_j r_{B,j} \exp\left[ i\delta_{B,j} \right] F_j(p)}{\sum_j c_j F_j(p)} \right) \, , \label{eq:delta}
\end{eqnarray}
where the $r_{B,j}$, $\delta_{B,j}$ values are allowed to differ for each $\Kp\pim$ resonance, and $r_{B,j} = 0$ for $D\pim$ resonances.
(The $r_B$, $\delta_B$ notation without the $j$ subscript is retained for the parameters associated with the $\Bz \to D\Kstar(892)^0$ decay.)
In the limit that there is no amplitude (either resonant or nonresonant) contributing within the $\Kstar(892)^0$ selection window other than those associated with the $\Bz\to D\Kstar(892)^0$ decay, one finds $\left| A_{ub}(p) \right| \longrightarrow r_B \left| A_{cb}(p) \right|$ and $\delta(p) \longrightarrow \delta_B$, and hence $\kappa \longrightarrow 1$, $\bar{r}_B \longrightarrow r_B$ and $\bar{\delta}_B \longrightarrow \delta_B$.
In order to reduce correlations between  $\bar{r}_B$ and $r_B$ and between $\bar{\delta}_B$ and $\delta_B$, it is convenient to introduce the parameters
\begin{eqnarray}
   \bar{R}_B & = & \frac{\bar{r}_B }{r_B} \, , \\
   \Delta \bar{\delta}_B & = & \bar{\delta}_B - \delta_B \, ,
\end{eqnarray}
which are obtained by replacing all $r_{B,j}$ by $r_{B,j}/r_B$ and all $\delta_{B,j}$ by $\delta_{B,j} - \delta_B$ in Eqs.~(\ref{eq:Acb})--(\ref{eq:delta}).

\begin{figure}[!tb]
  \centering
    \includegraphics[width=0.48\textwidth]{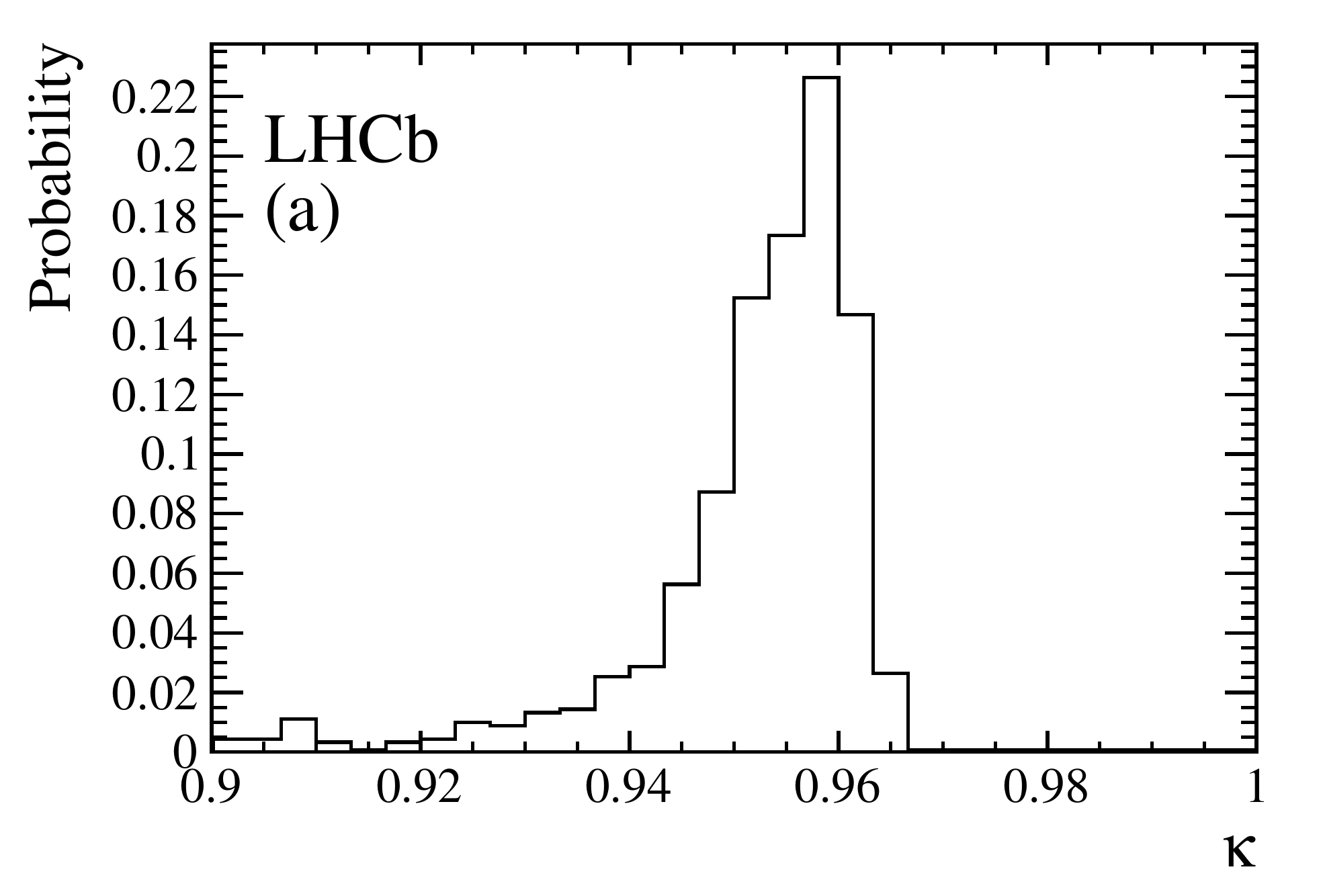}
    \includegraphics[width=0.48\textwidth]{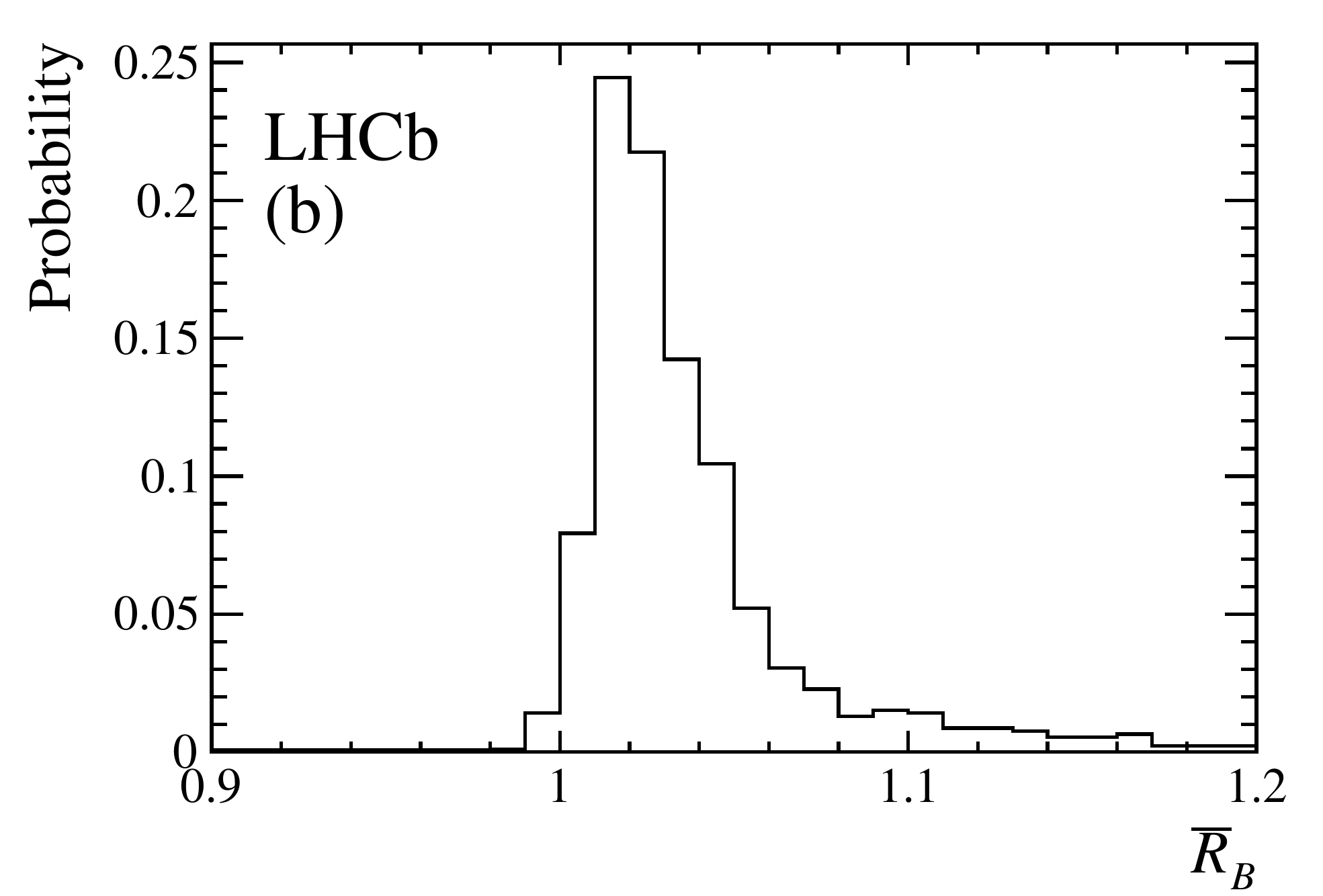}
    \includegraphics[width=0.48\textwidth]{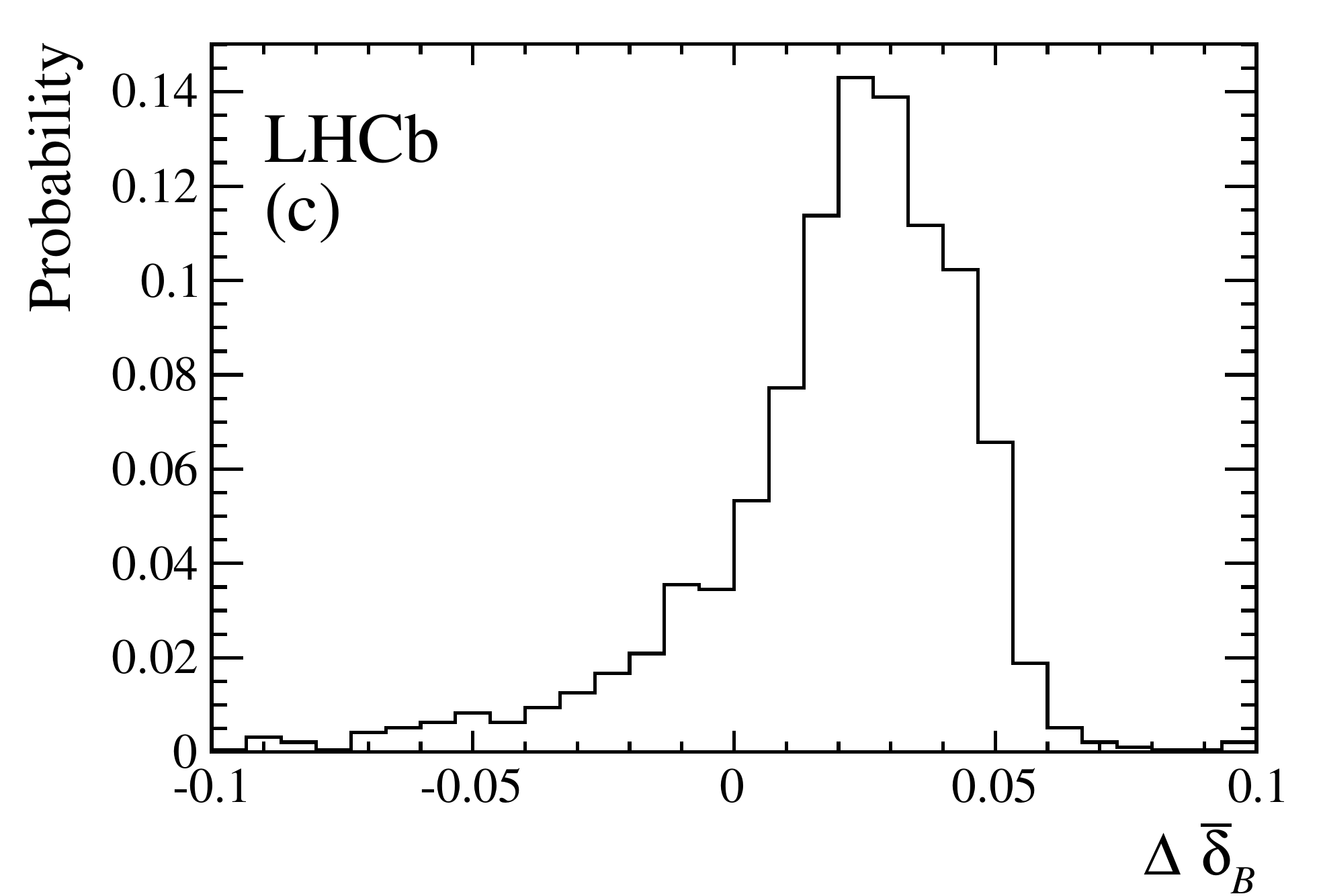}
  \caption{\small
    Distributions of (a) $\kappa$, (b) $\bar{R}_B$ and (c) $\Delta \bar{\delta}_B$, obtained as described in the text.
  }
  \label{fig:q2b-alt}
\end{figure}

These quantities are determined from the results of the Dalitz plot analysis.
An alternative fit is performed with $x_{\pm,\,j} + iy_{\pm,\,j}$, defined in Eq.~(\ref{eq:param1}), replaced by $r_{B,\,j} \exp \left[ i \left(\delta_{B,\,j} \pm \gamma\right) \right]$.
The results of this fit are consistent with the values for $\gamma$, $r_B$ and $\delta_B$ obtained from the fitted $x_\pm$ and $y_\pm$, and are used to evaluate $\left| A_{cb}(p) \right|$, $\left| A_{ub}(p) \right|$ and $\delta(p)$ at many points inside the selection window and thereby to determine $\kappa$, $\bar{R}_B$ and $\Delta \bar{\delta}_B$.
The procedure is repeated many times with both $V_{cb}$ and $V_{ub}$ amplitude model parameters varied within their statistical uncertainties from the fit, leading to the distributions shown in Fig.~\ref{fig:q2b-alt}.
Since the transformations from the fitted model parameters to the quasi-two-body parameters are highly non-linear, the reported central values correspond to the peak positions of these distributions, while positive and negative uncertainties are obtained by incrementally including the most probable values until $68\,\%$ of all entries are covered.

Sources of systematic uncertainty are accounted for by evaluating their effects on the quasi-two-body parameters.
The dominant sources are from the use of an alternative description of the $\Kp\pim$ S-wave, and from changing the treatment of \CP violation in the $D_{s1}^*(2700)^+$ component and the $\Kp\pim$ S-wave.
Most systematic uncertainties are symmetrised for consistency with the rest of the analysis, but asymmetric systematic uncertainties are reported on $\kappa$ since this quantity is $\le 1$ by definition.

\clearpage

\ifx\mcitethebibliography\mciteundefinedmacro
\PackageError{LHCb.bst}{mciteplus.sty has not been loaded}
{This bibstyle requires the use of the mciteplus package.}\fi
\providecommand{\href}[2]{#2}

\clearpage

\centerline{\large\bf LHCb collaboration}
\begin{flushleft}
\small
R.~Aaij$^{39}$,
C.~Abell\'{a}n~Beteta$^{41}$,
B.~Adeva$^{38}$,
M.~Adinolfi$^{47}$,
A.~Affolder$^{53}$,
Z.~Ajaltouni$^{5}$,
S.~Akar$^{6}$,
J.~Albrecht$^{10}$,
F.~Alessio$^{39}$,
M.~Alexander$^{52}$
% ...

{\footnotesize \it
% ...
$ ^{5}$Clermont Universit\'{e}, Universit\'{e} Blaise Pascal, CNRS/IN2P3, LPC, Clermont-Ferrand, France\\
$ ^{6}$CPPM, Aix-Marseille Universit\'{e}, CNRS/IN2P3, Marseille, France\\
% ...
$ ^{10}$Fakult\"{a}t Physik, Technische Universit\"{a}t Dortmund, Dortmund, Germany\\
% ...
$ ^{38}$Universidad de Santiago de Compostela, Santiago de Compostela, Spain\\
$ ^{39}$European Organization for Nuclear Research (CERN), Geneva, Switzerland\\
% ...
$ ^{41}$Physik-Institut, Universit\"{a}t Z\"{u}rich, Z\"{u}rich, Switzerland\\
% ...
$ ^{47}$H.H. Wills Physics Laboratory, University of Bristol, Bristol, United Kingdom\\
% ...
$ ^{52}$School of Physics and Astronomy, University of Glasgow, Glasgow, United Kingdom\\
$ ^{53}$Oliver Lodge Laboratory, University of Liverpool, Liverpool, United Kingdom\\
% ...
}
\end{flushleft}

\end{document}